\documentclass[nofootinbib,eqsecnum,tightenlines,superscriptaddress,11pt]{revtex4}

\usepackage{hyperref}
\usepackage{graphicx}
\usepackage{amsmath,amssymb,amsfonts,amsthm,stmaryrd,mathtools,bm}
\usepackage{mathrsfs}
\usepackage{color}
\usepackage{multirow,bigdelim}
\usepackage{hyperref}
\usepackage{dsfont}
\usepackage{booktabs}
\usepackage{float}
\allowdisplaybreaks[0]
\usepackage[dvipsnames]{xcolor}

\usepackage{pstricks}
\usepackage{tikz,tikz-3dplot}
\usepackage{pgfplots}
\usepackage{ulem}

\def\be#1\ee{\begin{align}#1\end{align}}

\def\ba{\begin{eqnarray}}
\def\ea{\end{eqnarray}}
\def\nn{\nonumber}
\def\q{\quad}

\begin{document}

\title{Lorentzian quantum cosmology goes simplicial}

\author{Bianca Dittrich}
\email{bdittrichATperimeterinstitute.ca} 
\affiliation{Perimeter Institute, 31 Caroline Street North, Waterloo, ON, N2L 2Y5, CAN}
\author{Steffen Gielen}     
\email{s.c.gielenATsheffield.ac.uk}
\affiliation{School of Mathematics and Statistics, University of Sheffield, Hicks Building, Hounsfield Road, Sheffield S3 7RH, United Kingdom}
\author{Susanne Schander}
\email{sschanderATperimeterinstitute.ca}
\affiliation{Perimeter Institute, 31 Caroline Street North, Waterloo, ON, N2L 2Y5, CAN}

\begin{abstract}
We employ the methods of discrete (Lorentzian) Regge calculus for analysing Lorentzian quantum cosmology models with a special focus on discrete analogues of the no-boundary proposal for the early universe. We use a simple 4-polytope, a subdivided 4-polytope and shells of discrete 3-spheres as triangulations to model a closed universe with cosmological constant, and examine the semiclassical path integral for these different choices. We find that the shells give good agreement with continuum results for small values of the scale factor and in particular for finer discretisations of the boundary 3-sphere, while the simple and subdivided 4-polytopes can only be compared with the continuum in certain regimes, and in particular are not able to capture a transition from Euclidean geometry with small scale factor to a large Lorentzian one. Finally, we consider a closed universe filled with dust particles and discretised by shells of 3-spheres. This model can approximate the continuum case quite well. Our results embed the no-boundary proposal in a discrete setting where it is possibly more naturally defined, and prepare for its discussion within the realm of spin foams.\end{abstract}

\maketitle

\vspace{0.5cm}
\tableofcontents

\vspace{1cm}
\section{Introduction}

Many approaches to quantum gravity are based on defining a path integral, but there are quite a number of different proposals for the precise form of this definition. The first systematic approach was the continuum Euclidean quantum gravity programme \cite{EuclideanQG} which, in analogy to what is done for most usual quantum field theories, aimed to define a sum over {\it Riemannian} metrics on a manifold of given topology, perhaps supplemented with a sum over topologies. The full continuum path integral in this approach is badly ill-defined due to the unboundedness of the action from below \cite{unbounded}, but some progress could be made at the semiclassical level and in particular for cosmological models, where initially one only integrates over highly symmetric (usually homogeneous and isotropic) configurations. Perhaps the culmination of this line of research was the no-boundary proposal of Hartle and Hawking \cite{HartleHawk} which defined a path integral for cosmological models for gravity with an inflaton field based on the hypothesis that there is no past boundary to the spacetime manifold. Rather than ``beginning'' in the classical Big Bang singularity the geometries appearing in the path integral can then be pictured as originating in a Riemannian four-sphere. While rooted in the Euclidean approach, the path integral is then usually defined by complex contour integration in order to identify the leading saddle point contributions, which cannot be characterised as purely Lorentzian or Riemannian \cite{HalliwellLouko}. These saddle point contributions will, inasmuch as they contain a Riemannian part, lead to a semiclassical exponential factor that either enhances or suppresses these configurations. The question of which of these saddle points should be included is the focus of ongoing debate \cite{Debate}. Given that the Euclidean path integral is not well-defined, one can also start from an initially Lorentzian definition of the path integral for quantum cosmology \cite{BrownMartinez,Feldbrugge}.

Other approaches are often based on discrete geometries in order to overcome the difficulties of the continuum setting. Here, a crucial development was the definition of Regge calculus \cite{Regge} as a discrete version of general relativity: the usual continuum manifold is replaced by a simplicial complex made up of discrete building blocks (simplices) glued together, in four dimensions, along their shared boundary tetrahedra. Curvature then arises by the appearance of a ``deficit angle'' at lower-dimensional subspaces. The Regge action defines the analogue of the Einstein--Hilbert action in this setting, and can in principle be used as a starting point for a gravitational path integral. There are a number of rather nontrivial technical issues when doing this, such as defining a measure, deciding how to implement triangle inequalities, or choosing a Riemannian or Lorentzian approach. In this work we will particularly focus on the last question, and study a {\it Lorentzian} setting for Regge calculus based on the Lorentzian notion of dihedral angles detailed in \cite{Sorkin2019}. We will do so in the context of homogeneous, isotropic geometries, studying triangulations that approximate a closed Friedmann--Lema\^{i}tre--Robertson--Walker (FLRW) universe, the setting of the usual no-boundary path integral. In this sense, we aim to provide an extension of the no-boundary proposal into Lorentzian Regge calculus. We will encounter a number of peculiarities specific to the Lorentzian setting: we will observe exponential enhancement or suppression of particular configurations when no real Lorentzian solution exists and the integration contour is therefore deformed into the complex plane.  The action evaluated along such configurations picks up an imaginary part that can be of either sign. Understanding the r\^{o}le of such configurations is essential for understanding the proper definition of a Lorentzian path integral for quantum gravity.  

Another feature peculiar to Lorentzian Regge calculus is the appearance of configurations with irregular causal structure \cite{Sorkin2019}. Such configurations are associated with a complex action (even for real variables), which can also lead to an exponential enhancement or suppression of such configurations. Not much is known about the prevalence of such configurations.

Our study of cosmological ``minisuperspace'' models in Regge calculus has several motivations. The first is to understand better the dynamics of Regge calculus within Lorentzian geometry, which have not been explored very much so far. The second is to embed the no-boundary proposal into a discrete setting where it may be more naturally defined -- after all, we know that the continuum path integral on which the original no-boundary proposal is based is not well-defined. The issues and technical discussions surrounding the no-boundary proposal can then be studied from a different angle. Having simplices as regulators may also provide an easier route towards the inclusion of inhomogeneities later on. Perhaps most importantly, Regge calculus is related to the spin foam approach to quantum gravity \cite{Perez} in which one uses a path integral, based on simplicial building blocks, whose dynamics reduce in certain limits to those of Regge calculus. In spin foam models too, one can ask what configurations are included in the path integral, and what the relative contributions are of Lorentzian and Riemannian configurations. Our work here, working first within the simpler setting of homogeneous cosmology, can be seen as a preparation for extending this discussion from Regge calculus into the spin foam setting.

Our work follows some previous studies on cosmological models in Regge calculus, in particular \cite{Collins1973,Hartle1985,Liu2015}. In contrast to \cite{Hartle1985}, which advocated a Regge version of Euclidean quantum gravity, here we concentrate on the Lorentzian version of the path integral.  The works \cite{Collins1973,Brewin,Liu2015,Tsuda} explore the classical dynamics of Regge calculus and thus do not discuss important issues for the path integral, in particular for the construction of an analogue of the no-boundary proposal.

We will see that even the simplest classical Regge models require numerical tools for their analysis. The Regge path integral, even if reduced to minisuperspace, involves an infinite integration range, preventing a straightforward numerical evaluation.
In this work we will therefore rely on semiclassical aspects and leave the study of the fully non-perturbative path integral for future work. One framework in which such  non-perturbative path integrals for Lorentzian quantum cosmology can be discussed are the recently introduced effective spin foams \cite{EffSF1,EffSF2,EffSF3}. Effective spin foams work more directly with the Regge action than usual spin foams, which results in a much enhanced numerical efficiency.  These improvements would then allow a treatment of minisuperspace path integrals as discussed here. One nevertheless needs to be aware of the numerical resources required for such a task. We will therefore discuss a range of discrete models from very simple ones  to more involved ones. For each of these models we identify to which extent they mirror continuum physics, and  provide useful models for a non-perturbative quantisation. 

We start with a discussion of the continuum minisuperspace path integral in Section \ref{Sec:Cont}. We provide a short overview of (Lorentzian) Regge calculus and the associated path integral in Section \ref{Sec:Regge}. This allows us to discuss our first discrete models for the beginning of the universe, based on simple discretisations of the 4-ball in Section \ref{Sec:Ball}. Here we will discuss a closed universe with a positive cosmological constant, which starts in a Euclidean phase and then transitions to a Lorentzian geometry. 
We will see that the discretisations employed in Section \ref{Sec:Ball} can only model the early Euclidean phase of the universe. We therefore introduce in Section \ref{Sec:Shells} a discretisation based on spherical shells (of topology $S^3\times[0,1]$) which can model the evolution of the universe also at later times. The dynamics simplify very much in the limit of infinitesimally small time steps, that is infinitesimally thin shells, and can in this limit more directly be compared with the continuum FLRW dynamics. Whereas Sections \ref{Sec:Ball} and \ref{Sec:Shells} consider a closed universe with positive cosmological constant, in Section \ref{Sec:Dust} we will consider a (closed) universe filled with dust. We close with a discussion and outlook in Section \ref{Sec:Discuss}.

\section{Continuum minisuperspace path integral} \label{Sec:Cont}

One of the main goals of this paper is to compare the quantum properties of cosmological models based on the simplicial formalism of Regge calculus to their continuum analogues. To facilitate this comparison, in this section we briefly review the usual discussion of path integral methods for minisuperspace models in continuum general relativity, in particular in the context of the no-boundary proposal of Hartle and Hawking \cite{HartleHawk}.

The models we are interested in are based on the dynamics of a closed $k=1$ homogeneous, isotropic universe, whose metric can be written in hyperspherical coordinates as
\ba
{\rm d}s^2 &=& -N^2(t){\rm d}t^2+a^2(t)\left({\rm d}\chi^2+\sin^2\chi({\rm d}\theta^2+\sin^2\theta\,{\rm d}\varphi^2)\right)
\label{flrwmetric}
\ea
where the coordinates $\chi$ and $\theta$ run from 0 to $\pi$ and $\varphi$ runs from 0 to $2\pi$. $N(t)$ is the lapse function and $a(t)$ is the cosmological scale factor, which here corresponds to the radius of the three-sphere representing the universe at a given value of $t$.

Thinking of such a universe as evolving from some initial time $t_0$ to some final time $t_1$, the boundary geometries at $t_0$ and $t_1$ are 3-spheres of radius $a(t_0)$ and $a(t_1)$, with three-dimensional volume $V(t_{0,1})=2\pi^2 a(t_{0,1})^3$ and intrinsic (three-dimensional) curvature ${\bf R}_3(t_{0,1})=6/a(t_{0,1})^2$. In Section \ref{Sec:Cells} we will use these expressions when matching discrete and continuum descriptions.

The four-dimensional Ricci scalar corresponding to the metric (\ref{flrwmetric}) is given by
\ba
{\bf R}_4 = {\bf R}_3 + 6\left(\frac{\ddot{a}}{N^2a}+\frac{\dot{a}^2}{N^2 a^2}-\frac{\dot{a}\dot{N}}{N^3a}\right)
\ea
where $\dot{}$ denotes derivatives with respect to $t$. 

We can then calculate the gravitational Einstein--Hilbert action for this universe, finding
\ba
S_{{\rm bulk}} & := & \frac{1}{16\pi G}\int {\rm d}^4 x\,\sqrt{-g}\,\left({\bf R}_4-2\Lambda\right)\nonumber
\\& = & \frac{3\pi}{4G}\int_{t_0}^{t_1} {\rm d}t\left(Na+\frac{a\dot{a}^2}{N}-\frac{a^2 \dot{a}\dot{N}}{N^2}+\frac{a^2\ddot{a}}{N}-\frac{\Lambda}{3}Na^3\right)
\ea
plus a boundary term which will determine the type of boundary value problem one is studying (see, e.g., \cite{DiTucci} for a recent discussion of possible choices). The most common choice is to demand a boundary value problem for which the metric, but not its time derivative, is held fixed at the initial and final times. The required boundary term is then the Gibbons--Hawking--York term \cite{GHY}
\ba
S_{{\rm GHY}} & := & -\frac{1}{8\pi G}\int {\rm d}^3 x\sqrt{q}K =  \left[-\frac{3\pi}{4 G}\frac{a^2\dot{a}}{N}\right]^{t_1}_{t_0} = -\frac{3\pi}{4 G} \int_{t_0}^{t_1} {\rm d}t \left(2\frac{a\dot{a}^2}{N}+\frac{a^2\ddot{a}}{N}-\frac{a^2\dot{a}\dot{N}}{N^2}\right)
\ea
where $K=\frac{3\dot{a}}{Na}$ is the trace of the extrinsic curvature evaluated on the initial and final hypersurfaces.

The total action with these boundary conditions is then
\ba
S_{{\rm GR}}:=S_{{\rm bulk}}+S_{{\rm GHY}} = \frac{3\pi}{4G}\int_{t_0}^{t_1} {\rm d}t \left(Na-\frac{a\dot{a}^2}{N}-\frac{\Lambda}{3}Na^3\right)
\label{GRaction}
\ea
and one can derive the equations of motion (or Friedmann equations)
\ba
\frac{1}{a^2}+\frac{\dot{a}^2}{N^2a^2} & = & \frac{\Lambda}{3}\,,
\label{friedma1}
\\\frac{1}{a^2}+\frac{\dot{a}^2}{N^2a^2}+2\frac{\ddot{a}}{N^2a}-2\frac{\dot{a}\dot{N}}{N^3a} & = & \Lambda \,.
\label{friedma2}
\ea
One would now like to formulate the path integral for the corresponding quantum problem of going from a geometry characterised by scale factor $a(t_0)$ to one with scale factor $a(t_1)$; of particular interest is the ``no-boundary'' case $a(t_0)=0$. One has to deal with the gauge symmetry under reparametrisations of the time coordinate $t$. Following Halliwell \cite{Halliwell}, one way of defining the path integral is to gauge-fix reparametrisation invariance but ensure that the resulting path integral does not depend on this gauge choice by applying the Batalin--Fradkin--Vilkovisky (BFV) formalism \cite{BFV}. For the gauge-fixing choice $\dot{N}=0$ one obtains a path integral
\ba\label{PICont}
\mathcal{G}(a(t_1)|a(t_0)) & := & \int \mathcal{D}p \,\mathcal{D}a\,\mathcal{D}\Pi\,\mathcal{D}N\,\mathcal{D}\rho\,\mathcal{D}\bar{c}\,\mathcal{D}\bar\rho\,\mathcal{D}c\,\exp(\imath S_T)
\ea
where the notation suggests that this defines a two-point function or propagator, and the total action $S_T$ is
\ba
S_T & := & \int_{t_0}^{t_1} {\rm d}t\left(p \dot{a} - N \mathcal{H} + \Pi \dot{N} +\bar\rho\dot{c} +\bar{c}\dot\rho-\bar\rho\rho\right)
\label{totalaction}
\ea
where $\Pi$ can be seen as a Lagrange multiplier enforcing the gauge-fixing condition $\dot{N}=0$ or  as a conjugate momentum to $N$, and $\rho,\bar{c},\bar\rho$ and $c$ are ghost fields added to ensure that the action has a global Becchi--Rouet--Stora (BRS) symmetry under the transformations
\be
\delta a=\alpha c\frac{\partial \mathcal{H} }{\partial p}\,,\;\;\delta p=-\alpha c\frac{\partial \mathcal{H} }{\partial a}\,,\;\;\delta N= \alpha\rho\,,\;\;\delta\bar{c}=-\alpha\Pi\,,\;\;\delta\bar\rho=-\alpha \mathcal{H}
\label{brstransf}
\ee
where $\alpha$ is an anticommuting constant transformation parameter; the path integral then does not depend on the gauge-fixing condition used. Finally the Hamiltonian constraint $\mathcal{H}$ appearing in (\ref{totalaction}) and (\ref{brstransf}) is
\ba
\mathcal{H}  &:=& -\frac{Gp^2}{3\pi a}+\frac{3\pi}{4G}\left(\frac{\Lambda}{3}a^3-a\right)\,.
\ea
Halliwell shows that the integrals over the ghosts can be done analytically; the $\Pi$ integral enforces the condition $\dot{N}=0$ which reduces the $N$ integral to an ordinary integral. One then finally obtains
\ba
\mathcal{G}(a(t_1)|a(t_0)) &= & \int {\rm d}N(t_1-t_0)\int \mathcal{D}p \,\mathcal{D}a\,\exp\left(\imath \int_{t_0}^{t_1} {\rm d}t\left(p \dot{a} - N \mathcal{H}\right)\right)\nonumber
\\& = & \int {\rm d}N(t_1-t_0)\int \mathcal{D}a\,\exp\left(\imath S_{{\rm GR}}\right)\,.
\ea
The ordinary integral is an integral over the total proper time $N(t_1-t_0)$ between the initial and final configuration. Depending on whether one chooses to integrate this proper time over the entire real line or just over positive numbers, the resulting path integral then defines either a solution to the Wheeler--DeWitt equation or a Feynman propagator-like Green's function.

\begin{figure}[htp]
\begin{center}
\includegraphics[scale=1]{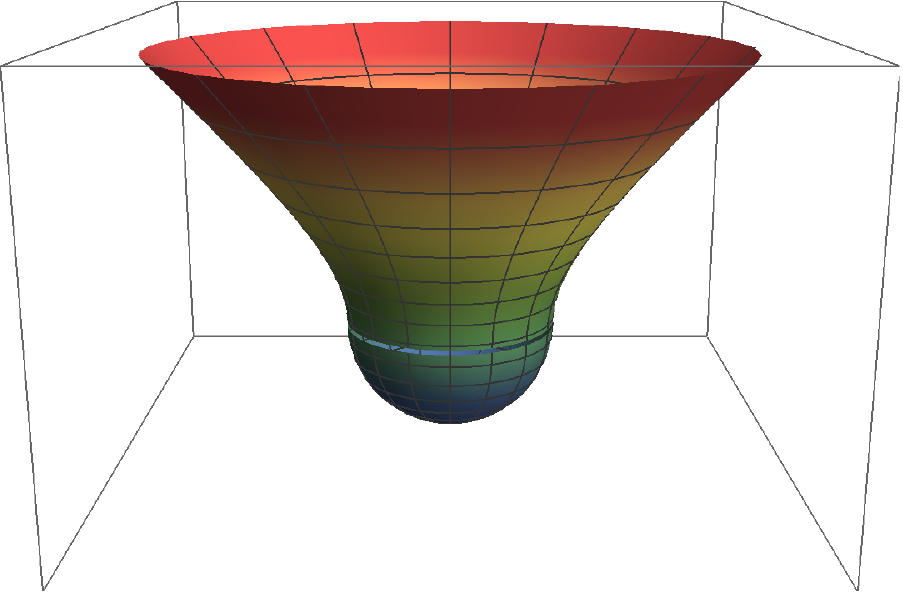}
\end{center}
\caption{Representation of a no-boundary saddle point for the path integral: the geometry starts off as a Riemannian 4-sphere which is then glued to one half of de Sitter spacetime.}
\label{fig1}
\end{figure}

It turns out that the path integral over the scale factor can also be evaluated analytically if one works in different variables and chooses a slightly different gauge fixing \cite{HalliwellLouko}. Namely, starting from (\ref{GRaction}) change variables by setting $N=\mathfrak{N}/a$ so that the action becomes 
\ba
S_{{\rm GR}}&=&\frac{3\pi}{4G}\int_{t_0}^{t_1} {\rm d}t \left(\mathfrak{N}-\frac{a^2\dot{a}^2}{\mathfrak{N}}-\frac{\Lambda}{3}\mathfrak{N}a^2\right)
\label{newGRaction}
\ea
and the Friedmann equations expressed in terms of $\mathfrak{N}$ are
\ba
\frac{1}{a^2}+\frac{\dot{a}^2}{\mathfrak{N}^2} & = & \frac{\Lambda}{3}\,,
\\\frac{\dot{a}^2}{\mathfrak{N}^2}+\frac{a\ddot{a}}{\mathfrak{N}^2}-\frac{a\dot{a}\dot{\mathfrak{N}}}{\mathfrak{N}^3} & = & \frac{\Lambda}{3} \,.
\label{friedmann3}
\ea
These clearly encode the same dynamical information as the equations (\ref{friedma1})-(\ref{friedma2}). One can now work with the gauge-fixing condition $\dot{\mathfrak{N}}=0$\footnote{As an attempt to give a more direct physical interpretation to the gauge $\dot{\mathfrak{N}}=0$, notice that the infinitesimal area element of two-dimensional timelike surfaces is $Na\,{\rm d}t\,{\rm d}\sigma$ for a time-independent ${\rm d}\sigma$; the gauge $N\cdot a={\rm const}$ is the one in which this area element is time-independent.}; if one also introduces a new variable $q= a^2$, the action (\ref{newGRaction}) becomes quadratic in the variable $q$. The path integral over $q$ is then just Gaussian and can be done analytically.\footnote{There are various technical subtleties in this argument, such as the change of the path integral measure under passing from $a$ to $q$, and the fact that the variable $a$ or $q$ should be restricted to the positive half-line only. We will ignore these here since we are really only interested in the contributions from stationary points of the action.} The result is of the form $e^{\imath S_{{\rm GR}}(a(t_0),a(t_1);\mathfrak{N})}$, the exponential of the action (\ref{newGRaction}) evaluated on the classical solution connecting the prescribed boundary values $a(t_0)$ and $a(t_1)$, multiplied by a prefactor resulting from integrating over fluctuations over the classical solution. For the purposes of our summary here, we are only interested in the phase factor, which can be seen as the purely classical contribution to the path integral. To evaluate this, one solves the equation of motion (\ref{friedmann3}) for $a$ (or equivalently for $q$) to obtain
\ba
a_{{\rm sol}}(t) &:=& \sqrt{a(1)^2 t - a(0)^2(t-1)+\frac{\Lambda}{3}\mathfrak{N}^2t(t-1)}
\ea
where we have set $t_0=0$ and $t_1=1$ to lighten the notation a little; with this choice we have $\mathfrak{N}(t_1-t_0)=\mathfrak{N}$ and $\mathfrak{N}$ represents the total time (in our gauge) between initial and final states. The classical action (\ref{newGRaction}) evaluated on $a(t)= a_{{\rm sol}}(t)$ is 
\ba
S_{{\rm GR}}(a(0),a(1);\mathfrak{N}) &=& \frac{3\pi}{8G}\left(2\mathfrak{N}-\frac{(a(0)^2-a(1)^2)^2}{2\mathfrak{N}}-\frac{\Lambda}{3}\mathfrak{N}(a(0)^2+a(1)^2)+\frac{\Lambda^2}{54}\mathfrak{N}^3\right)\,.
\label{Naction}
\ea
In this expression for the minisuperspace action we have integrated out the variable $a(t)$, but still have a dependence on the parameter $\mathfrak{N}$. The only remaining step is to evaluate the ordinary integral over $\mathfrak{N}$. This integral can no longer be done exactly, but one can perform a stationary phase approximation to identify the leading contributions in a semiclassical limit, potentially making use of Picard--Lefshetz theory as advocated in \cite{Feldbrugge}. Such a stationary phase approximation requires analytic continuation in $\mathfrak{N}$ in the no-boundary case $a(0)=0$ for which the action (\ref{Naction}) does not have any real stationary points. In the general case, there are four stationary points in the complex $\mathfrak{N}$ plane located at
\ba
\mathfrak{N}_{{\rm cr}}(a(0),a(1))&:=&\frac{3}{\Lambda}\left(\pm\sqrt{\frac{\Lambda}{3}a(0)^2-1}\pm\sqrt{\frac{\Lambda}{3}a(1)^2-1}\right)
\label{lapseexp}
\ea
where both signs can be chosen freely. Clearly these possible values for $\mathfrak{N}$ correspond to the total time (again, in our chosen gauge) to go from $a(0)$ to $a(1)$ on a classical solution. Notice that for both the initial and final values of $a$, choosing a value below $a_{\Lambda}:=\sqrt{\frac{3}{\Lambda}}$ (in particular, $a(0)=0$) means that this time picks up an imaginary part whereas values above $a_{\Lambda}$ increase this time by a real amount. This is because the classical Lorentzian solution to the equations of motion is de Sitter spacetime, for which $a(t)\ge a_{\Lambda}$ everywhere; conversely, a Riemannian solution to the same theory (obtained from choosing an imaginary lapse) is given by the 4-sphere, with $a(t)\le a_{\Lambda}$. The famous picture commonly associated to the no-boundary proposal, in which a universe transitioning from $a(0)=0$ to some large $a(1)$ is obtained by gluing half of a 4-sphere to a section of de Sitter spacetime (Fig.~\ref{fig1}) is represented in (\ref{lapseexp}): a classical solution starting at the classically forbidden value $a(0)=0$ has to first go into imaginary time before reaching $a=a_{\Lambda}$ and hence the classically allowed Lorentzian regime, where it can continue along the real $\mathfrak{N}$ axis.    

Substituting the four possible values (\ref{lapseexp}) into (\ref{Naction}) gives the final expression
\ba\label{HJCont}
S_{{\rm HJ}}(a(0),a(1)) &:=& \frac{3\pi}{2G\Lambda}\left(\pm\left(\frac{\Lambda}{3}a(0)^2-1\right)^{3/2}\pm\left(\frac{\Lambda}{3}a(1)^2-1\right)^{3/2}\right)
\ea
which now represents the Hamilton--Jacobi function for this minisuperspace model, dependent only on the boundary values for $a(t)$ and on the choice of one out of four possible classical solutions. Notice that if $a(0)=0$ one term gives $\pm\imath\frac{3\pi}{2G\Lambda}$ and hence an exponential enhancement or suppression factor $e^{\pm \frac{3\pi}{2G\Lambda}}$. The preferred sign for this factor (and hence, the question of which of the four possible stationary points should be included in evaluating the path integral approximately) has been the focus of intense debate over the last years \cite{Debate}. 

Note that the expression for the 
Hamilton--Jacobi function (\ref{HJCont}), i.e., the value of the 
classical action evaluated along a solution, can also be obtained more 
straightforwardly: one can simply solve the first order (constraint) 
Friedmann equation in a convenient gauge for the given boundary data, 
given that the value of the Hamilton--Jacobi function cannot depend on 
the choice of gauge. If we again work in the particularly convenient 
gauge $N=\mathfrak{N}/a$ where $\mathfrak{N}$ is a constant, the first order Friedmann equation in (\ref{friedmann3}) is
\ba
\frac{\dot{a}^2 }{\mathfrak{N}^2} & = & \frac{\Lambda }{3} -\frac{1}{a^2}\,.
\ea
The two 
possible solutions (corresponding to the two possible signs for 
$\dot{a}$) starting from a given $a(0)$ are
\begin{equation}
a(t)=\sqrt{a(0)^2+\frac{1}{3}\mathfrak{N}t\left(\Lambda\mathfrak{N} t + 
6\sigma_1\sqrt{\frac{\Lambda}{3}a(0)^2-1}\right)}
\end{equation}
where $\sigma_1=\pm 1$. One can now eliminate $\mathfrak{N}$ in favour of the final boundary value $a(1)$; this yields 
\begin{equation}
a(t)=\sqrt{a(0)^2(t-1)^2+a(1)^2t^2-\frac{6}{\Lambda}t(t-1)\left(1+\sigma_1\sigma_2\sqrt{\frac{\Lambda}{3}a(0)^2-1}\sqrt{\frac{\Lambda}{3}a(1)^2-1}\right)}
\end{equation}
where $\sigma_2$ is another sign that can be chosen freely, coming from 
the ambiguity in eliminating $\mathfrak{N}$ in favour of $a(1)$. 
Evaluating the action (\ref{newGRaction}) on this solution again leads 
to (\ref{HJCont}), with the two signs $\sigma_1$ and $\sigma_2$
free as before.

Below we will compare these continuum results for the stationary phase approximation to those obtained in the discrete setting of Regge calculus.
The sign ambiguities in (\ref{HJCont}) will also be important in the discussion there, so a few more comments regarding their interpretation might be useful. First of all, note that if one chooses either $a(0)=a_\Lambda$ or $a(1)=a_\Lambda$, there is only a single global sign to choose.  Furthermore, the Hamilton--Jacobi function is, for any choice of signs, purely imaginary if $a(0)\leq a_\Lambda$ and $a(1)\leq a_\Lambda$ (so that all classical solutions are Euclidean) and real if $a(0)\geq a_\Lambda$ and $a(1)\geq a_\Lambda$ (so that all classical solutions are Lorentzian). For the remaining case where $a(0)\leq a_\Lambda$ and $a(1)\geq a_\Lambda$, which is considered in the no-boundary proposal, the two sign ambiguities amount to a choice of sign for the imaginary part (resulting from evolution from $a(0)$ to $a_\Lambda$) and the sign for the real part (resulting from evolution from $a_\Lambda$ to $a(1)$), respectively.

In the purely Euclidean case where $a(0)\leq a_\Lambda$ and $a(1)\leq a_\Lambda$ there are still four saddle point solutions: there is one ambiguity which can be associated to the choice of orientation or sign of the lapse, but the other ambiguity is associated to the geometry of the 4-sphere. For any two boundary values of the radius $a(t)$, there is a solution connecting these values ``directly'' by staying within the same hemisphere, and another solution which passes through the equator, so that the boundary values are associated to two different hemispheres. Similarly, in the purely Lorentzian case the two boundary values can be on the same side or on opposite signs of the ``throat'' where de Sitter spacetime takes its minimal radius $a=a_\Lambda$.

\section{Regge calculus}\label{Sec:Regge}

\subsection{Regge calculus for Lorentzian triangulations}
\label{Lorentzian-Regge}

Regge calculus \cite{Regge} defines general relativity on a lattice and accordingly allows for a coordinate free formulation of gravity. It relies on regarding four-dimensional spacetime as a piecewise flat\footnote{One can also choose homogeneously curved building blocks \cite{Improved,NewRegge}, which are in particular appropriate when working with a cosmological constant. The (four-dimensional) Regge action, more precisely the volume term, for such curved building blocks is however not known in explicit form as a function of the length variables.} complex, i.e., a collection of four-dimensional simplicial building blocks that are glued together at their flat three-dimensional tetrahedral subsimplices.

While in continuum Einstein relativity geometry is represented by the metric tensor, Regge calculus employs the edge lengths $l_e$ of the simplicial building blocks as the geometric degrees of freedom.\footnote{Other variables can be used, e.g., angles and areas \cite{AreaAngle,NewRegge}.} The geometry of a flat simplex is completely determined by the totality of these lengths, and every geometric quantity, such as an area, volume or angle, is hence a function of the edge lengths in this formulation. 

The curvature resides on subspaces of co-dimension two (here triangles), also denoted as ``hinges'' or ``bones''. The amount of curvature depends on the hypothetical angular gap between the $4$--simplices meeting at such a triangle $t$ that would be obtained if these simplices were flattened out in four--dimensional Minkowski space. To compute this so-called deficit angle $\epsilon_t$ at a bulk triangle $t$, it is convenient to project the adjacent $4$-simplices to the plane orthogonal to the triangle, which itself reduces to a point. The solid angles  resulting from the projection of the 4-simplices $\sigma$ then correspond to the dihedral angles $\theta^\sigma_t$ at $t\subset\sigma$.

Since one can use this projection, it is sufficient to define deficit (or dihedral) angles in any two-dimensional plane. In a four-dimensional Lorentzian geometry, this plane is of Euclidean geometry, and the deficit angle is Euclidean, if the triangle $t$ is time-like.  If the triangle $t$ is space-like, the plane is of Minkowskian geometry and we have a Lorentzian angle.

In case of an Euclidean angle, the deficit angle is given by the difference between the flat angle and the sum of the dihedral angles, that is 
\be
\epsilon_t=2\pi-\sum_{\sigma \supset t} \theta^\sigma_t\,.
\ee

The definition of Minkowskian angles is more involved, see \cite{Sorkin74,Sorkin2019} for a thorough discussion and \cite{EffSF3} for explicit formulas for the computation of the dihedral angles.  

We want to mention one peculiarity of angles in the Minkowskian plane, which will be important later on: these angles can have an imaginary part, which is a multiple of $\pi/2$. These imaginary parts can appear due to the fact that the Minkowskian equivalent to the Euclidean circle is given by the union of four disconnected hyperbolae. Whereas the Euclidean angle parametrises the distance between two points on the circle, the Minkowskian angle has to parametrise the distance between points which may lie on different hyperbolae; if they do, the angle includes an imaginary part. More precisely, in the conventions of \cite{Sorkin2019} that we will be using, one associates $-\imath \pi/2$ for each ``jump''  from one hyperbola to a neighbouring hyperbola in an anti-clockwise direction. Note that these jumps cross one of the four light rays originating from the point representing the projected triangle in the Minkowskian plane.  Going around the full circle the flat Lorentzian angle has an imaginary part of $-2\pi \imath$, and the Lorentzian deficit angle is therefore defined as
\be
\epsilon_t=-2\pi\imath-\sum_{\sigma \supset t} \theta^\sigma_t\,.
\ee
 Now if the union of the dihedral angles at $t$ include four ``jumps'', that is two light cones, we obtain a real Lorentzian angle. But it may also happen that the union of the dihedral angles includes more or less than two light cones, in which case the deficit angle will include imaginary parts.  Hence, deficit angles with imaginary parts occur for configurations which include points with more or less than two light cones attached. We will refer to such configurations as having an irregular light cone structure.

Deficit angles associated to bulk triangles of a given triangulation provide a measure for the integrated intrinsic curvature. If we consider a triangle $t$ in the boundary of a triangulation, a similar concept leads to the extrinsic curvature angle 
\be
\psi_t := \Psi-\sum_{\sigma \supset t} \theta^\sigma_t\,.
\ee
Here $\Psi$ is the flat angle for the half-plane\footnote{For different types of boundaries, for instance if one has a corner, one can also associate a different ``flat'' value. In this work we will only need the values for the half-plane.}, $\Psi=\pi$ for Euclidean angles and $\Psi=-\imath \pi$ for Lorentzian angles.

The equivalent of the Einstein--Hilbert action (with Gibbons--Hawking--York boundary term)  in Regge calculus is a functional of the edge lengths. Supplemented by a cosmological constant term, it is given by \cite{Regge}
\begin{equation} \label{eq:Definition Regge Action}
S_{\text{R}} := \frac{1}{8\pi G} \left(\sum_{t \subset \text{bulk} } A_t(l) \epsilon_t(l)  +\sum_{t \subset \text{bdry} } A_t(l) \psi_t(l)   - \Lambda \sum_\sigma V_\sigma(l)\right) 
\end{equation}
where $A_t(l)$ and $V_\sigma(l)$ are the area of a triangle $t$ and the volume of a $4$--simplex $\sigma$ respectively. These quantities depend on the set of edge lengths, which we denote collectively by $l$. Similarly, all angles need to be implicitly defined as functions of the lengths $l$ using the procedures detailed above. In fact, this commonly used notation is somewhat ambiguous for a Lorentzian geometry, where edges can be time-like or space-like. It is more useful to think of the Regge action as a function of length square assignments $l^2$ to each edge, where $l^2$ can take either sign; this is what we will usually do in the following.

The first term in (\ref{eq:Definition Regge Action}) corresponds to the standard curvature term in the continuum theory. It converges to the continuum Regge action in the sense of measures \cite{Cheeger1984}. The equivalent of Einstein's equations is similarly recovered by the principle of least action, varying $S_{\text{R}}$ with respect to the bulk edge lengths $l_e$. Using Regge's finding \cite{Regge} that the variation of the deficit angles gives zero when summed over each simplex (due to Schl\"afli's differential identity), the Regge equation reads
\begin{equation}
\frac{1}{8\pi G} \left(\sum_{t \supset e} \frac{\partial A_t(l)}{\partial l_e} \epsilon_t(l) - \Lambda \sum_{\sigma \supset e} \frac{\partial V_\sigma(l)}{\partial l_e} \right) =0 \,.
\end{equation}
In the following, we will construct the Regge action \eqref{eq:Definition Regge Action} for various types of simplicial complexes adapted to the cosmological model of a closed FLRW universe.

\subsection{Quantum Regge calculus} 

Having defined the (Lorentzian) Regge action, we can now proceed to the Regge-based path integral. Here we will be interested in the Lorentzian path integral for the Euclidean quantum gravity version, see \cite{WilliamsReview,Hartle1985}.  We define the path integral as
\ba
{\cal G}(\{l_e\}_{e \subset \text{bdry}} ) \,\,:=\,\, \int {\cal D}\mu(l)    \exp( \imath S_{\text{R}}(l)) 
\ea
where the  integral is over all bulk edge lengths.  In this work we will only be interested in a (semi)-classical evaluation of (a symmetry reduced version of) the path integral. We will therefore not need the explicit form for the measure ${\cal D}\mu(l)$, but it can be further specified by demanding a discrete version of diffeomorphism invariance \cite{Menotti,DittrichSteinhausM}. 

An important choice is however in the support of the measure. A standard requirement is to restrict the integral to configurations of length (square) variables such that each 4-simplex satisfies the Lorentzian generalised triangle inequalities. These inequalities demand that the signed volume square of the (Lorentzian) 4-simplex, which can for instance be defined through the Cayley--Menger determinant \cite{Sorkin74}, is negative, and that the signature of the volume squares of a given sub-simplex corresponds to the signature of this sub-simplex. For example, the signed area square of a space-like triangle is negative and the signed area square of a time-like triangle is positive. Furthermore, a simplex containing a time-like sub-simplex has to be time-like or null. Imposing all these inequalities in the path integral will pose many practical difficulties.

Another choice, which is specific to the Lorentzian case, is whether to include configurations with irregular light cone structure or not.   The issue with such configurations is that, as we saw in the previous discussion of the Regge action, such configurations come with complex deficit angles leading to a complex action. In particular, configurations with deficit angles with negative imaginary part will be enhanced in the path integral. With our choice of conventions (see \cite{Sorkin2019} for a justification for this choice), such deficit angles with negative imaginary parts appear if we have points with less than two light cones attached. 

There has so far not been much research on the prevalence of such configurations. In this work we will consider some quite simple symmetry reduced configurations, but we will already find that each example with bulk edges does admit configurations with irregular light cone structure. These can easily dominate the path integral \cite{EffSF3} and moreover, for our examples, do not correspond to any continuum configurations. 

To prevent the path integral from being dominated by such seemingly unphysical configurations, we can restrict to configurations with regular light cone structure. A very similar restriction is imposed for the Causal Dynamical Triangulation approach \cite{CDTReview, LollJordan}.  There, this restriction  does lead to a much better behaved continuum limit for the path integral than in (Euclidean) Dynamical Triangulations. 

The last choice we will mention here is whether to sum over the orientations of $4$-simplices or not. Such a sum over orientations arises naturally in state sum models, like spin foams \cite{Perez}, which derive from a connection representation. The (Regge) action for a negatively oriented 4-simplex is defined to be minus the one for the positively oriented 4-simplex, and one can therefore absorb the sum over orientations into a replacement of the simplex amplitude $\exp(\imath S_\sigma)$, where $S_\sigma$ is the action for the (positively oriented) simplex $\sigma$, by $\cos(S_\sigma)$. 

The change of sign for the Regge action under a change of orientation is similar to the change of overall sign of the continuum action (\ref{GRaction}) if one changes the time orientation, that is the sign of the lapse. Indeed the sum over orientations can be seen as the discrete analogue of a continuum integral over positive and negative lapse \cite{OritiFeynman}, and thus the version of the path integral that defines a solution to the Wheeler--DeWitt equation \cite{satisfyconstraints}. In contrast, just allowing for one orientation leads to a Green's function analogous to the Feynman propagator. The latter choice is more standard in Regge calculus, and we will for simplicity use this choice in this work. Since the result of an integral over both orientations can be recovered from the real part of the integral over one orientation only, there is no loss of generality in using only  a single orientation.

\subsection{Triangulations of the three-sphere}\label{Sec:Cells}

Regge calculus requires choosing a triangulation of the manifold under consideration. Here we will discuss triangulations of three-dimensional spheres. These will serve as a basis for the construction of the four-dimensional triangulations which we will consider in the subsequent sections. 

In order to model a homogeneous and isotropic geometry, we choose the triangulation of the three-sphere to be the boundary of a regular convex 4-polytope; this will be true regardless of whether the full spacetime is modelled as a 4-polytope or has a more complicated structure. Alternative choices are possible; one can, e.g., start from a construction of regular 4-polytopes and refine these \cite{Brewin} or define a more abstract notion of discretisation \cite{Tsuda}. Having the quantum gravitational path integral in mind, here we focus on the simplest choice of regular 4-polytopes.

Regular polytopes are polytopes of maximal symmetry. There are six regular convex 4-polytopes \cite{Coxeter}, but only three of these have a triangulation as boundary. (The remaining three define a more general polyhedral subdivision.)  The three 4-polytopes are known as 5-cell (or pentachoron or 4-simplex), 16-cell (or hexadecachoron), and 600-cell (or hexacosichoron). 
The $X$ in $X$-cell refers to the number of tetrahedra in the three-dimensional boundary of the 4-polytope.  We  denote the number of tetrahedra, triangles, edges and vertices in the three-dimensional boundary with $n_\tau,n_t, n_e$ and $n_v$, respectively. Table \ref{Ntable1} lists the values for the 5-cell, the 16-cell and the 600-cell.

\begin{table}[H]
\centering
\begin{tabular}{|l||c|c|c|c|}\hline
&\; $n_\tau$ \;& \;$n_t$\; &\; $n_e$\; & \;$n_v$\; \\ \hline
5-cell & 5 & 10 & 10 & 5\\ \hline
16-cell & 16 & 32 & 24 & 8\\ \hline
600-cell & 600 & 1200 & 720 & 120 \\ \hline
\end{tabular}
\caption{Number of tetrahedra $n_\tau$, triangles $n_t$, edges $n_e$ and vertices $n_v$ in the boundary of the X-cell. Note that $n_t=2n_\tau$ in all cases, which can be used to simplify formulae containing both $n_t$ and $n_\tau$. \label{Ntable1}}
\end{table}

\begin{figure}[!htb]
\minipage{0.32\textwidth}
  \includegraphics[width=\linewidth]{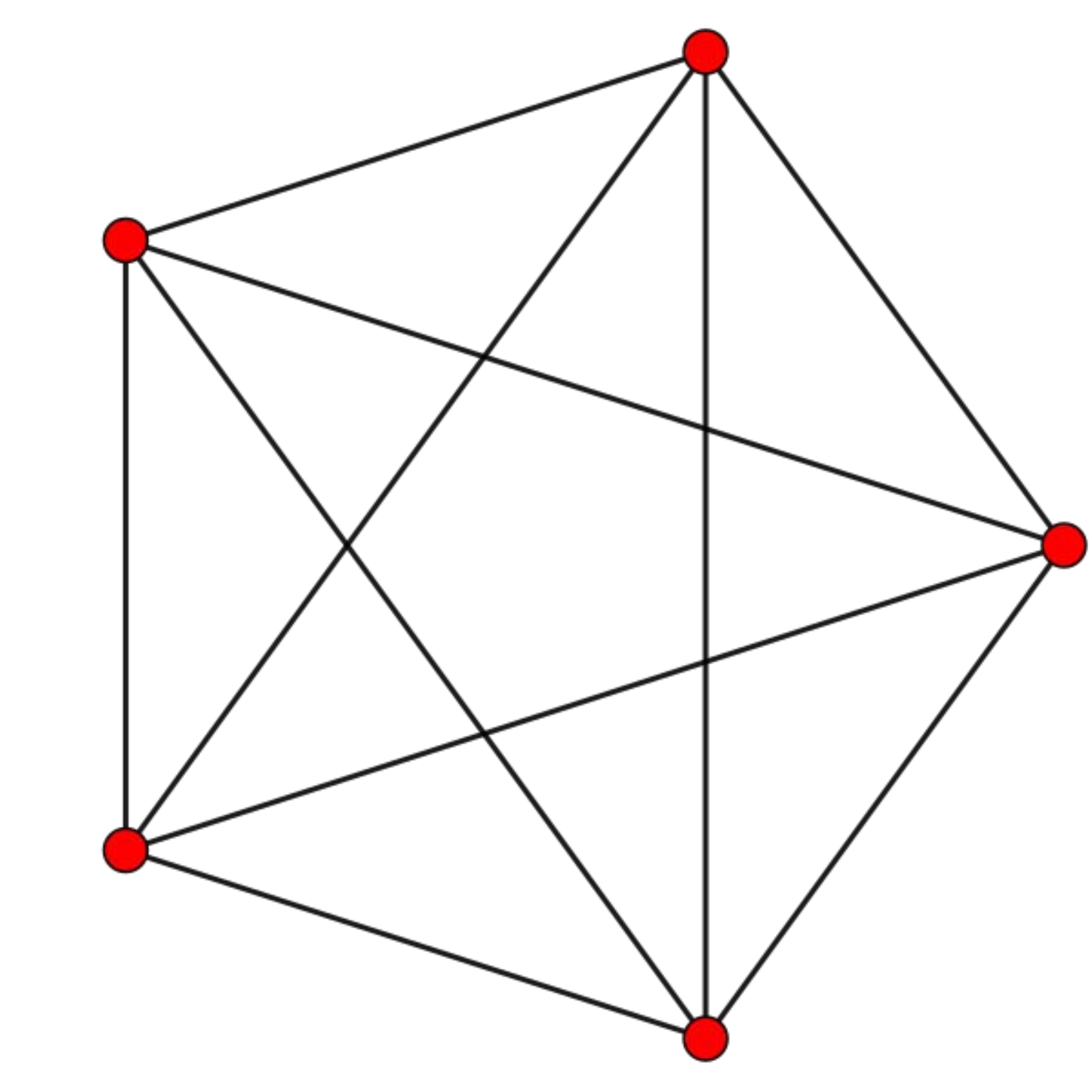}
  \caption{1-skeleton and dual 1-skeleton of the 5-cell. Credit: en.wikipedia.org/wiki/5-cell}\label{fig:5-cell-1-skeleton}
\endminipage\hfill
\minipage{0.32\textwidth}
  \includegraphics[width=\linewidth]{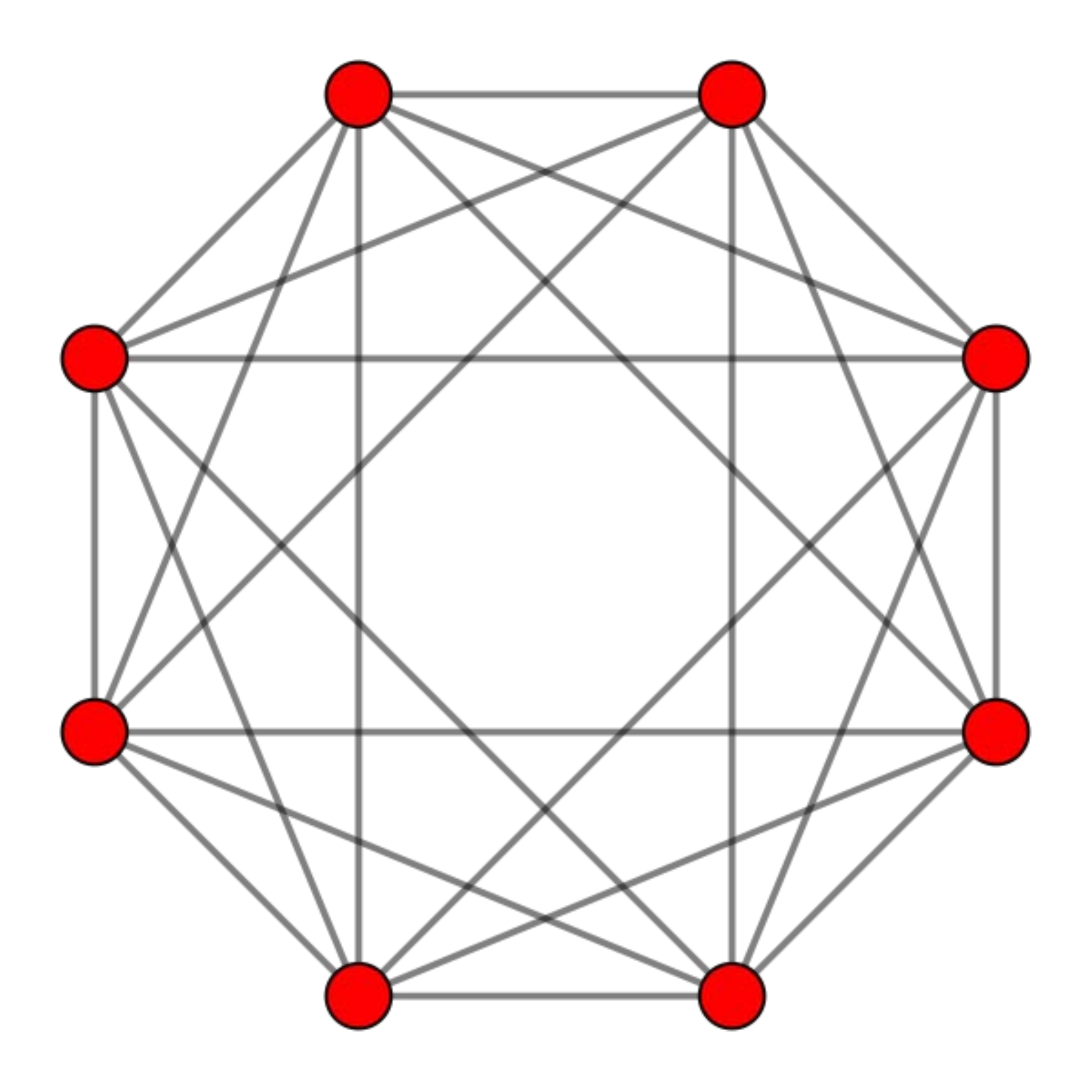}
  \caption{1-skeleton of the 16-cell. Credit: en.wikipedia.org/wiki/16-cell }\label{fig:16-cell-1-skeleton}
\endminipage\hfill
\minipage{0.32\textwidth}%
  \includegraphics[width=\linewidth]{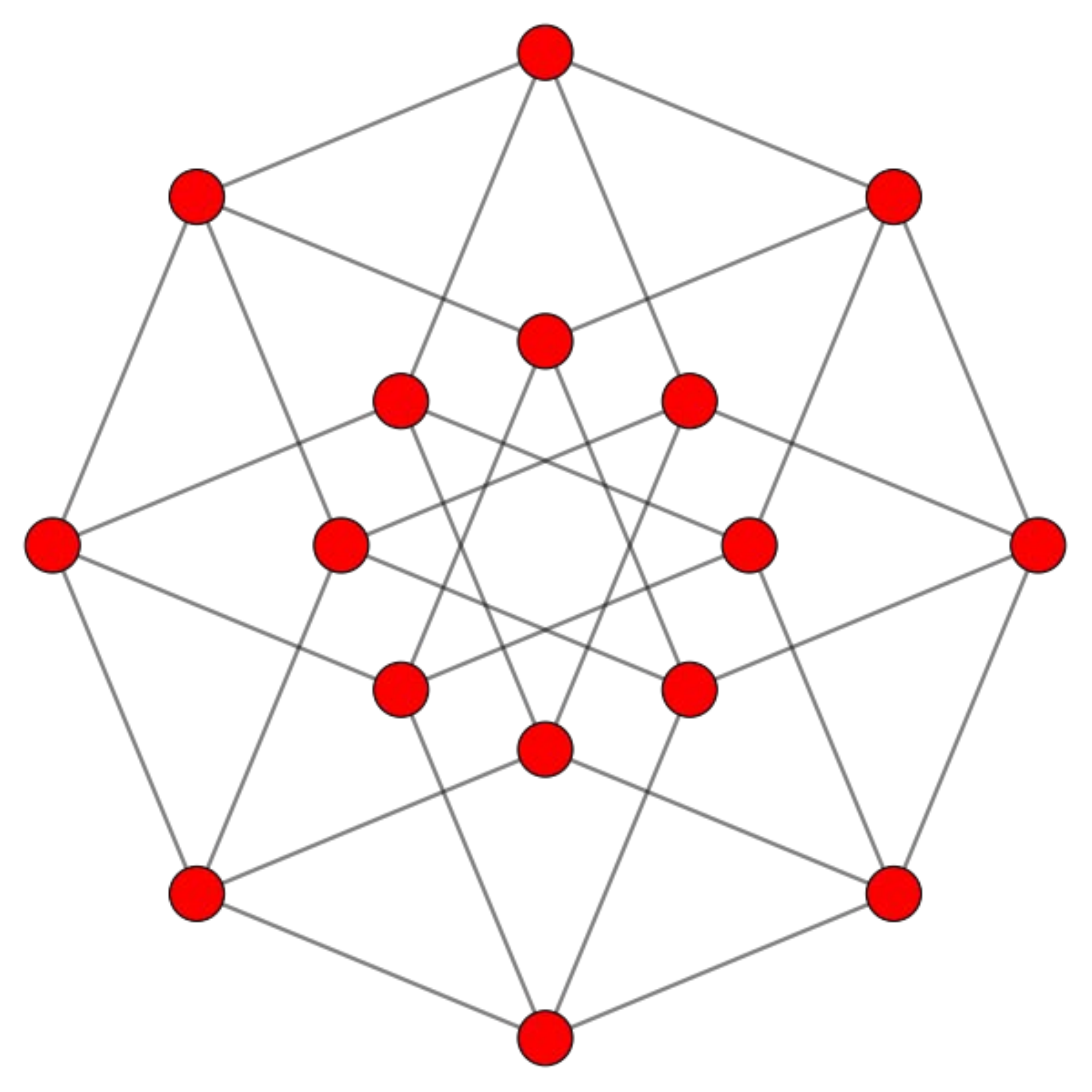}
  \caption{Dual 1-skeleton of the 16-cell. Credit: en.wikipedia.org/wiki/Tesseract}\label{fig:16-cell-dual-1-skeleton}
\endminipage
\end{figure}

For illustrative purposes Figures \ref{fig:5-cell-1-skeleton}--\ref{fig:16-cell-dual-1-skeleton} show the 1-skeleton and the dual 1-skeleton of the 5-cell, the 1-skeleton of the 16-cell, and the dual 1-skeleton of the 16-cell. For the (dual) 1-skeleton of the 600-cell we refer to the {\it Wikipedia} pages of the 600-cell \cite{600cell} and of its dual, the 120-cell \cite{120cell}.

In order to approximate a homogenous three-dimensional geometry, we choose all edge lengths in the triangulations of the three-sphere to be the same. The edge length, which we will denote by $l$, then completely specifies the piecewise flat geometry of the triangulations. It clearly provides a measure for the size of the triangulated three-sphere. In order to compare this variable $l$ with the radius $a$ of a continuum homogeneous three-sphere (see (\ref{flrwmetric})), we can compare $(i)$ the three-volume and $(ii)$ the (three-dimensional) Ricci scalar  of the discretised and the continuum three-sphere. Requiring that either $(i)$ or $(ii)$ are equal for continuum and discrete geometries will give us different relations between $l$ and $a$, and the mismatch provides a measure for how well a given $X$-cell can approximate the continuum geometry.

To compare the three-volumes we note that the volume of an equilateral tetrahedron with edge length $l$  is given by  $V_\tau=l^3/(6\sqrt{2})$ and the volume of the three-sphere with radius $a$ is $V_{\rm S}=2\pi^2 a^3$. Equating $V_{\rm S}=n_\tau V_\tau$ we obtain the relation 
$l = : \nu(n_\tau) a$ between the length $l$ and the scale factor, where
\ba
\nu(n_\tau) =2^{5/6}\left(\frac{3\pi^{2}}{n_\tau}\right)^{1/3}  \,.
\label{nurelation}
\ea

The integrated three-curvature of the three-sphere triangulated with equilateral tetrahedra is given by $R_{\text{D}} = 2 l n_e \epsilon_e$   where $\epsilon_e=2\pi-  \frac{6n_\tau}{n_e}\cos^{-1}(1/3)$ is the deficit angle at an edge. Equating this quantity with the integrated three-curvature $R_{\rm C}=12\pi^2 a$ for the continuum three-sphere we obtain the relation $l =:  \xi(n_\tau,n_e) a$, where
\ba
  \xi(n_\tau,n_e) =\frac{3 \pi^2}{n_e \pi-3 n_\tau \cos^{-1}(1/3) } \,.
  \label{xirelation}
\ea

In Table \ref{Table2} we compare the numerical values for $\nu$ and $\xi$ for the 5-cell, the 16-cell and the 600-cell. Whereas $\nu$ is by about 41 percent larger than $\xi$ for the 5-cell, the mismatch for the 600-cell is only about 2 percent. We will see at various points in the following that the 600-cell can provide a reasonably good approximation to a continuum three-sphere whereas the cruder discretisations by  a 5-cell or 16-cell can show very different behaviour.

\begin{table}[!]
\begin{tabular}{|l||c|c|}\hline
&\; $\nu$ \;& \;$\xi$\; \\ \hline
5-cell & 3.224 & 2.286 \\ \hline
16-cell & 2.188 & 1.815 \\ \hline
600-cell & 0.654 & 0.641\\ \hline
\end{tabular}
\caption{Different definitions of the ratio $l/a$, based on respectively comparing the three-volume ($\nu$) and the integrated three-curvature of the continuum three-sphere ($\xi$) with the boundary of the X-cell. \label{Table2}}
\end{table}

When formulating a cosmological model in Regge calculus, the basic variable characterising the geometry is the edge length $l$, whereas in cosmology we are used to the scale factor $a$. In the subsequent sections we will often derive equations for the dynamics of the Regge universe which are given in terms of $l$, and then translate these into an effective scale factor $a$ afterwards using (\ref{nurelation}) or (\ref{xirelation}). Given that the conversion factors $\nu$ and $\xi$ are different for the different discretisations, all effective cosmological equations in Regge calculus that are expressed in terms of  a scale factor $a$ then also depend on the choice of $X$-cell. The results will also be different depending on whether one uses $\nu$, $\xi$, or another quantity (e.g., a geometric mean of the two). In this work we will mostly use $\nu$ to translate the length $l$ to the scale factor $a$.  The differences between the different choices are only small for the 600-cell, in the other cases they represent the poor approximations to the continuum that these discretisations give.

Below we will introduce definitions of the Regge action and other functions for various discretisations based on different $X$-cells. These functions will depend on the particular $X$-cell used, but to lighten notation we will mostly suppress this dependence in our notation.

\section{A simplicial model for the beginning of the universe}\label{Sec:Ball}

\subsection{Discretisation with one 4-polytope}\label{NSD}

The simplest triangulation to describe the evolution of the universe from a 3-sphere with vanishing scale factor to a 3-sphere with finite scale factor is given by the 4-simplex itself. Allowing for more general discretisations, we can also use 16-cell or 600-cell polytopes. Since there are no unspecified bulk variables, the path integral associated to these discretisations does not come with any variables to integrate. Hence, ignoring measure factors,  the path integral is just given by the amplitude $\exp(\imath S_\text{P})$ resp.~$\cos(S_\text{P})$, where $S_\text{P}$ is the Regge action   associated to the 4-polytope. Indeed, various works \cite{DittrichHoehn,DittrichSteinhaus13,GozVid} have proposed to use such an amplitude to model the very beginning of the universe. 

In the simplest model, our Regge universe would then be a 4-simplex whose edges are all space-like and have equal length. In the context of Lorentzian Regge calculus, however, this is a priori not possible, since such a triangulation is forbidden by the generalised triangle inequalities for simplices with Lorentzian signature.\footnote{A simpler example are Lorentzian triangles in the Minkowskian plane. Assuming all edges are space-like the edge lengths have to satisfy $l_1+l_2\leq l_3$ or $l_1+l_3\leq l_2$ or $l_2+l_3\leq l_1$. Such triangles can therefore not be equilateral.} Likewise, there do not exist 16-cell or 600-cell 4-polytopes whose edges are all space-like with equal length and whose internal geometry is Minkowskian flat. The Lorentzian amplitude for the triangulation based on these 4-polytopes would therefore vanish.

This actually mirrors the fact that, as we have seen in Section \ref{Sec:Cont}, there is no Lorentzian solution for a closed FLRW universe (without matter) below the minimal scale factor $a_\Lambda$. But  we have also seen that there is a nonvanishing saddle point contribution to the path integral for such ``forbidden'' boundary data. This contribution can be associated to a Euclidean geometry, resulting from a deformation of the path integral contour for the lapse into the complex plane. 

This observation in the continuum case provides an argument to modify the prescription for the path integral in (Lorentzian) quantum Regge calculus, and to allow for contributions resulting from 4-simplices whose edge lengths define a Euclidean geometry. Indeed, the spin foam path integral, which results from a connection representation of general relativity, includes contributions from such Euclidean simplices \cite{BarrettFoxon, HanLiu}. These contributions come with an exponentially suppressed amplitude $\exp(-|S_{\sigma,\text{E}}|)$ where $S_{\sigma,\text{E}}$ is the Regge action for a 4-simplex with Euclidean geometry.\footnote{The 4-simplex wavefunction in \cite{GozVid} however relies on a different mechanism, namely the fact that the connection representation does also admit a so-called sector of vector geometries, from whose data one can reconstruct an Euclidean geometry for the 4-simplex. For these geometries one obtains an amplitude which is not exponentially suppressed. But one can argue that the sector of vector geometries should be excluded from the path integral \cite{Engle}.}

A different reason to include such Euclidean contributions in the Lorentzian quantum Regge path integral is based on requiring discretisation independence, which is deeply related to implementing a notion of diffeomorphism symmetry \cite{DittrichDiff}. As we will see below, instead of  a 4-polytope we can choose a subdivided 4-polytope as discretisation. For the subdivided 4-polytope the path integral does involve a proper integration and, similarly to the continuum case, this integration leads to a nonvanishing amplitude, which can be associated to a Euclidean configuration. One can thus argue that a 4-polytope whose boundary data allow for a Euclidean flat geometry should carry a nonvanishing amplitude. We propose this amplitude to  be  given by $\exp(-S_{\text{P},\text{E}})$  with the Euclidean\footnote{Here we use the Regge action for a triangulation with Euclidean geometry \cite{Regge}. Note that it differs by a global sign from the action usually used for Euclidean quantum gravity \cite{Hartle1985}.} Regge action
\ba\label{ActionP0}
8\pi G S_{\text{P,E}}(l^2)\,=\,&& n_t \, \frac{\sqrt{3}}{4} l^2  \left( \pi  -  2\cos^{-1}\left(   \frac{l}{  2\sqrt{2}\sqrt{-l^2+3m^2_\text{flat}} }  \right)\right)
-n_\tau \Lambda \frac{l^3}{96}\sqrt{-3l^2+8m_\text{flat}^2}\,~~~~~~\;
\ea
where 
\ba\label{ActionP0B}
m_\text{flat}^2=\frac{l^2}{2}\cdot \frac{2\cos(\pi \frac{n_e}{3n_\tau})-1}{3\cos(\pi \frac{n_e}{3n_\tau})-1} \, .
\ea
Here $n_t,n_e$ and $n_\tau$ denote the number of triangles, edges and tetrahedra in the boundary of the 4-polytope, cf.~Table~\ref{Ntable1}. The action (\ref{ActionP0}) is a  Regge action for flat $X$-cells. To construct this action for non-simplicial 4-polytopes, i.e., the $16$-cell and the $600$-cell, we  made use of the  simplicial Regge action for $X$-cells subdivided into $X$ 4-simplices, given in (\ref{SBall1E}) in Section \ref{Sec:S-P} below. This subdivision introduces bulk edges with length square $m^2$. In (\ref{ActionP0})-(\ref{ActionP0B}) we determine this length square $m^2$ by demanding that the deficit angle at the bulk edges vanishes\footnote{This is in difference to Section \ref{Sec:S-P}, where  $m^2$ will become a dynamical variable. The equations of motion will determine $m^2$ to be different from $m^2_\text{flat}$, since the presence of the cosmological constant demands a non-vanishing deficit angle.}, thus obtaining a flat $X$-cell.  In the case of the 5-cell the action (\ref{ActionP0}) is equal to  the Regge action for a (Euclidean) 4-simplex whose edges all have length $l$, given that in this case $m_\text{flat}^2=2l^2/5$ so that the extrinsic curvature angle appearing in (\ref{ActionP0}) is $\pi-2\cos^{-1}(\sqrt{5/8})=\pi-\cos^{-1}(1/4)$ and the total volume in the second term reduces to the 4-simplex volume $\sqrt{5}l^4/96$.

As we are approximating a constantly curved geometry with a flat building block, we can then only expect to obtain a valid picture if the length $l$ is much smaller than  $1/\sqrt{\Lambda}$. In this case the first term in (\ref{ActionP0}) dominates over the cosmological constant term, and $S_{\text{P},\text{E}}$ is positive, so that such configurations are exponentially suppressed.
To be more precise we compare the behaviour of the (Euclidean) discrete action  (\ref{ActionP0}) as a function of $a(X):=l/\nu(X)$ with the Hamilton--Jacobi function (\ref{HJCont}) for the continuum case. We want to describe an evolution from $a(0)=0$ to $a(1)= a_\Lambda$, and choose the signs  (and roots) such that the continuum Hamilton--Jacobi function is given by 
\ba \label{eq:4-polytope-euclidean-continuum}
S_E(a):=-\imath S_{\rm HJ}(a)= \frac{3\pi}{2G\Lambda}\left(1  - \left(1- \frac{\Lambda}{3}a^2\right)^{3/2}\right)    \,.
\ea

\begin{figure}[H]
\centering
\includegraphics[scale=1]{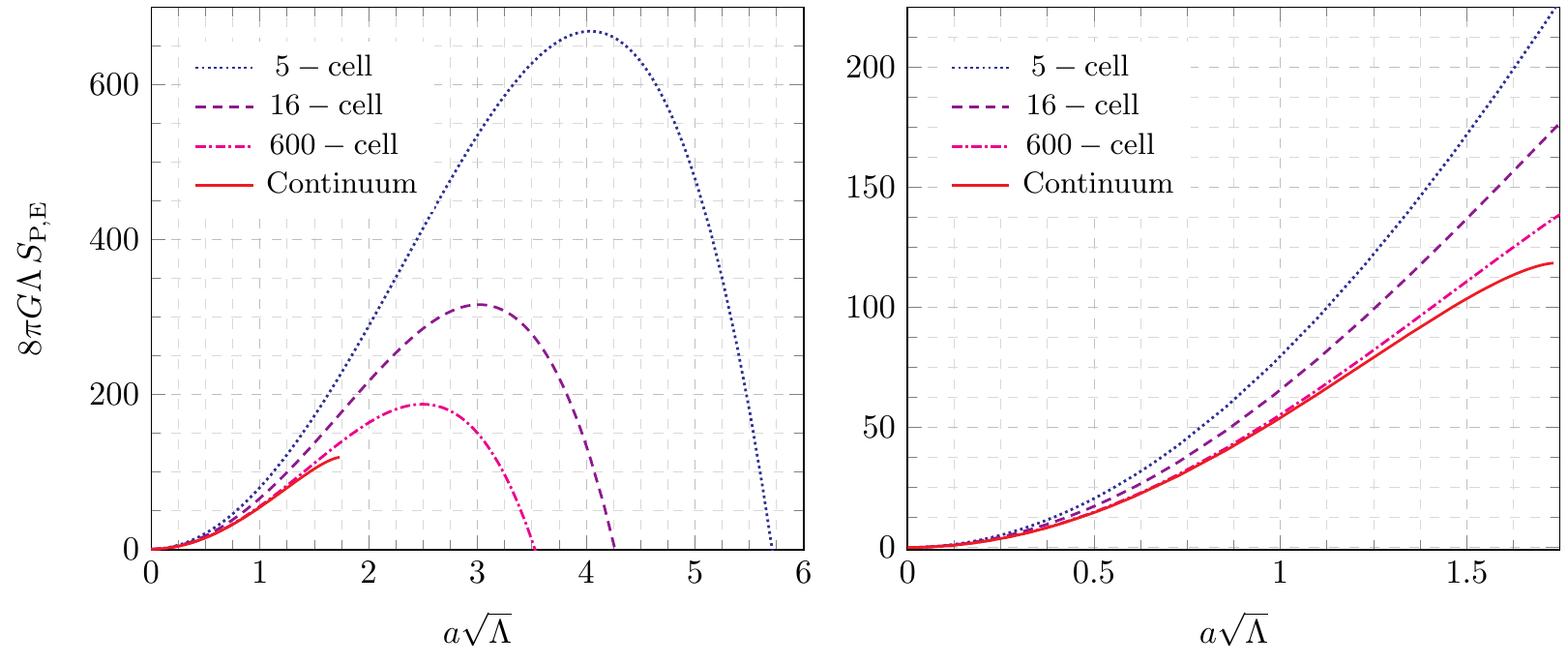}
\caption{Left panel: Rescaled Euclidean Regge action $ 8\pi G\Lambda\, S_{\text{P,E}}$ as function of $a\sqrt{\Lambda}=\sqrt{\Lambda} l/\nu(X)$ for the 5-, 16- and 600-cell as defined in \eqref{ActionP0}, and the continuum Euclidean action as defined in \eqref{eq:4-polytope-euclidean-continuum}. This continuum Euclidean action is only shown for the range $a\sqrt{\Lambda}<\sqrt{3}$, where it is real-valued. Right panel: Zoom of the left panel.} \label{fig:4-Polytope-Simple}
\end{figure}

These signs are chosen such that $S_E(a)$ vanishes for $a=0$ and monotonically increases until it reaches $a=a_\Lambda$, see Figure \ref{fig:4-Polytope-Simple}. Note that the derivative $\partial S_E(a)/\partial a$ vanishes for $a=a_\Lambda$. 

The discrete action $S_{\text{P},\text{E}}$ in (\ref{ActionP0}) is a quadratic polynomial in $l^2$. It 
 starts at zero for $l^2=0$, goes to a maximal positive value at a certain point $l^2_s$ (where $\partial S_{\text{P},\text{E}} /\partial l^2=0$) and then, with the cosmological constant term taking over, decreases  and eventually goes to negative values, see Figure \ref{fig:4-Polytope-Simple}.

The values for $a_s(X)^2=l^2_s/\nu(X)^2$ at which the discrete action takes its maximal value are approximately given by $16.31/\Lambda$, $9.10/\Lambda$ and $6.22/\Lambda$ for the $5$-cell, $16$-cell and $600$-cell, respectively. The corresponding maximal values for the action $S_{\text{P},\text{E}}$ are $669.16/(8\pi G\Lambda)$, $315.827/(8\pi G\Lambda)$ and $187.13/(8\pi G\Lambda)$, respectively. See also Table \ref{Table4} below, where we collect these values and compare them to values obtained from more refined discretisations  discussed later in the paper.

Note that, as in the continuum, we can rescale all dynamical variables by powers of $\sqrt{\Lambda}$, introducing a dimensionless length variable $\tilde l:=l\sqrt{\Lambda}$ and a dimensionless action $8\pi G\Lambda S_{\text{P},\text{E}}$, which can be expressed as a function of $\tilde l$ only. These dimensionless quantities, which use the natural units in the presence of a cosmological constant, will be employed in the following in particular when we give plots of the action and related dynamical quantities.

Comparing the qualitative behaviour of the discrete action $S_{\text{P},\text{E}}$ with the continuum action $S_E$, we might identify the value $a_s(X)$ in the discrete as the analogue of the continuum value $a_\Lambda$, since in both cases these are the points where the Euclidean action becomes maximal. As $S_E$ is only real for $a<a_\Lambda$, in this comparison we should only take the regime seriously where $l<l_s$. Comparing the quantitative values in Figure \ref{fig:4-Polytope-Simple} (see also Table \ref{Table4}), we see however large differences both for the values of $a_s(X)$ and $a_\Lambda$ and for the actions evaluated on these scale factors. The 600-cell does lead to the best approximation, but the $a_s^2$-value for the 600-cell still overestimates the continuum value $a_\Lambda^2$ by more than $100$ percent.

This model, in which we approximate the evolution of the universe up to a certain time with just one 4-polytope, is clearly very crude. If we allow for a nonvanishing contribution to the path integral from the Euclidean sector, it captures the ``Euclidean'' phase for very small scale factors of the continuum minisuperspace model discussed in Section \ref{Sec:Cont}. But for larger scale factors it does not even match the qualitative features of the continuum.  To achieve that, we would  need a more refined discretisation.

\subsection{Discretisation with a subdivided 4-polytope}
\label{Sec:S-P}

\subsubsection{Configuration with regular light cone structure }
\label{RegularLightCone}

In the following we will consider such more refined discretisations. To begin with, we consider a subdivision of a 4-polytope obtained by inserting an edge from each of the boundary vertices to the centre of the original polytope. The subdivided polytope then consists of 5, 16 or 600 4-simplices, depending on which $X$-cell we started from.  We assume that the added bulk edges all have the same length square $m^2$, and that all edges in the boundary of the 4-polytope have length square $l^2>0$.  We thus have one variable to integrate over in the path integral.  The length square $m^2$ can be negative or positive, as the bulk edges may be time-like or space-like. More precisely, we can differentiate between the following cases:
\begin{itemize}
\item[$(a)$] time-like bulk edges, $m^2<0$\,,
\item[$(b)$] space-like bulk edges but time-like bulk triangles, $0<m^2<\tfrac{1}{4} l^2$,
\item[$(c)$] space-like bulk triangles but time-like bulk tetrahedra, $\tfrac{1}{4} l^2<m^2<\tfrac{1}{3}l^2$,
\item[$(d)$] space-like bulk tetrahedra, $\tfrac{1}{3}l^2<m^2<\tfrac{3}{8} l^2$. 
\end{itemize}
Cases where building blocks are null arise from replacing these inequalities by equalities.

To begin with we will discuss the cases $(a)$ and $(b)$. These can be understood to represent a discretisation of the minisuperspace metric (\ref{flrwmetric}), using only one time step from the scale factor $a=0$ to some finite scale factor encoded in the length $l$ of the boundary edges. We will later also analyse the cases $(c)$ and $(d)$, but will see that these describe configurations with irregular light cone structure and therefore do not constitute a discretisation of the minisuperspace metric (\ref{flrwmetric}).

Restricting to the cases $(a)$ and $(b)$, the path integral for the subdivided 4-polytopes, reduced to configurations allowed by our assumptions on (maximal) isotropy and homogeneity, is given by
\ba\label{SBallPI}
\mathcal{G}_{\text{S-P}}(l\,| 0) := \int_{m^2<\tfrac{1}{4} l^2} {\rm d}m^2 \mu(l^2,m^2)\, \, \exp\left(\imath S_{\text{S-P}}(l,m)\right)  \q 
\ea
where $\mu(l^2,m^2)$ is a measure factor.

For these cases $(a)$ and $(b)$ the Regge action is given by\footnote{We could also use as boundary triangulation a 3-spherical complex 
obtained by gluing two tetrahedra along all four triangles to each 
other. The corresponding action is obtained by setting $n_\tau=2,n_t=4$ 
and $n_e=6$ in (\ref{SBall1}). The Hamilton--Jacobi function for this 
case deviates even more from the continuum than in the case of a 5-cell 
based triangulation. This choice of boundary triangulation is even less 
useful for the family of (non-subdivided) 4-polytopes: the corresponding 
4-polytope would be degenerate and therefore have vanishing 4-volume.}
\ba\label{SBall1}
8\pi G S_{\text{S-P}}(l^2,m^2)&:=& -n_t \, \frac{\sqrt{3}}{2} l^2 \sinh^{-1}\left(   \frac{l}{2\sqrt{2}\sqrt{l^2-3m^2}}  \right)\nn \\&&
+n_e \frac{1}{4} l\sqrt{l^2-4 m^2}
\left( 2\pi -\frac{6n_\tau}{n_e} \cos^{-1}\left( \frac{l^2-2m^2}{2l^2-6m^2}\right)        \right) \nn\\&&
-n_\tau \Lambda \frac{l^3}{96}\sqrt{3l^2-8m^2}\,.
\ea
The path integral (\ref{SBallPI}) can be understood as an approximation to the continuum path integral (\ref{PICont}) (with $t_0=0$ and $t_1=1$) in the following way. Using the partial gauge $N=\text{const.}$  we can identify the integration over $-m^2$ with the integration over the (squared) lapse variable $N^2$.  We do however allow values in the range $-\tfrac{1}{4}l^2< -m^2<\infty$ whereas in the continuum we would use the range $0<N^2<\infty$ (recall that we are restricting to a single orientation, so our starting point in the continuum would be to integrate over positive $N$ only).  

An alternative possibility is to use the height square of the 4-simplices, given by 
\ba
h^2:=m^2-\frac{3}{8}l^2 \,,
\ea
as integration variable, where the condition of time-like bulk triangles translates to $-h^2>\tfrac{1}{8} l^2$. In either case, the integration range in the discrete and continuum cases would be different.

We can perform a $(3+1)$-decomposition of the 4-polytope \cite{DittrichLoll2}, allowing us to define a time variable $t\in (0,1]$ and associated triangulated hypersurfaces. The triangulations of these hypersurfaces are given by the boundary of an $X$-cell, with all edge lengths equal to each other and given by $l(t)=l\cdot t$. We can see this construction as approximating the (quantum) solution $a(t)$ of the continuum equations with the linear function $l(t)=l\cdot t$. The path integral then only involves  an integration over the lapse parameter $N$, which in the discrete is represented by $m^2$ (or $h^2$). 

As in the discussion for the continuum case in Section \ref{Sec:Cont}, we will only be interested in a (semi-)classical evaluation of the path integral. We therefore need to understand  the stationary points for the action (\ref{SBall1}). Unfortunately, the stationary point equation for (\ref{SBall1}) cannot be solved analytically but regarding stationary points in the range $-m^2>-\frac{1}{4} l^2$ (corresponding to stationary points along the real lapse axis) we can say the following:  
\begin{itemize}
\item There is a critical value $l_{\text{crit}}$ such that for $0<l<l_{\text{crit}}$ we do not have stationary points for  $-m^2>-\frac{1}{4} l^2$. The value for $l_{\text{crit}}$ depends on the choice of $X$-cell. 
\item There is one stationary point $-m^2_s>-\frac{1}{4} l^2$ for $l> l_{\text{crit}}$. This stationary point moves towards $-m^2=-\frac{1}{4} l^2$ as we take $l^2 \rightarrow \infty$. 
\end{itemize}

\begin{figure}[H]
\centering
\includegraphics[scale=0.95]{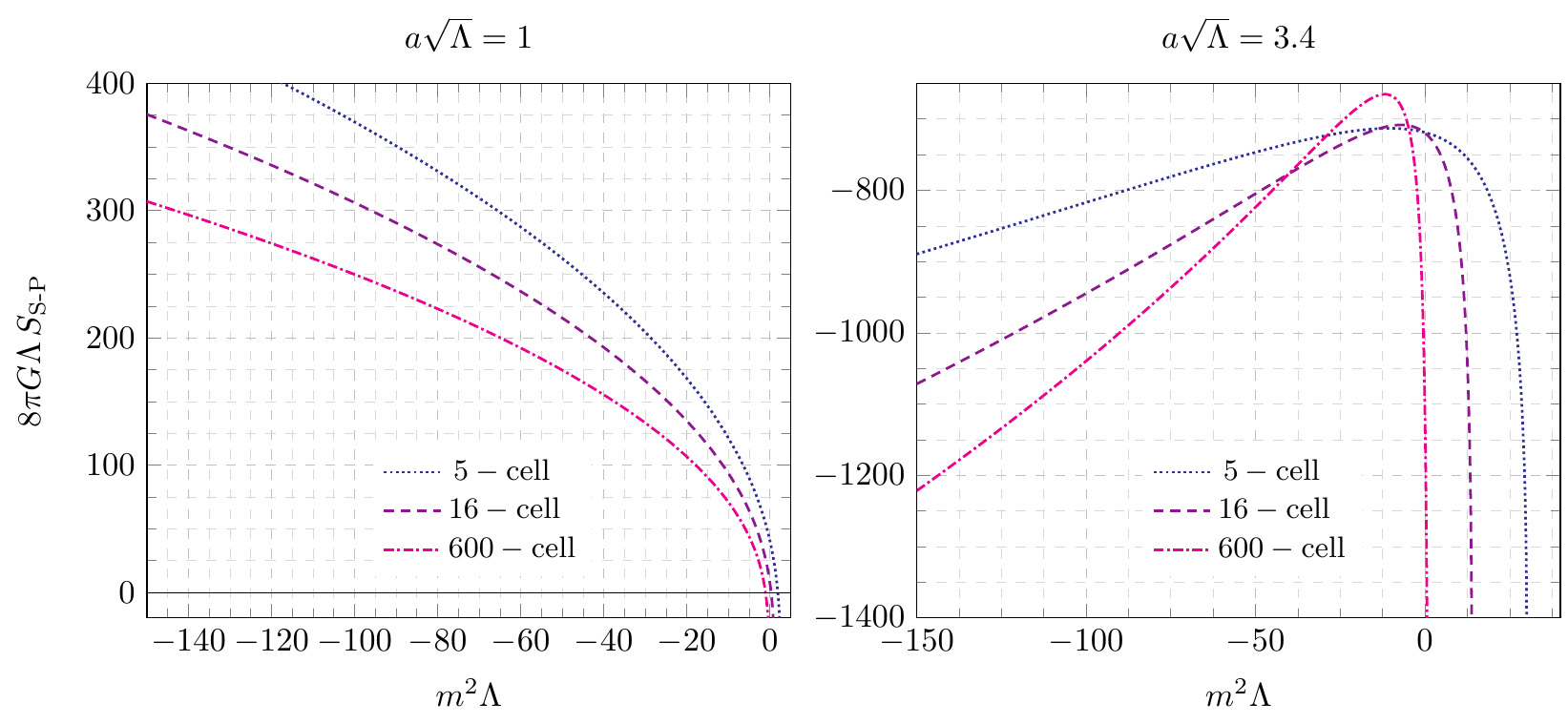}
\caption{Rescaled Lorentzian Regge action $8 \pi G \Lambda \,S_{\text{S-P}}$ as a function of $m^2 \Lambda$ for two different values of $a=l/\nu(X)$ for the different $X$-cell based triangulations. Left panel: $a=1/\sqrt{\Lambda}$. Right panel: $a=3.4/\sqrt{\Lambda}$. } \label{fig:4-Polytope-Lorentzian}
\end{figure}

Figure \ref{fig:4-Polytope-Lorentzian} shows plots of the action for $S_{\text{S-P}}(l,m^2)$ for the cases $(i)$ $l<l_{\text{crit}}$ and for $(ii)$ $l>l_{\text{crit}}$. As can be seen from these figures, the change from case $(i)$ to $(ii)$ occurs because the behaviour of the function for large negative $m^2$ changes from growing to decaying.   This is due to the cosmological constant term competing against the  curvature term coming from the bulk triangles for large negative $m^2$, whereas the (extrinsic) curvature term for the boundary triangles goes to zero. 

We can estimate $l_{\text{crit}}$ by considering the leading term in the expansion of  $S_{\text{S-P}}(l,m^2)$ around $m^2\rightarrow -\infty$. 
     This leading term vanishes  for  
\ba\label{lcrit1}
l^2_{\text{crit}}&=&24\sqrt{2}\,\,\frac{\pi n_e-3 n_\tau\cos^{-1}(1/3) }{n_\tau \Lambda} \,.
\ea

The numerical values for  $\sqrt{\Lambda} a_{\text{crit}}:=\sqrt{\Lambda}l_{\text{crit}}/\nu(X)$ (with $\nu(X)$ given in Table \ref{Table2}) are given by (approximately) $2.909$ for the subdivided 5-cell, by $2.689$ for the subdivided 16-cell and by $2.474$ for the subdivided 600-cell. 

The fact that there are no stationary points for $l<l_{\text{crit}}$ reflects the situation in the continuum, where there are also no stationary points with real lapse if either $a(0)$ or $a(1)$ are smaller than $a_\Lambda \approx 1.732/\sqrt{\Lambda}$. But for the subdivided 4-polytopes we do have real solutions for $l>l_{\text{crit}}$, which deviates from the continuum case, given that these would still be configurations with initial value $l(0)=0$. Again, we are approximating a (homogeneously) curved space with piecewise flat simplices and we expect that this approximation becomes more and more unreliable for $l$ much larger than $1/\sqrt{\Lambda}$. We therefore see these real ``Lorentzian'' solutions as discretisation artefacts.

In Section \ref{Sec:Cont} we saw that in the continuum case there are stationary points for complex values of the lapse. For $a<a_\Lambda$ these stationary values for the lapse are purely imaginary, and describe a Riemannian (``Wick-rotated'') geometry. In fact, it is possible to identify saddle points for complex values of $m^2$ in (\ref{SBall1}), more precisely there is such a point for positive $m^2>\frac{3}{8} l^2$. This saddle point only exists for $l<l_{\text{crit}}$. Note that if we identify $m^2-\frac{3}{8} l^2$ as the analogue of $-N^2$ in the continuum, this corresponds to a solution with purely imaginary lapse; also, $m_{s}^2>\frac{3}{8} l^2$  summarises the generalised triangle inequalities for the Euclidean 4-simplices in our triangulation. Given that the square roots in the action (\ref{SBall1}) all take negative arguments for $m^2>\frac{3}{8} l^2$, we see that this stationary point is situated on the branch cut for these square roots\footnote{One can rotate the branch cut for the square roots, such that the saddle points do not lie on the branch cut. The rotation can be done in two different directions leading to the sign ambiguity discussed below.}. The value of the action evaluated on this critical point therefore depends on whether we consider the analytical continuation of the action in $m^2$ through the lower or upper complex half-plane\footnote{See also \cite{CDTReview} for a discussion of the analytical continuation for the Lorentzian Regge action, and its relation to the usual notion of Wick rotation.}. The analytically continued Lorentzian action leads then to $\pm S_{\text{S-P,E}}$ where 
\ba\label{SBall1E}
8\pi G S_{\text{S-P,E}}(l^2,m^2)\,:=\,&& n_t \, \frac{\sqrt{3}}{4} l^2  \left( \pi  -  2\cos^{-1}\left(   \frac{l}{  2\sqrt{2}\sqrt{-l^2+3m^2} }  \right)\right)\nn \\&&
+n_e \frac{1}{4} l\sqrt{-l^2+4 m^2}
\left( 2\pi -\frac{6n_\tau}{n_e} \cos^{-1}\left( \frac{l^2-2m^2}{2l^2-6m^2}\right)        \right) \nn\\&&
-n_\tau \Lambda \frac{l^3}{96}\sqrt{-3l^2+8m^2}\, .
\ea
 For this Euclideanised geometry we have $l^2>0$ and $m^2\geq \tfrac{3}{8} l^2$. Changing variables from $m^2$ to the height (square) $h^2=m^2-\tfrac{3}{8}l^2$, the second inequality simply becomes $h^2 \geq 0$.

\begin{figure}[H]
\centering
\includegraphics[scale=0.95]{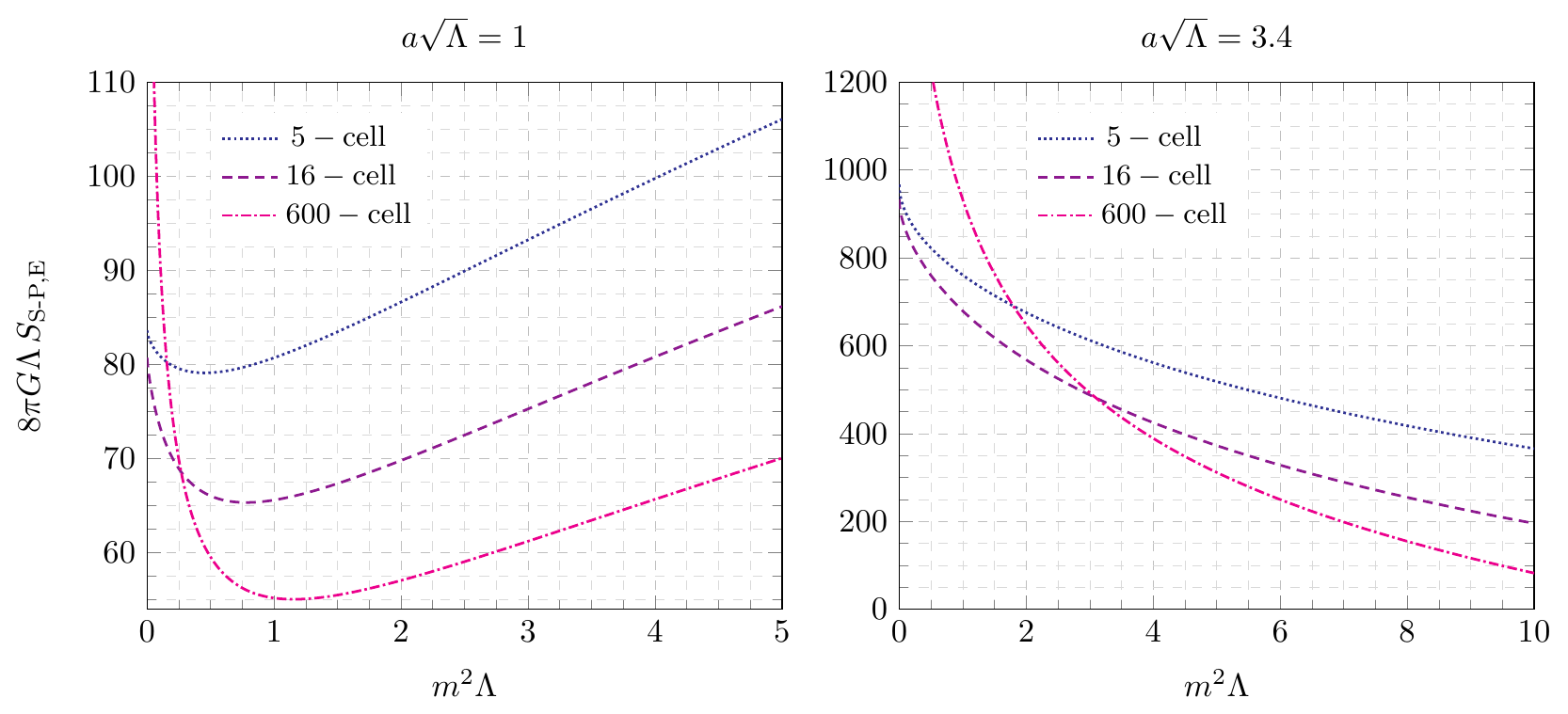}
\caption{Rescaled Euclidean Regge action $8 \pi G \Lambda\,S_{\text{S-P},E}$ as a function of $m^2 \Lambda$ for two different values of $a=l/\nu(X)$ for the different $X$-cell based triangulations. Left panel: $a=1/\sqrt{\Lambda}$. Right panel: $a=3.4/\sqrt{\Lambda}$.
 } \label{fig:4-Polytope-Euclidean}
\end{figure}

Figure \ref{fig:4-Polytope-Euclidean} shows plots of the action (\ref{SBall1E}) as a function of $m^2$, for different values of $a=l/\nu(X)$. We see that the existence of a stationary point in $m^2$ depends on the asymptotic behaviour of the action for large $m^2$. This asymptotic behaviour is characterised, as for the Lorentzian action (\ref{SBall1}), by the competition between the cosmological constant term and the curvature term for the bulk triangles. It leads to the same critical value $l_{\text{crit}}$ defined in (\ref{lcrit1}), and we now have a stationary point  $h^2_s>0$ (or $m^2_s>\frac{3}{8}l^2$) for $l<l_{\text{crit}}$. This stationary point moves to infinity if $l\rightarrow l_{\text{crit}}$.

\begin{figure}[H]
\centering
\includegraphics[scale=1]{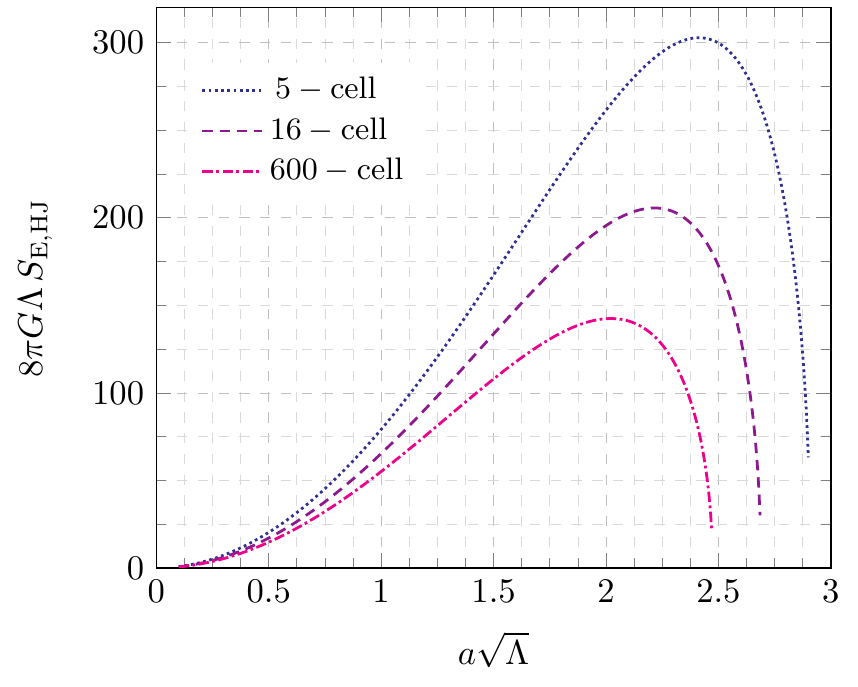}
\caption{The rescaled Hamilton--Jacobi function $8\pi G \Lambda\, S_{\text{E,HJ}}:= 8\pi G \Lambda\, S_{\text{S-P,E}}(l^2,m_s^2)$ as a function of $a \sqrt{\Lambda}$ with $a=l/\nu(X)$ for the three $X$-cell based triangulations.} \label{fig:4-Polytope-HJ-Function}
\end{figure}

We therefore conjecture that for $l<l_{\text{crit}}$, the semiclassical approximation of the path integral (\ref{SBallPI}) leads to an amplitude $\sim \exp( \pm S_{\text{S-P,E}}(l^2,m^2_s))$. The Hamilton--Jacobi function $S_{\text{S-P,E}}(l^2):=S_{\text{S-P,E}}(l^2,m^2_s)$ (i.e., the action evaluated on the solution $m^2_s$) can be computed numerically and is shown in Figure \ref{fig:4-Polytope-HJ-Function}. We see a similar behaviour to the action $S_{\text{P,E}}(l^2)$ for the 4-polytope without subdivision: the Hamilton--Jacobi function vanishes for $l=0$ and increases until a maximal positive value attained for a certain length $l_s<l_{\text{crit}}$. (In particular, since the Hamilton--Jacobi function is positive an amplitude $\exp( -S_{\text{S-P,E}}(l^2,m^2_s))$ leads to exponential suppression.) In contrast to the continuum version $S_E$, the discrete Hamilton--Jacobi function  $S_{\text{S-P,E}}(l^2)$ then decreases for $l>l_s$  until it reaches $0$ for $l=l_{\text{crit}}$. Again this happens in a regime where we would no longer trust a discretisation based on flat building blocks.

For the range $l<l_s$ 
the Hamilton--Jacobi function $S_{\text{S-P,E}}(l^2)$ increases monotonically and therefore behaves  similarly to the continuum version $S_E(a)$ in the regime $a<a_\Lambda$. Hence, we can again identify the discretisation-dependent value $a_s=l_s /\nu$ as the analogue of $a_\Lambda$. Table \ref{Table4} in Section \ref{Sec:NBP} below lists the values for $a_s$, as well as the value of the Hamilton--Jacobi function at $a_s$, for the three different subdivided 4-polytopes. Again we  see that the discretisation error decreases for finer triangulations.

In summary, the model based on a subdivided 4-polytope only gives a reasonable approximation to the continuum for scale factors smaller than the threshold $a_\Lambda$.  
In particular, it can not describe the change  from the Euclidean phase to a Lorentzian phase that is essential part of the ideas behind the continuum no-boundary proposal. Describing this transition requires a subdivision into several time steps. We will therefore consider a discretisation of the 4-ball by shells in Section \ref{Sec:Shells}.

Using different frameworks, the triangulation based on the subdivided 5-cell has been previously discussed in \cite{Hartle1985} and \cite{Mikov}, with the aim to describe the emergence of a macroscopic universe. We here provided a more complete analysis of the classical solutions and the Hamilton--Jacobi function, which revealed that the triangulations considered in this section can only be expected to describe the early Euclidean phase of the de Sitter quantum universe.

\subsubsection{Configuration with irregular light cone structure }

In the previous section we defined the path integral as an integral over the bulk length  square $m^2$, but restricted to the cases where the bulk triangles are time-like. The extrinsic curvature angles and the deficit angle at the bulk triangle were real in these cases, which led to a real action (\ref{SBall1}).  As discussed in Section \ref{Lorentzian-Regge}, real angles in Lorentzian Regge calculus describe a configuration with regular light cone structure. 

In the following we will consider the cases $(c)$ and $(d)$ defined in Section \ref{RegularLightCone}, where the bulk triangles are space-like. Note that this is a different configuration from the ``Wick rotated'' geometries we considered above, which can be understood to result from a deformation of the integration contour. There, the resulting geometry had Euclidean signature.  In this section, we will consider a geometry of Lorentzian signature,  but we will see that the resulting triangulations have an irregular light cone structure. These configurations do not  have an analogue  in the continuum. 

For the case $(c)$ where $\tfrac{1}{4} l^2<m^2<\tfrac{1}{3}l^2$ the Regge action is given by
\ba\label{SBallc}
8\pi G S^{(c)}_{\text{S-P}}(l^2,m^2)&=&- n_t \, \frac{\sqrt{3}}{2} l^2 \sinh^{-1}\left(   \frac{l}{2\sqrt{2}\sqrt{l^2-3m^2}}  \right)\nn \\&&+n_e \frac{1}{4} l\sqrt{-l^2+4 m^2}
\left( -2\pi\imath +\frac{6n_\tau}{n_e} \cosh^{-1}\left( \frac{l^2-2m^2}{2l^2-6m^2}\right)        \right) \nn\\&&
-n_\tau \Lambda \frac{l^3}{96}\sqrt{3l^2-8m^2}\, .
\ea
We see explicitly the appearance of an imaginary part $-2\pi\imath$ in the deficit angle, corresponding to triangles that have no light cones attached to them.
If we would use the action $S^{(c)}_{\text{S-P}}$ to define an amplitude $\exp(\imath S^{(c)}_{\text{S-P}})$ we would get exponentially enhanced configurations, leading to a result very different from the continuum. We can demand a regular light cone structure and exclude such configurations from the path integral. 

Similarly, for the case $(d)$ where $\tfrac{1}{3}l^2<m^2<\tfrac{3}{8} l^2$, we obtain
\ba\label{SBalld}
8\pi G S^{(d)}_{\text{S-P}}(l^2,m^2)&=&
 n_t \, \frac{\sqrt{3}}{4} l^2 \left( -\pi\imath - 2 \cosh^{-1}\left(   \frac{l}{2\sqrt{2}\sqrt{-l^2+3m^2}}  \right) \right) \nn \\&&
+n_e \frac{1}{4} l\sqrt{-l^2+4 m^2}
\left( \pi\imath +\frac{6n_\tau}{n_e} \cosh^{-1}\left( \frac{l^2-2m^2}{6m^2-2l^2}\right)        \right) \nn\\&&
-n_\tau \Lambda \frac{l^3}{96}\sqrt{3l^2-8m^2}\, .
\ea
Here the deficit angle at the bulk triangles includes three light cones, whereas the extrinsic curvature angle at the boundary triangles includes no light cone. Overall the action (\ref{SBalld}) has a negative imaginary part, which would also lead to exponentially enhanced configurations. As before, demanding a regular light cone structure excludes these configurations from the path integral.

\section{Modelling the universe with discrete shells}\label{Sec:Shells}

In the previous section we studied a discrete implementation of the no-boundary proposal in which the early stages of the universe's history are described by either a polytope or a subdivided polytope whose only boundary is a single triangulation of a three-sphere. We saw that such a discretisation captures the relevant dynamics of geometry in a purely Euclidean regime where the radius of the final three-sphere is small, but that it has no analogue of the continuum picture of the emergence of a Lorentzian geometry. To obtain a discrete model that better approximates the continuum FLRW dynamics, we now consider a triangulation that allows for arbitrarily many time steps instead of a single one. To achieve this, we subdivide the 4-ball into shells of topology $S^3\times[0,1]$. The ``shell universe'' is then obtained from gluing many such shells.

\subsection{Discretisation of the shells}

\begin{figure}[H]
\centering
\begin{tikzpicture}[scale=1.8,tdplot_main_coords]
    \coordinate (O) at (0,0,0);

     \draw node[at={(1.02,0,0.05)},anchor=north east]{$A$};
	 \draw node[at={(1.35,0,2.95)},anchor=north east]{$A'$};
	 \draw node[at={(0,1.8,0.25)},anchor=north east]{$B$};
	 \draw node[at={(-0.35,1.7,3.15)},anchor=north east]{$B'$};
	 \draw node[at={(-1.05,0.8,0.55)},anchor=north east]{$C$};
	 \draw node[at={(-1.42,0.8,3.52)},anchor=north east]{$C'$};
	 \draw node[at={(-0.1,0.7,1.49)},anchor=north east]{$D$};
	 \draw node[at={(-0.06,0.7,4.85)},anchor=north east]{$D'$};
     \draw[<->] (1.067,0.2,0.05) -- (0,0.95,-0.05);
	 \draw node[at={(0.38,0,-0.19)},anchor=north east]{$l_0$};
	 \draw[<->] (1.38,0.2,2.92) -- (-0.17,0.95,2.79);
	 \draw node[at={(0.38,0,2.65)},anchor=north east]{$l_1$};
	 \draw[<->] (1.075,0.2,0.05) -- (1.41,0.2,2.95);
	 \draw node[at={(1.2,0,1.43)},anchor=north east]{$m_0$};
	 \draw node[at={(0.53,0,1.3)},anchor=north east]{$d_0$};

    \tdplotsetcoord{A}{1}{90}{0}    
    \tdplotsetcoord{B}{1}{90}{90}   
    \tdplotsetcoord{C}{1}{80}{180}  
    \tdplotsetcoord{D}{3.5}{45}{225}  
    \tdplotsetcoord{E}{1}{0}{0}     
	\tdplotsetcoord{F}{4.3}{0}{0}
	\tdplotsetcoord{G}{3.5}{300}{100}
	\tdplotsetcoord{H}{2.5}{350}{0}

    \draw (A) -- (B) -- (C);
    \draw (E) -- (A) ;
    \draw (E) -- (B) ;
    \draw (E) -- (C) ;
	\draw[thick,dashed] (C) -- (D);
	\draw[thick,dashed] (E) -- (F);
	\draw[thick,dashed] (G) -- (A);
	\draw (G) -- (F);
	\draw (D) -- (F);
	\draw[thick,dashed] (H) -- (B);
	\draw (D) -- (H) -- (G);
	\draw (H) -- (F);
	\draw[thick,dotted] (D) -- (G);
    \draw[thick,dotted] (C)  -- (A);
	\draw[dashdotted] (B) -- (G);
\end{tikzpicture}
\caption{Illustration of a four-dimensional building block which is bounded by a base and a top tetrahedron $(A,B,C,D)$ and $(A',B',C',D')$ with edge lengths squared $l_0^2$ and $l_1^2$ respectively. The struts as well as the diagonals connecting the two tetrahedra have all the same length squared $m_0^2$ and $d_0^2$ respectively.} \label{fig:4D-Block}
\end{figure}
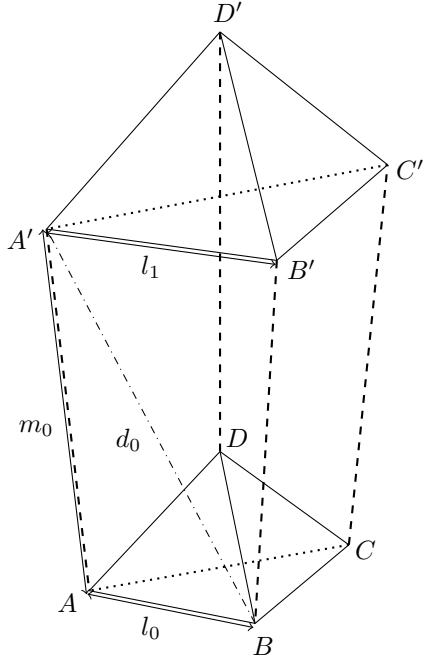

The shells we are interested in are bounded by an ``inner'' triangulation of the 3-sphere and an ``outer'' triangulation of a different 3-sphere. We will assume that both these triangulations are given by either the 5-cell, the 16-cell or the 600-cell, as discussed in Section \ref{Sec:Cells}. We then introduce auxiliary four-dimensional building blocks from which to build these shells, as illustrated in Fig. \ref{fig:4D-Block}. These auxiliary building blocks have topology  $\tau \times [0,1]$, where $\tau$ denotes a tetrahedron. Gluing these four-dimensional building blocks following the connectivity of the tetrahedra in the triangulation of the three-spheres, we obtain a triangulation of a four-dimensional spherical shell.
The auxiliary four-dimensional building block has three types of edges: six edges of the tetrahedron $\tau \times \{0\}$, six edges of the tetrahedron $\tau\times \{1\}$ and four edges $v\times[0,1]$ for each of the vertices $v$ of $\tau$.

We aim to model a homogeneous universe and therefore choose the length square of the edges in $\tau \times \{0\}$, resp.  $\tau \times \{1\}$, to be all  equal and given by $l_0^2>0$, resp. $l_1^2>0$.  Likewise we choose the length square of the four edges $v \times [0,1]$ to be equal and given by $m^2_0$. These edges can be time-like, null or space-like and are referred to as struts.   Demanding that the faces with edge length squares $(l_0^2,m_0^2,l_1^2, m_0^2)$ are isosceles trapeziums\footnote{To avoid confusion, here and in the following ``trapezium'' refers to the British English (and original Greek) terminology; in American English this would be called a trapezoid.} and that the  bulk geometry of the building block is (Lorentzian) flat fixes the geometry of the building block uniquely.   The building blocks now have the geometry of frusta -- the base and top tetrahedra are parallel to each other with respect to the internal flat geometry of the building block. 

\tdplotsetmaincoords{70}{165}

Alternatively, we can subdivide the building block into four  4-simplices. This can be done by introducing six additional edges, each of which diagonally subdivides one of the six quadrilateral faces into two triangles, see Appendix \ref{Tshells} for details. One then has additional parameters, given by the edge lengths $d^2_{0,j}$ of the diagonals, which one can also choose to be all equal $d^2_{0,j}=d^2_0$. This subdivision allows defining a Regge action for the now fundamentally simplicial building blocks  and therefore (after gluing the building blocks) for the entire shell.  
Choosing the lengths squares of the diagonals to be $d_0^2=l_0l_1+m_0^2$ forces the quadrilateral faces to be isosceles trapeziums. It also leads to the vanishing of all deficit angles associated to the triangles inside the building blocks, i.e., the building blocks again become frusta. 
Setting the length squares of the diagonals to $d^2_0=l_0l_1+m_0^2$  can be seen as an approximation to the exact solutions to the equations which would be obtained by varying the simplicial Regge action with respect to the lengths of the diagonals. In particular, flat building blocks will again only be a good approximation below length scales set by the cosmological constant in our model.

Inserting the ``flat'' values $d^2=l_0l_1+m^2_0$ for the length squares of the diagonals into the simplicial Regge action, we obtain a generalised Regge action for the discretisation of the shells with frusta. This discretisation has been originally proposed by Collins and Williams \cite{Collins1973} and also used in \cite{Liu2015}, but none of these works actually provide the action for these discretisations. 

We will consider the case of thin shells, that is we assume that $l^2_0 \gg |m^2_0|$ and $l^2_1 \gg |m^2_0|$. The struts, which constitute the non-parallel pair of opposite sides of the trapeziums, can be either time-like $m^2_0<0$, null $m^2_0=0$ or space-like $m_0^2>0$. For this reason $m^2_0$ does not provide an adequate measure for time \cite{Collins1973}.  We therefore introduce the height squared in the frusta, defined as the distance squared, as measured within the flat geometry of the frustum, between the bottom and top tetrahedron:
\ba\label{height1}
h^2_0:=-\frac{3}{8} (l_1-l_0)^2+m^2_0  \,.
\ea
The height squared $h_0^2$ is always negative for a Lorentzian frustum.

\subsection{ Signatures of the building blocks}\label{Sec:Signatures}

Depending on the relation between the height square and the growth rate $(l_1-l_0)^2$ for the spatial lengths, the building blocks can have the following signatures: 
\begin{itemize}
\item[$(a)$] Time-like struts, $m_0^2<0$ and so $3(l_1-l_0)^2<8|h_0^2|$.  All higher-dimensional building blocks containing the struts are also time-like.
\item[$(b)$] Space-like struts but time-like trapeziums containing the struts, $\frac{8}{3}|h_0^2|<(l_1-l_0)^2<8|h_0^2|$.
\item[$(c)$] Space-like struts and space-like trapeziums, but time-like three-dimensional building blocks (three-dimensional frusta) containing the trapeziums, $8|h_0^2|<(l_1-l_0)^2<24|h_0^2|$.
\item[$(d)$]  Space-like three-dimensional frusta, $(l_1-l_0)^2>24|h_0^2|$. 
\end{itemize}
Again, limiting cases where some building blocks are null are obtained by replacing inequalities with equalities.

These different cases may lead to different types of dihedral angles attached to the trapeziums. Recall from Section \ref{Lorentzian-Regge} that when defining the dihedral angle at a given (two-dimensional) face ${\cal F}$, the dihedral angle is Euclidean for a time-like face, whereas if the face is space-like the angle will be Lorentzian. In the latter case, the signatures of the two three-dimensional building blocks ${\cal T}_1, {\cal T}_2 $ hinging at the face are passed on to their projections. These projections give the edges of the wedge which defines the dihedral angle.  

We then see that the cases $(a)$ and $(b)$ do not differ in the types of dihedral angles, and the action for these two cases can be defined via the same function of the edge lengths. Going from $(b)$ to $(c)$ however, the signature of the trapeziums changes, and with it also the signature of the associated deficit angles. In $(c)$ we also encounter an irregular light cone structure associated to the trapeziums: projecting the frusta around a given trapezium to the planes orthogonal to the trapezium, we obtain a number of two-dimensional wedges which we can glue to a cone. The geometry of this cone is (away from the tip) Lorentzian. But in this case, there are no light cones attached to the tip itself, which defines an irregular light cone structure. This leads to an imaginary part for the action, as we will see below.  We will therefore argue that the case $(c)$ should be excluded from the path integral. The case $(d)$ also leads to an imaginary part of the Regge action, and should likewise be excluded from the path integral.

\subsection{ Generalised Regge action for shells and limit of continuous time evolution}\label{ThinShells}

We start by considering the cases $(a)$ and $(b)$, where we have time-like trapeziums.
The generalised Regge actions for the shells based on the  $X$-cell are  given by
\ba\label{ReggeFrust1}
8 \pi G \,S_{\text{Shell}}&:=&
-\frac{n_\tau}{96} \Lambda  \left(l_0+l_1\right) \left(l_0^2+l_1^2\right) \sqrt{3
   \left(l_0-l_1\right){}^2-8 m_0^2}
   \nn\\&&
   +\frac{n_e}{4} (l_0+l_1)  \sqrt{ \left(l_0-l_1\right){}^2-4
   m_0^2} \left(2\pi-\frac{6 n_\tau}{n_e} \sec ^{-1}\left(\frac{2 m_0^2}{2
   m_0^2-\left(l_0-l_1\right){}^2}+2\right) \right)
   \nn\\&&
   -\frac{n_t}{8} \sqrt{3} \Bigg(\, l_0^2 \Bigg(3
   \sinh ^{-1}\left(\frac{l_0-l_1}{2 \sqrt{2} \sqrt{\left(l_0-l_1\right){}^2-3
   m_0^2}}\right)
   \nn\\&&
\q \q \q \q\q-\sinh ^{-1}\left(\frac{1}{2} \sqrt{-\frac{ \left(2 l_0^2+3
   l_1^2-5 l_0 l_1-6 m_0^2\right)^2}{\left(3
   m_0^2-\left(l_0-l_1\right){}^2\right) \left(-l_0^2+3 l_0 l_1+3 m_0^2\right)}}\right)
   \nn\\&&
\q\q\q  \q\q +\cosh ^{-1}\left(\frac{l_0+3 l_1}{2 \sqrt{-2 l_0^2+6
   l_0 l_1+6 m_0^2}}\right)\Bigg)
   \nn\\&&
  \q\q \,\,\,\,\,+\;l_1^2 \Bigg(3 \sinh ^{-1}\left(\frac{l_1-l_0}{2 \sqrt{2}
   \sqrt{\left(l_0-l_1\right){}^2-3 m_0^2}}\right)
   \nn\\&&
  \q\q\q\q \q
  -\sinh ^{-1}
  \left(\frac{1}{2}
   \sqrt{-\frac{ \left(3 l_0^2+2 l_1^2-5 l_0 l_1-6 m_0^2\right)^2  }{\left(3 m_0^2-\left(l_0-l_1\right){}^2\right) \left(-l_1^2+3 l_0 l_1+3
   m_0^2\right)}}\right)
   \nn\\&&
  \q\q\q\q\q +\cosh ^{-1}\left(\frac{3
   l_0+l_1}{2 \sqrt{-2 l_1^2+6 l_0 l_1+6 m_0^2}}\right)\Bigg)\Bigg) \,,
\ea
where for each choice of $X$-cell the coefficients $n_\tau,n_t$ and $n_e$ are given in Table \ref{Ntable1}.

Stacking $T$ shells and using an action
\ba
S_T= \sum_{i=0}^{T-1} S_{\text{Shell}}(l_i,l_{i+1},m^2_i)
\ea
 we can now consider the overall evolution from a universe with edge length  $l_0$ to one with edge length $l_T$. Fixing these boundary lengths we would need to solve for the lengths $l_i$  and the length squares $m^2_i$ for $i=1,\ldots , T-1$.  The sum of the heights $\chi_i:=\sqrt{-h^2_i(l_i,l_{i+1},m^2)}$ from $i=0$ to $i=T-1$ would provide us with a measure of time -- indeed this sum gives the proper time for a particle travelling along a trajectory along the central axis of the frusta.  

Comparing with continuum general relativity, we would expect that this system displays a gauge symmetry, which describes the freedom to reparametrise the time coordinate. If such a symmetry was realised, the solutions to the system would not be unique; instead we would be able to choose the heights $\chi_i$ freely and would need to only solve for the lengths $l_i$.

Discretisations however typically break such reparametrisation symmetries \cite{BahrDittrich09a}. This is also the case for the stacked shells: a numerical investigation of the solutions shows that these are unique, but also that there are directions around these solutions where the  action is almost constant. This broken symmetry can be restored by two procedures:

$(i)$ Keeping the discrete time evolution we can iteratively subdivide one time step into two smaller time steps and integrate out the bulk variables \cite{BahrDittrich09a, PerfectPI}.  The action for one time step then converges to the Hamilton--Jacobi function of an underlying Lagrangian. Hamilton--Jacobi functions are so-called perfect actions \cite{Improved,DittrichDiff}: they define a discrete time evolution which exactly mirrors the time evolution of the underlying continuum system. 
 
 $(ii)$ Alternatively we can take the limit of infinitesimal time steps $\chi_i$ for the shell actions $S_{\text{Shell}}(l_i,l_{i+1},\chi_i)=S_{\text{Shell}}(l_i,l_{i+1},m_i^2(l_i,l_{i+1},\chi_i))$. This will result in a Lagrangian which can be used to define continuum dynamics. These continuum dynamics will feature a gauge symmetry given by time reparametrisations. 
 
We will follow $(ii)$. Assuming an infinitesimal time step we parametrise the height of the frusta as $\chi_0=N\, {\rm d}t$, where $N$ plays the role of a lapse corresponding to the freedom to redefine the variable $t$. The labels $l_i$ representing edge lengths at different time steps are replaced by a differentiable function $l(t)$, and the difference between the length parameter at two successive times is then $(l_{i+1}-l_i)\rightarrow \dot l \,{\rm d}t$. Taking the limit ${\rm d}t\rightarrow 0$ we then obtain the Lagrangian
\ba
8\pi G\, L_{\text{Shell}} &:=& \lim_{{\rm d}t \to 0} \frac{8\pi G\,S_{\text{Shell}}( l, l+ \dot l\, {\rm d}t, N\, {\rm d}t)}{{\rm d}t}\nn
\\&=& -\frac{n_\tau}{6\sqrt{2}} \,N\Lambda  l^3 +\frac{n_e}{\sqrt{2}}\, Nl \sqrt{8 -(\dot{l}/N)^2} \left(\pi -\frac{3n_\tau}{n_e} \cos
   ^{-1}\left(\frac{8+(\dot{l}/N)^2 }{24-({\dot l}/N)^2}\right)\right)
   \nn\\&&
   -2\sqrt{3}n_\tau \, Nl \,(\dot{l} /N)\sinh
   ^{-1}\left(\frac{\dot{l}/N}{\sqrt{24 -({\dot l}/N)^2}}\right)\,,
   \label{LShell}
\ea
where $n_\tau$ and $n_e$ are listed in Table \ref{Ntable1} and we have now used $n_t=2n_\tau$ to simplify notation.

Again, we could introduce dimensionless length and lapse variables $\tilde{l}=l\sqrt{\Lambda}$ and $\tilde N=N\sqrt{\Lambda}$ and a dimensionless Lagrangian $8\pi G\Lambda L_{\text{Shell}}$ (a function of $\tilde l$ and $\tilde N$) which does not depend on $\Lambda$.

The Lagrangian (\ref{LShell}) describes a continuous time evolution of triangulated 3-spherical shells, i.e., a spatially discretised system which evolves continuously in time. The spatially discretised system is described by only one configuration variable $l$, which gives the edge lengths of  the maximally symmetric triangulation of the shells. As in the continuum, we can choose a lapse parameter $N$ which determines the relation between the time coordinate and proper time (measured along the trajectories of the centres of the tetrahedra). 

The Lagrangian (\ref{LShell}) has some similarities with the minisuperspace Lagrangian defined by (\ref{GRaction}), but a key difference is that the shell Lagrangian (\ref{LShell}) includes arbitrarily high powers of $\dot l$, whereas the minisuperspace Lagrangian (\ref{GRaction}) only includes a quadratic power of $\dot a$.  But in a regime where $\dot l/N$ is small\footnote{In our conventions, $l$ and $N$ have units of length, so the time parameter $t$ does not carry units. The expression $\dot l/N$ is then dimensionless.}, we can truncate $L_{\text{Shell}}$ to quadratic order in $\dot l$, which results in
\ba\label{Lshelltr}
8 \pi G\, L^{\text{Tr}}_{\text{Shell}}(l,\dot l, N;\Lambda)&:=&  c_1N l  - c_2  \frac{l \dot{l}^2}{N} - c_3 \Lambda N l^3 \q 
\ea
where
\ba\label{eq57}
c_1:= n_e\left( 2  \pi- \frac{6n_\tau}{n_e} \cos^{-1}\left(\tfrac{1}{3}\right) \right)\, ,\q
c_2:= \frac{n_\tau}{2\sqrt{2}}+\frac{n_e}{8}\pi - \frac{3 n_\tau}{8} \cos^{-1}\left(\tfrac{1}{3}\right)\, ,\q
c_3:= \frac{n_\tau}{6\sqrt{2}} \, .
\ea
The Lagrangian $L^{\text{Tr}}_{\text{Shell}}$ now has the same form as the minisuperspace version (\ref{GRaction}), just that we use the edge length $l$  instead of the scale factor $a$ as variable. We can match this length $l$ to the scale factor by either equating the volume or the integrated curvature of the triangulated 3-shells with the ones of the 3-sphere, see Section \ref{Sec:Cells}.  Replacing $l$ with the scale factor $a$, we obtain
\ba\label{Lshelltr2}
L^{\text{Tr}}_{\text{Shell}}(a,\dot a, N;\Lambda)&=& \frac{3\pi}{4G}\left( \tilde c_1N a  - \tilde c_2  \frac{a \dot{a}^2}{N} - \tilde c_3 \Lambda N a^3\right) 
\ea
and can compare the coefficients $\tilde c_1,\tilde c_2, \tilde c_3$ with the ones from the minisuperspace version.  The numerical values are listed in Table \ref{Table3}, which shows that the truncated Lagrangian for the 600-cell matches reasonably well with the minisuperspace version (\ref{GRaction}). We also generally see that matching the three-volumes of boundary spheres gives a better approximation to the continuum than if we match the integrated intrinsic curvature.

\begin{table}[!]
\begin{tabular}{|l||c|c|c|c||c|c|c|c||}\hline
&\; $\tilde c_1^\nu $ \;& \;$\tilde c_2^\nu $\; & \;$\tilde c_3^\nu $\;& \;$(a_\text{min}^\nu)^2\Lambda $\;& \;$\tilde c_1^\xi $\;& \;$\tilde c_2^\xi $\;& \;$\tilde c_3^\xi $\;& \;$(a_\text{min}^\xi)^2 \Lambda $\;\\ \hline
5-cell & 1.410& 1.916&1/3& 4.230 &1&0.683&0.119&8.411 \\ \hline
16-cell & 1.205 & 1.360&1/3& 3.616&1& 0.777&0.190&5.251\\ \hline
600-cell & 1.020 & 1.027& 1/3&3.061&1&0.967&0.314& 3.186  \\ \hline\hline
Cont.&1&1&1/3&3&1&1&1/3&3\\ \hline
\end{tabular}
\caption{Coefficients $(\tilde c_1,\tilde c_2, \tilde c_3)$ appearing in the truncated Lagrangian (\ref{Lshelltr2}), obtained by either comparing the 3-volume $(\tilde c_1^\nu,\tilde c_2^\nu, \tilde c_3^\nu)$ or the integrated curvature $(\tilde c_1^\xi,\tilde c_2^\xi, \tilde c_3^\xi)$ of the triangulated 3-spheres with the volume or curvature of the continuum 3-sphere, respectively. We furthermore give the values $(a_\text{min})^2\Lambda$, where $a_\text{min}$ is the minimal scale factor for a solution to exist, for the various discretisations. We list the corresponding values for the continuum  Lagrangian in the last row. \label{Table3}}
\end{table}

\subsection{Classical dynamics in the continuum time limit}

The non-truncated Lagrangian (\ref{LShell}) differs from the truncated version, and therefore also from the minisuperspace continuum theory, by terms that are higher order in $(\dot l/N)$. These higher order terms can be understood as a result of the spatial discretisation. Indeed, we started our construction by approximating continuum spacetime with a discretisation built from flat building blocks. This approximation can be expected to fail when the building blocks become large compared to the curvature radius of the continuum spacetime. 

To see when the higher order terms in $(\dot l/N)$ become relevant, we can use the equation of motion of the shell Lagrangian (\ref{LShell}) obtained from varying with respect to the lapse variable.  This equation of motion corresponds to the Hamiltonian constraint and determines  $(\dot l/N)$ via the equation
\ba\label{detdotl}
F((\dot l/N)^2) :=\left(8-(\dot l/N)^2 \right)^{-1/2} \left(  48\frac{n_e}{n_\tau} \pi-       144 \cos^{-1}\left(\frac{8+(\dot l/N)^2}{24-(\dot l/N)^2} \right) \right)  &=&  \Lambda l^2 \,.
 \ea
The left-hand side of (\ref{detdotl}) defines a function  $F$ of $(\dot l/N)^2$ which is well defined for $(\dot l/N)^2<8$. 

\begin{figure}[H]
\centering
\includegraphics[scale=0.96]{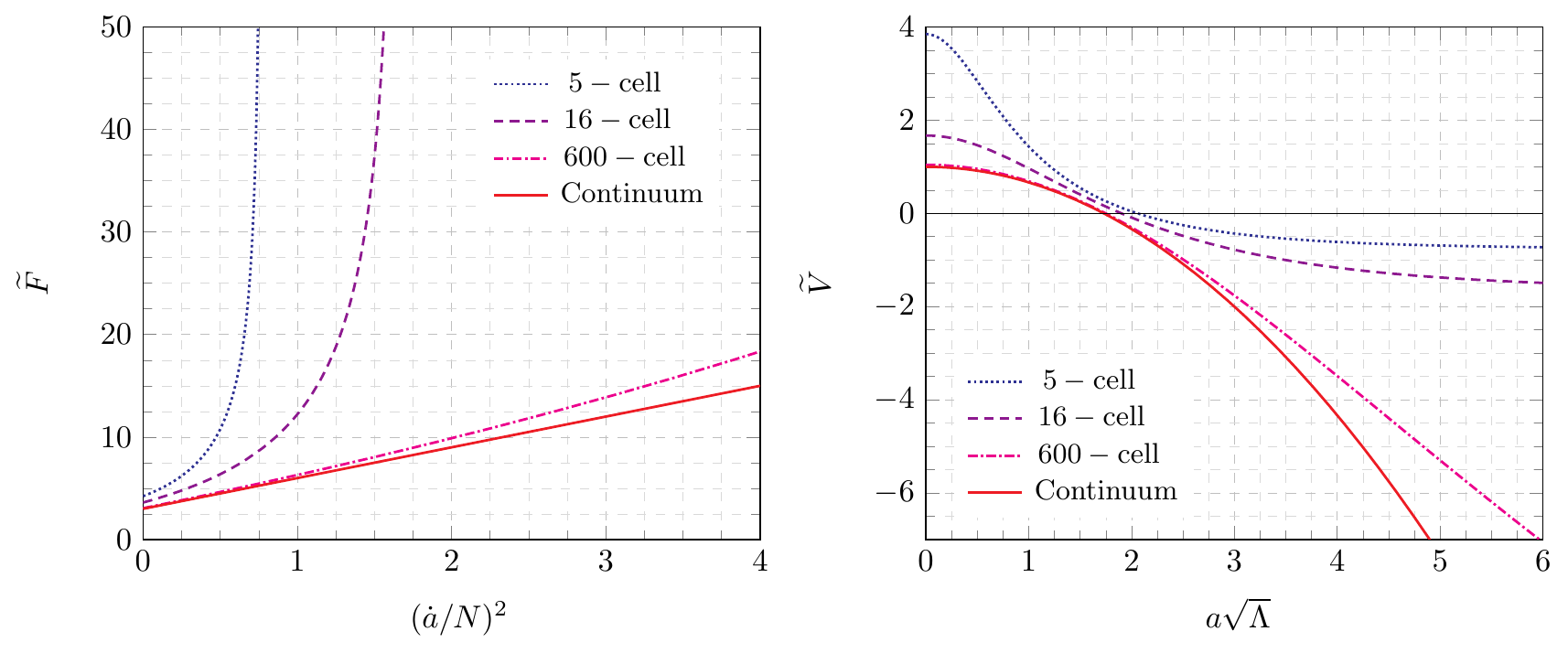}
\caption{Left panel: Rescaled versions $\widetilde{F}= \widetilde{F}(X)$ of $F$ as functions of $(\dot{a}/N)^2$ given by $\widetilde{F} \left((\dot{a}/N)^2 \right) := F \left((\nu(X) \dot{a}/N)^2\right)/(\nu(X)^2)$. Recall that $X=5,16,600$ is the number of tetrahedra in the inner resp. outer $3$-spheres of the four-dimensional blocks, and $\nu(X)$ is the volume conversion factor for the $X$-cell given in Table \ref{Table2}. $a = l/\nu(X)$ is the triangulation-dependent scale factor. The continuous red line represents 
$\widetilde{F} \left(( \dot{a}/N)^2\right)$ for the general relativistic continuum given by $\widetilde{F} = 3\left(1+ \left(\dot{a}/N \right) \right)$. 
Right panel: The potential $\widetilde{V} = \widetilde{V}(X)$ as a function of 
$a \sqrt{\Lambda}$ given by $\widetilde{V}(a \sqrt{\Lambda}) = - \widetilde{F}^{-1}( a \sqrt{\Lambda} )$. 
The continuous red line is the potential for the general relativistic case given by $\widetilde{V} = - ((a \sqrt{\Lambda} )^2/3 -1)$.
} \label{fig:F-and-Potential-Full}
\end{figure}

\begin{figure}[H]
\centering
\includegraphics[scale=0.96]{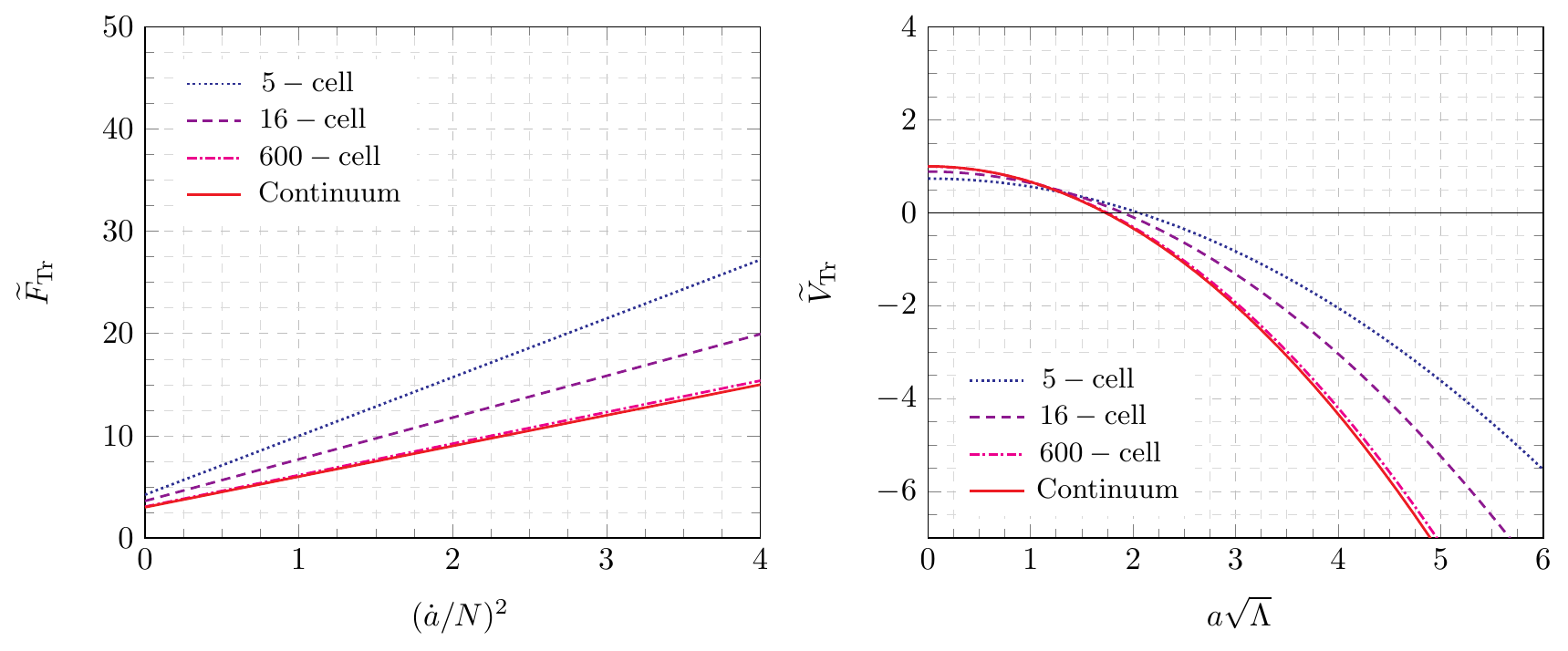}
\caption{Right and left panel: See caption of Fig.~\ref{fig:F-and-Potential-Full} but with $\widetilde{F}(X)$ replaced by $\widetilde{F}_{\text{Tr}}(X)$ which is the truncation of $\widetilde{F}(X)$ to second order in $(\dot{a}/N)$, and $\widetilde{V}(X)$ replaced by $\widetilde{V}_{\text{Tr}}(X)$ obtained in the same manner as in Fig.~\ref{fig:F-and-Potential-Full} but using $\widetilde{F}_{\text{Tr}}(X)$. In both panels, the continuous red lines represent the continuum case and are hence identical to the corresponding curves in Fig.~\ref{fig:F-and-Potential-Full}.} \label{fig:F-and-Potential-Truncated}
\end{figure}

Starting from $F(0)=( 48\pi n_e/n_\tau -144 \cos^{-1}(1/3))/\sqrt{8}>0$, $F$ increases monotonically for growing $(\dot l/N)^2$ and diverges for $(\dot l/N)^2\rightarrow 8$, see Figure \ref{fig:F-and-Potential-Full} where we have converted $F$ into a function that depends on the effective discretisation dependent scale factor $a=l/\nu(X)$ in order to facilitate comparison with the continuum. We see that $(i)$ there is a threshold value for $l$, given by $l_{\text{min}}:=\sqrt{F(0)/\Lambda}$, below which the solutions for $\dot l/N$ cannot be real, and that $(ii)$ $(\dot l/N)$ is small for small differences $(l-l_{\text{min}})$, but grows with this difference. Table \ref{Table3} details the (dimensionless) values $a^2_{\text{min}}\Lambda$, using the two different translations of $l$ to $a$ discussed in Section \ref{Sec:Cells}.

These two effects are also captured by the truncated Lagrangian (\ref{Lshelltr}), which leads to the truncated Hamiltonian constraint 
\ba
8\pi G\,H_{\text{Shell}}^{\text{Tr}}:= -c_1 l - c_2 l (\dot l/N)^2 + c_3 \Lambda l^3 
\ea
which agrees in this general form (though not in the exact values of the coefficients $c_i$) with the continuum minisupersace Hamiltonian.  The Hamiltonian constraint determines $(\dot l/N)^2$ to be
\ba\label{detdotl2}
F_{\text{Tr}}((\dot l/N)^2) \,:=\, c_3^{-1}(c_2(\dot l/N)^2+c_1  )    &=&\Lambda l^2 \, .
\ea
The threshold values $l^{\text{Tr}}_{\text{min}}:=\sqrt{F_{\text{Tr}}(0)/\Lambda}$  agree with $l_{\text{min}}=\sqrt{F(0)/\Lambda}$ (as this is where $(\dot l/N)=0$ and the truncation leads to an exact result), but in the truncated version we no longer see the bound $(\dot l/N)^2<8$ on the extrinsic curvature (see Figure \ref{fig:F-and-Potential-Truncated}).

\begin{figure}[H]
\centering
\includegraphics[scale=1]{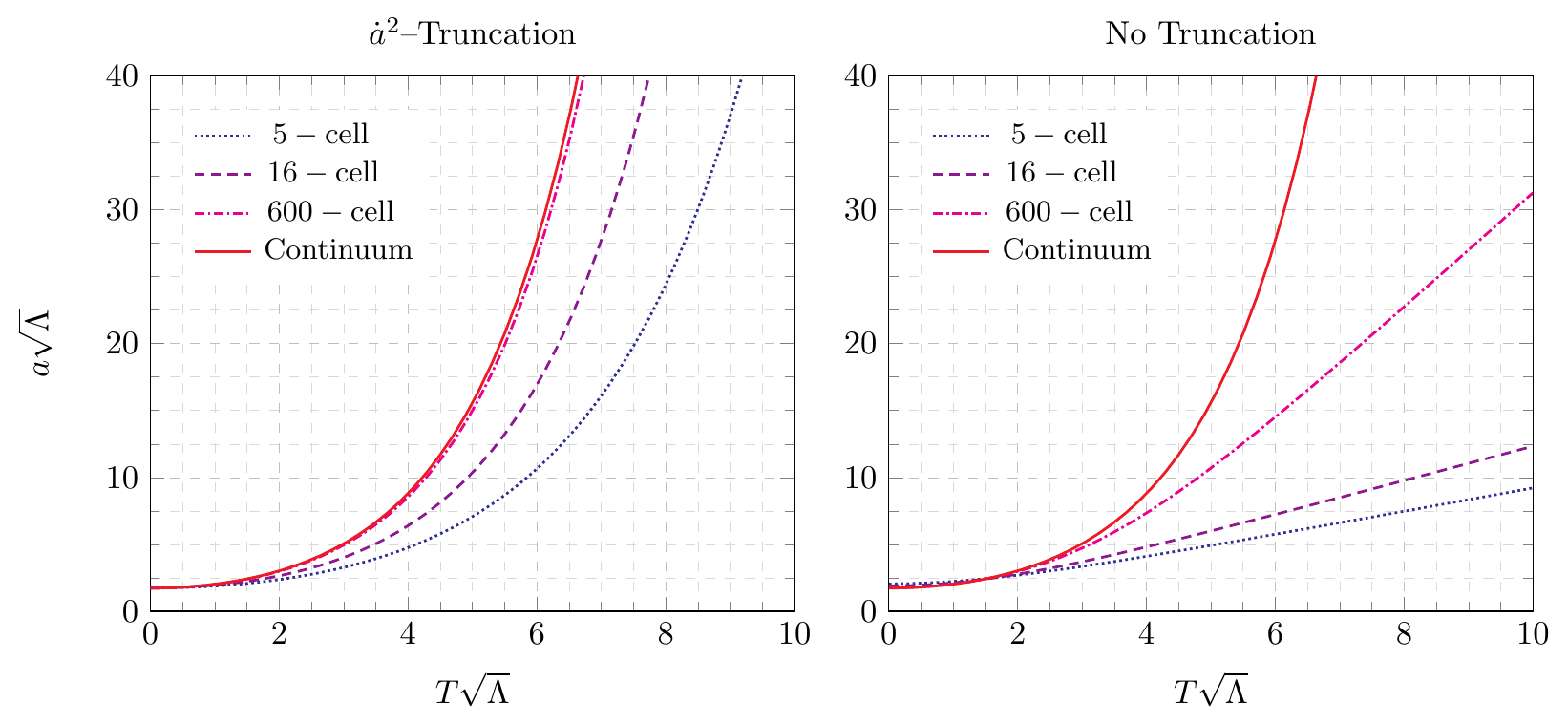}
\caption{Left panel: Analytical solutions $a(T)$ (where $T$ is proper time) of the equations of motion obtained from the Regge Lagrangians $L_{\text{Shell}}^{\text{Tr}}$ for different $X$-shell based triangulations, with cosmological constant $\Lambda$ and truncated at second order in $(\dot{a}/N)$. Right panel: Numerical solutions $a(T)$ of the equations of motion obtained from the full Regge Lagrangians $L_{\text{Shell}}$ with cosmological constant $\Lambda$. The continuous red lines in both panels represent the exact general relativistic solutions $a(T) = a_\Lambda \cosh(T/a_\Lambda)$. As in previous plots, we rescale $T$ and $a$ with $\sqrt{\Lambda}$ to obtain dimensionless quantities.} \label{fig:Lambda-Solutions}
\end{figure}

From (\ref{detdotl}) and (\ref{detdotl2}) we obtain $(\dot l/N)^2 = F^{-1}(\Lambda l^2)$, respectively $(\dot l/N)^2 = F^{-1}_{\text{Tr}}(\Lambda l^2)$. We can consider $\dot l/N$ as a (dimensionless) generalised velocity, and thus interpret $V(l)=-F^{-1}(\Lambda l^2)$ and $V_{\text{Tr}}(l)=-F^{-1}_{\text{Tr}}(\Lambda l^2)$ as (dimensionless) potentials for a dynamical system with total energy constrained to zero. 
Solutions to the equations of motion for $l(t)$ can be (in principle) obtained by direct integration (if we assume that $N=N(l)$ does not explicitly depend on $t$),
\ba
\int_{l_0}^{l_1} \frac{{\rm d}l}{N\sqrt{ -V(l)}} &=&\pm  \int_{t_0}^{t_1} {\rm d}t  \, .
\ea
See Fig.~\ref{fig:Lambda-Solutions} for plots of the resulting numerical solutions. Starting the system at a value $l_{\text{ini}} = l_{\text{min}}$ and choosing positive $\dot l$ and $N$, the decaying potential leads to a growing generalised velocity $\dot l/ N$. For the truncated version of our dynamics we have $V_{\text{Tr}}(l)\rightarrow-\infty$ for $l\rightarrow \infty$ so that the expansion keeps accelerating, but for the non-truncated version of the dynamics the potential asymptotes to $V(l)\rightarrow -8$. The generalised velocity $\dot l/N$ therefore also has a maximal value $|\dot l/N|_{\text max}=\sqrt{8}$ in the non-truncated version. 

Note that as $\dot l$ can be positive or negative and $N$ can also  be positive or negative, we will encounter the same sign ambiguities as in the discussion for the continuum in Section \ref{Sec:Cont}.  In Figure \ref{fig:Lambda-Solutions}, we only depict the case $\dot l\geq 0$ and $N>0$.

The continuum and discrete (non-truncated) versions differ in their evolution for late times and larger lengths $l$: the continuum solution continues to expand exponentially (in proper time) whereas solutions for the discrete systems asymptote to a linear growth. This discrepancy is due to the (spatial) size of the flat building blocks growing in time; we therefore approximate a constant curvature solution with a gluing of larger and larger flat  building blocks, leading to larger and larger discretisation errors.  This discretisation artefact could be avoided by refining the triangulation during time evolution \cite{DittrichSteinhaus13}. Such a refinement can be implemented (classically and quantum mechanically), at least if we use discrete time steps, by following the formalism of \cite{DittrichHoehn}. We leave exploration of this option to future work.

Next, we discuss the bound $(\dot l/N)^2<8$ on the generalised velocity, which appears in the non-truncated version but not in the truncated version of the Lagrangian. Interestingly, this bound does {\it not} arise from dihedral angles around triangles in the triangulations of the spatial three--spheres. These dihedral angles provide a measure for the extrinsic curvature of these spheres, and their value (being Lorentzian angles)  diverges in the limit $(\dot l/N)^2 \rightarrow 24$. 
 
We have seen that the potential $V(l)=-F^{-1}(\Lambda l^2)$ is bounded by $V(l)>-8$. As the Hamiltonian constraint imposes $(\dot l/N)^2=-V(l)$, the bound on the generalised velocity  $(\dot l/N)$ can be interpreted to result from the dynamics. 
 
But the maximal value $(\dot l/N)^2=8$ also has a kinematical interpretation: going back to the discussion of discrete building blocks in Section \ref{Sec:Signatures}  it corresponds to the  differentiation between the cases $(b)$ and $(c)$, described by $(l_1-l_0)^2\lessgtr 8 |h_0^2|$.  For $(l_1-l_0)^2= 8 |h_0^2|$ the trapeziums in the triangulation of the shell are null. If we consider dynamics defined by the continuum limit in the time parameter (given by $h_0$), we cannot change from the regime where the trapeziums are time-like to the regime where the trapeziums are space-like. We can however not exclude the possibility that such a transition appears for the dynamics defined by the fully discrete Regge action. 

Finally, let us note that whereas the threshold for the length derivative $(\dot l/N)^2=8$ is the same for all three $X$-cell based triangulations, the translation from $l$ to the scale factor variables, e.g., by defining a scale factor via comparing three-volumes as $a=l/\nu(X)$, leads to a triangulation-dependent threshold in terms of the scale factor $(\dot a/N)^2=8/ \nu^2(X) $.  This threshold increases from the 5-cell to the 600-cell from $(\dot a/N)^2\approx 0.77$ to $(\dot a/N)^2\approx 18.70$ (see also the divergence in Fig.~\ref{fig:F-and-Potential-Full}). In this sense one could argue that a refined truncation agrees more closely with the continuum, given that it allows for faster expansion rates in terms of the effective scale factor $a$.

 \subsection{Configurations with irregular causal structure}
 
 To close this discussion of the classical dynamics, we will consider the cases $(c)$ and $(d)$ discussed in Section \ref{Sec:Signatures}: here the trapeziums are space-like and the three-dimensional building blocks containing the trapeziums are time-like (for $(c)$) or space-like (for $(d)$). We can again compute the generalised Regge action. To obtain a simpler expression we perform the continuum limit in the time parameter in the same way as for the cases $(a)$ and $(b)$. We obtain for the case $(c)$
\ba\label{LShellCc}
8\pi G\,L^{(c)}_{\text{Shell}} &:=& -\frac{n_\tau}{6\sqrt{2}} \,N\Lambda  l^3 +\frac{n_e}{2\sqrt{2}}\, Nl \sqrt{(\dot{l}/N)^2-8} \left(-2\pi\imath +\frac{6n_\tau}{n_e} \cosh
   ^{-1}\left(\frac{8+(\dot{l}/N)^2 }{24-({\dot l}/N)^2}\right)\right)
   \nn\\&&
   -\sqrt{3}n_t \, Nl \,(\dot{l} /N)\sinh
   ^{-1}\left(\frac{\dot{l}/N}{\sqrt{24 -({\dot l}/N)^2}}\right)\, .
\ea
 We see that the deficit angle 
 \ba
 \epsilon_T=-2\pi\imath +\frac{6n_\tau}{n_e} \cosh
   ^{-1}\left(\frac{8+(\dot{l}/N)^2 }{24-({\dot l}/N)^2}\right)
 \ea
attached to the trapeziums has a negative imaginary part, $\text{Im}(\epsilon_T)=-2\pi$. As in the discussion of (\ref{SBallc}), this negative imaginary part represents triangles with no light cones attached to them. The action (and Lagrangian) then also has a negative imaginary part which would lead to an exponential enhancement of these configurations in the path integral \cite{Sorkin2019, EffSF3}. One can restrict the path integral to causally regular configurations, which would exclude the case $(c)$ and also $(d)$. 

For completeness, we give also the Lagrangian for the case $(d)$:
\ba\label{LShellCd}
8\pi G\,L^{(d)}_{\text{Shell}} &:=& -\frac{n_\tau}{6\sqrt{2}} \,N\Lambda  l^3 +\frac{n_e}{2\sqrt{2}}\, Nl \sqrt{(\dot{l}/N)^2-8} \left(\pi\imath +\frac{6n_\tau}{n_e} \cosh
   ^{-1}\left(\frac{8+(\dot{l}/N)^2 }{({\dot l}/N)^2-24}\right)\right)
   \nn\\&&
   +\frac{\sqrt{3}}{2}n_t \, Nl \,(\dot{l} /N)\left( -\pi\imath   - 2\cosh
   ^{-1}\left(\frac{\dot{l}/N}{\sqrt{({\dot l}/N)^2-24}}\right)    \right) \, .
   \ea
  
Here we have a positive imaginary part  resulting from deficit angles attached to the trapeziums and a negative imaginary part resulting from deficit angles attached to space-like triangles. These lead to a negative imaginary part for the Lagrangian, and therefore also to an exponential enhancement of these configurations if they are included in the path integral.

\subsection{The no-boundary proposal} \label{Sec:NBP}

In (\ref{LShell}) we defined a Lagrangian $L_\text{Shell}$ which describes the evolution of a triangulated three-sphere in continuum time. We have also discussed that this Lagrangian approximates well the continuum Lagrangian (\ref{GRaction}) in the regime of smaller edge lengths $l$.  This suggests that we can define the no-boundary path integral the same way as in the continuum discussion of Section \ref{Sec:Cont}. 

We define the  path integral as an integral over trajectories $(l(t),N(t))$ which start from $l(0)=0$ and end at $l(1)=l_1$, where $l_1$ is a free parameter.  By taking the continuous time limit of the Regge action for the shells we regained time reparametrisation invariance. We can therefore gauge-fix the integration over $N(t)$, e.g., by choosing $N(t)=N_0$, and remain with an ordinary integral over $N_0$ (which we can choose to restrict to $N_0>0$). 

However, there is an important difference to the continuum FLRW case:  for the dynamics of the discrete shells we saw that we had three different regimes, depending on whether $(\dot l/N)^2<8$ (case $(A)$), $8<(\dot l/N)^2<24$ (case $(B)$), or $24<(\dot l/N)^2$ (case $(C)$). These three cases resulted from the distinction between cases in which the trapeziums in the four-dimensional triangulations are $(A)$ time-like, $(B)$ space-like with time-like three-dimensional building blocks containing the trapeziums, or $(C)$ space-like with space-like three-dimensional building blocks containing the trapeziums.  The cases $(B)$ and $(C)$ define  configurations with irregular light cone structure and the action for these configurations includes negative imaginary parts. Using such an action in the path integral leads to an exponential amplification of these configurations which can easily lead to non-sensible results, see \cite{EffSF3}. To avoid this, we will exclude these configurations from the path integral and restrict to trajectories with $(\dot l/N)^2<8$.

We have seen that the Lagrangian $L_\text{Shell}$ does not admit a classical real solution for $l<l_\text{min}=\sqrt{F(0)/\Lambda}$. Thus, we will also not have a stationary point for the path integral along the real (and positive) axis for $l$ and $N$. This is similar to the continuum FLRW case, where we do not have a classical real solution for scale factors $a<a_\Lambda$. But in this case in the continuum, there are stationary points if we analytically continue $N\rightarrow \pm \imath N$.  This transformation leads to an Euclideanisation of the action (and therefore the Lagrangian) $S\rightarrow  \pm  \imath S_{E}$ where $S_E$ is the action for the Wick-rotated geometry. 

Let us therefore consider a ``Wick rotation'' for the simplicial geometry. We define this Wick rotation to be given by $h_0^2 \rightarrow -h_0^2$ for configurations with discrete time steps, which leads to $N\rightarrow \pm \imath N$ in the time continuum limit. We will here consider as starting point only Lorentzian configurations with regular light cone structure. 

The time continuum limit of the Regge action for the Euclidean configurations is given by (here $N$ is a positive variable)
\ba\label{LShell-E}
8\pi G\, L_{\text{Shell-E}} &:=& -\frac{n_\tau}{6\sqrt{2}} \,N\Lambda  l^3 +\frac{n_e}{2\sqrt{2}}\, Nl \sqrt{8 +(\dot{l}/N)^2} \left(2\pi -\frac{6n_\tau}{n_e} \cos
   ^{-1}\left(\frac{8-(\dot{l}/N)^2 }{24+({\dot l}/N)^2}\right)\right)
   \nn\\&&
   +\frac{\sqrt{3}}{2}n_t \, Nl \,(\dot{l} /N) \left( \pi - 2 \cos
   ^{-1}\left(\frac{\dot{l}/N}{\sqrt{24 +({\dot l}/N)^2}}\right)  \right)  \, ,
\ea
which (modulo a factor of $\pm \imath$) can be indeed seen to arise from an analytical continuation of the lapse square $N^2$  to $-N^2$  through the lower or upper half-plane. 

This Lagrangian for the Euclidean geometry leads, via the Hamiltonian constraint, to the following relation between the momentum $(\dot l/N)$ and the length $l$:
\ba\label{detdotlE}
F_\text{E} ((\dot l/N)^2) :=\left(8+(\dot l/N)^2 \right)^{-1/2} \left(  48\frac{n_e}{n_\tau} \pi-       144 \cos^{-1}\left(\frac{8-(\dot l/N)^2}{24+(\dot l/N)^2} \right) \right)  &=&  \Lambda l^2   \, .
 \ea
 
\begin{figure}[H]
\centering
\includegraphics[scale=0.96]{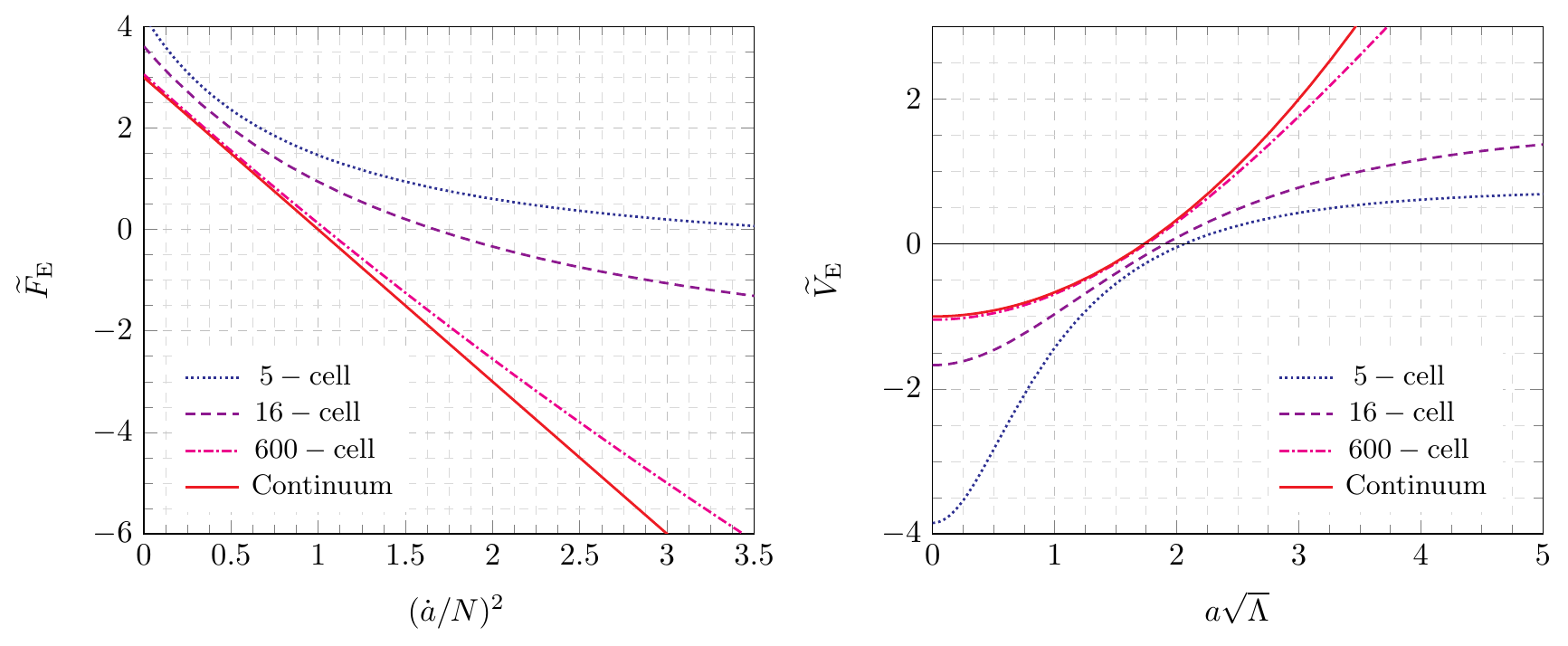}
\caption{Left panel: Rescaled versions $\widetilde{F}_{\text{E}} = \widetilde{F}_{\text{E}}(X)$ of $F_{\text{E}}$ as functions of $(\dot{a}/N)^2$ given by $\widetilde{F}_{\text{E}} \left((\dot{a}/N)^2 \right) := F_{\text{E}}\left((\nu(X) \dot{a}/N)^2\right)/(\nu(X)^2)$, where the discretisation of the inner and outer 3-spheres of the four--dimensional blocks is based on the $X$-cell and $\nu(X)$ is the volume conversion factor given in Table \ref{Table2}. $a = l/\nu(X)$ is the triangulation-dependent scale factor. The continuous red line represents $\widetilde{F}_{\text{E}} \left( \dot{a}/N\right)$ for the general relativistic continuum given by $\widetilde{F}_{\text{E}} = 3\left(1- \left(\dot{a}/N \right) \right)$. Right panel: The potential $\widetilde{V}_{\text{E}} = \widetilde{V}_{\text{E}}(X)$ as a function of $a \sqrt{\Lambda}$ given by $\widetilde{V}_{\text{E}} (a \sqrt{\Lambda}) = - \widetilde{F}_{\text{E}}^{-1}( a \sqrt{\Lambda} )$. The continuous red line is the potential for the general relativistic case given by $\widetilde{V}_{\text{E}} = (a \sqrt{\Lambda} )^2/3 -1$.} \label{fig:Figure-F-and-Potential-Euclidean}
\end{figure}

We can, as in the Lorentzian case, introduce the potential $V_\text{E}(l)=-F^{-1}_\text{E}( \Lambda l^2 )$, see Fig.~\ref{fig:Figure-F-and-Potential-Euclidean} for plots of the potential. We then have the relation $(\dot l/N)^2=-V_\text{E}(l)$, which we can integrate to a solution. The potential is negative for $0\leq l <l^E_\text{max}$ where $l^E_\text{max}=\sqrt{F_\text{E}(0)/\Lambda}$, and it has its maximal negative value, given by $-F_\text{E}^{-1}(0)$, at $l=0$.  Hence there exists a classical solution with initial values $l_\text{ini}=0, (\dot l/N)_\text{ini}=\sqrt{F_\text{E}^{-1}(0)}$ and final values $l_\text{fin}=l^E_\text{max}, (\dot l/N)_\text{fin}=0$.

\begin{figure}[H]
\centering
\includegraphics[scale=0.96]{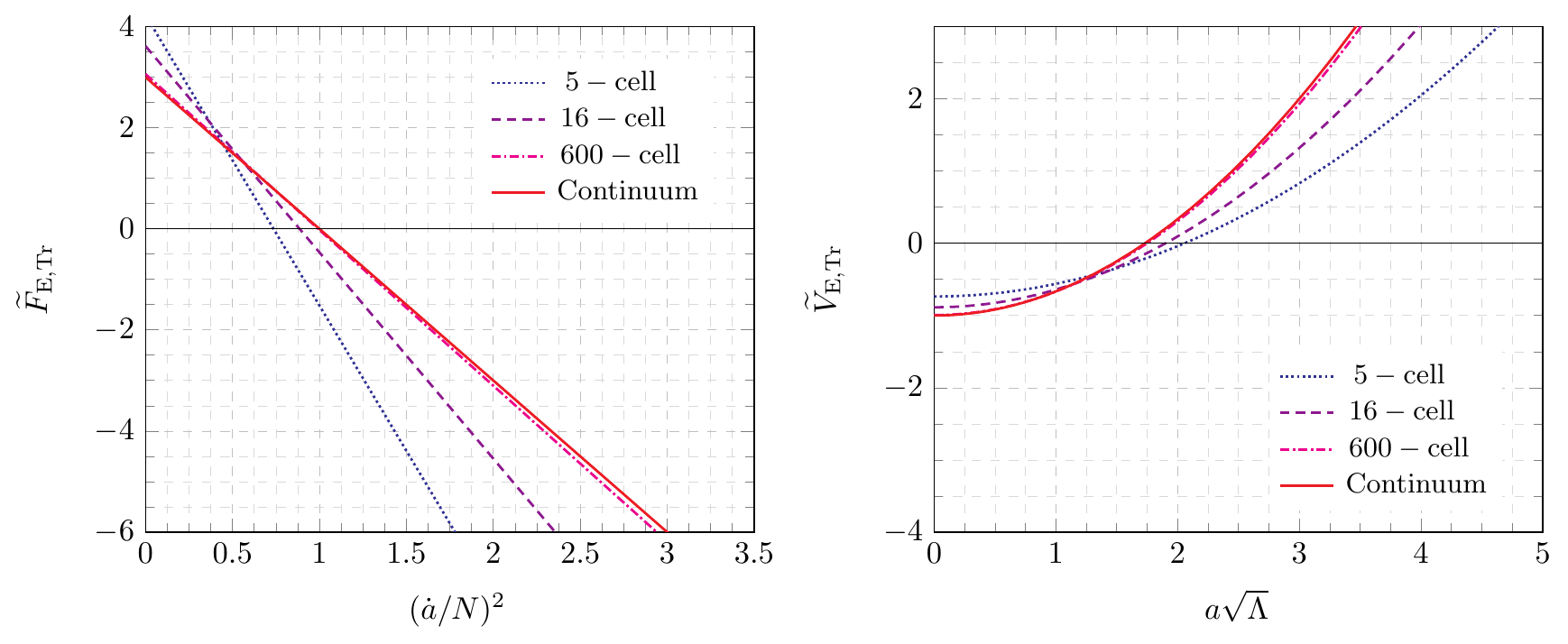}
\caption{Right and left panel: See caption of Fig.~\ref{fig:Figure-F-and-Potential-Euclidean} but with $\widetilde{F}_{\text{E}}$ replaced by $\widetilde{F}_{\text{E,Tr}}$ which is given by $\widetilde{F}_{\text{E}}$ truncated at second order in $(\dot{a}/N)$, and $\widetilde{V}_{\text{E}}$ replaced by $\widetilde{V}_{\text{E,Tr}}$ obtained in the same manner as in Fig.~\ref{fig:Figure-F-and-Potential-Euclidean} but using $\widetilde{F}_{\text{E,Tr}}$. In both panels, the continuous red lines represent the continuum case and are hence identical to the corresponding curves in Fig.~\ref{fig:Figure-F-and-Potential-Euclidean}.} \label{fig:Figure-F-and-Potential-Euclidean-Truncated}
\end{figure}

We also note that, as $F_\text{E}(0)=F(0)$, the allowed maximal length $l^\text{E}_{\text{max}}$ for the Euclidean dynamics coincides with the allowed minimal length $l_\text{min}$ for the Lorentzian dynamics. We thus have the same interpretation of an Euclidean phase for the saddle point contributions in the path integral as in the continuum FLRW case. Again, this maximal length is also visible if we truncate the dynamics to second order in $(\dot{l}/N)$ (see Figure \ref{fig:Figure-F-and-Potential-Euclidean-Truncated}).

These saddle point contributions will be characterised by the Hamilton--Jacobi function. For the no-boundary proposal we have to consider solutions that go from $l_0=0$ to some length $l_1$. For $l_1\leq l_\text{min}$  the integrand evaluates on these saddle points to $\exp(\pm S_\text{Shell-E,HJ}(l))$ where $S_\text{Shell-E,HJ}(l)$ is the Hamilton--Jacobi function for the Euclidean version of the shell Lagrangian (\ref{LShell-E}).

The Hamilton--Jacobi function is most easily computed by using the fact that for a totally constrained system it is given by 
\ba
S_\text{HJ}(l_0,l_1)&=& \int_0^{l_1} \!\! {\rm d}l\,P(l) \q \text{where} \q  P(l) := {p(l, (\dot l/N))}_{ | \frac{\partial L(l,\dot l, N)}{\partial N}=0   }
\ea
and 
\ba
 p(l, (\dot l/N)) :=\,\frac{\partial L(l,\dot l, N) }{\partial \dot l}
\ea
 is the canonical momentum conjugated to $l$: the Hamiltonian constraint allows us  to determine $(\dot l/N)$ and therefore the momentum as a function $P(l)$ of the lengths. We can then directly integrate $P(l)$ to obtain the Hamilton--Jacobi function for solutions $l(t)$ which are either monotonically increasing or decreasing from $l_0$ to $l_1$.

\begin{figure}[H]
\centering
\includegraphics[scale=1]{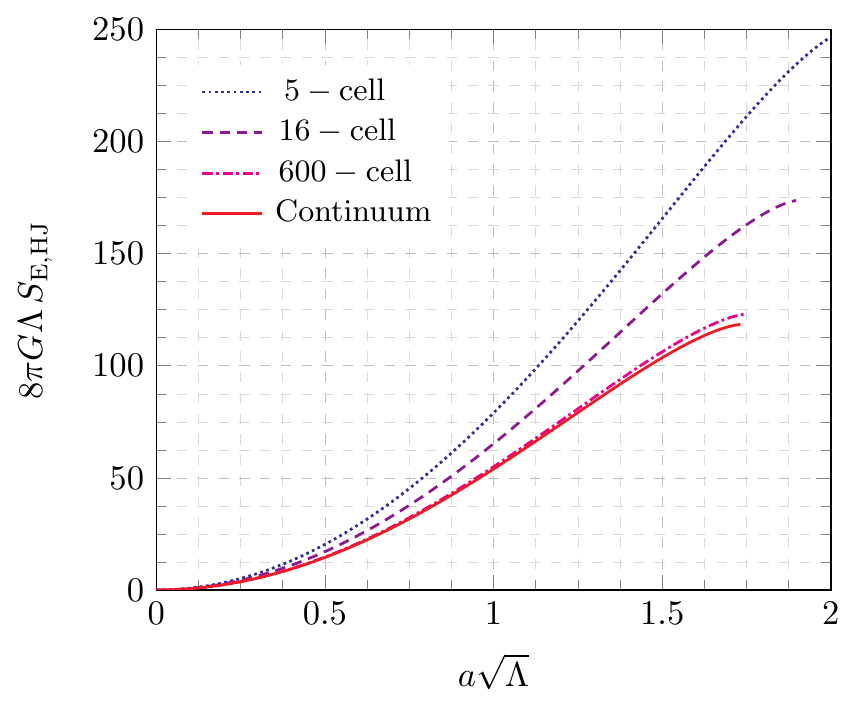}
\caption{The rescaled Hamilton--Jacobi function $8\pi G \Lambda\, S_{\text{Shell-E,HJ}}$ as a function of $a \sqrt{\Lambda}$ for the three $X$-cell based shell triangulations compared with the continuum (Euclidean) Hamilton--Jacobi function $8\pi G \Lambda S_\text{E}$ defined in (\ref{eq:4-polytope-euclidean-continuum}).} \label{fig:LambdaHJ}
\end{figure}

 In our case we can numerically solve the Hamiltonian constraint equation $F_\text{E} ((\dot l/N)^2) =\Lambda l^2$ for $(\dot l/N)$, and in this way obtain the momentum $P_\text{E}(l)$ as function of $l$. This function can then be integrated numerically. Fig.~\ref{fig:LambdaHJ} provides plots for the Hamilton--Jacobi function (with $l_0=0$ and $l_1\leq l_\text{min}$ translated to $a=l_1/\nu(X)$) for the triangulations based on the $X$-cells and compares these to the continuum result.  We see that, for this Euclidean phase, the  Hamilton--Jacobi function for the 600-cell gives a reasonable quantitative approximation to the continuum. Transitioning to the Lorentzian phase, we remark that the generalised velocity $(\dot l/N)$ is small for $l$ around $l_\text{min}.$ Thus, the truncation of the Lagrangian to second order in $(\dot l/N)$ is justified, and this truncation (defined in (\ref{Lshelltr})) approximates the continuum Lagrangian quite well in the case of the 600-cell, see Table \ref{Table3}. 
 The 600-cell shell model therefore gives a reliable model in which the quantum properties of the Euclidean phase and its transition to the Lorentzian phase can be studied.  
 
 On the other hand, concerning the Lorentzian phase at late times, we note that for the shell model studied here, where the discretisation for each of the shells representing different moments of time is the same, the dynamics do differ considerably from the continuum, see Fig.~\ref{fig:Lambda-Solutions}.  In particular, the discrete model shows a linear growth of the scale factor with respect to proper time at late times (due to the boundedness of the generalised velocity $(\dot l/N)^2\leq 8$) whereas it is exponential in the continuum. This issue can be alleviated by introducing a shell model in which the triangulation of the shells depends on the time step.

 ~\\
 
We have discussed a number of discretisations that describe minisuperspace models of a closed  universe with positive cosmological constant. The simplest choice, where the   evolution of the universe up to a certain scale factor is represented by one 4-polytope, can only approximate the continuum for very small scale factors deep in the Euclidean regime $a\ll a_\Lambda$.  As this model does not involve a proper integration it also relies on the postulate that Euclidean geometries are assigned a nonvanishing amplitude in the path integral. 
We obtained a more involved model by subdividing the 4-polytope into 4-simplices, introducing in this way bulk edges and an integration variable. This model does capture the feature of having complex solutions describing an Euclidean geometry. However it deviates even qualitatively from the continuum model for scale factors larger than $a_s$ where the discrete Euclidean Hamilton--Jacobi function is maximised. Thus, whereas this model does allow the study of quantum properties of the Euclidean phase, it cannot describe the transition to a Lorentzian phase. This transition can be studied with the shell model, and we have also seen that the 600-cell based model provides a reasonable quantitative agreement with the continuum as long as the scale factor is not too large. 
Table \ref{Table4} provides two of the key quantitative characteristics for the Euclidean phase of the various discrete models.

\begin{table}[!]
\begin{tabular}{|l||c|c||c|c||c|c||}\hline
&\; $ \text{P:}\, a^2_s\Lambda$ \;& \;$ 8\pi G \Lambda S_\text{P,E}( a^2_s )  $\; & \;$\text{S-P:} \, a^2_s \Lambda $\;& \;$8\pi G \Lambda S_\text{S-P,E}( a^2_s ) $\;& \;$\text{Shell:}\,a_\text{max}^2 \Lambda $\;& \;$8\pi G \Lambda S_\text{Shell-E,HJ}( a^2_\text{max} )$\;\;\\ \hline
5-cell & 16.31& 669.2&5.81& 303.8 &4.230&250.1\\ \hline
16-cell & 9.10 & 315.8 &4.93& 205.6&3.616&173.8\\ \hline
600-cell & 6.22 & 187.1 & 4.08&142.4&3.061&123.1 \\ \hline\hline
Cont.&&&&&3& $12 \pi^2\approx 118.4$ \\ \hline
\end{tabular}
\caption{Scale factor values $a^2_s$ resp. $a^2_\text{max}$ for which  the Euclidean action resp. Hamilton--Jacobi function is maximal  for the 4-polytope without subdivisions (P), the subdivided 4-polytope (S-P), and the shell model (Shell). We also provide the values of the action resp. Hamilton--Jacobi function for the solution going from $a=0$ to $a_s$ resp. $a_{\text{max}}$.  The last row lists the same values for the continuum. \label{Table4}}
\end{table}

\section{Coupling to dust}\label{Sec:Dust}

\subsection{Classical theory}

In this section we will consider (Regge) gravity without cosmological constant but coupled to a simple dust action, which has been also used in \cite{Collins1973}. This model will illustrate a behaviour quite different from the case with a positive cosmological constant and without matter.  

In the continuum there are various ways of defining an action for a perfect fluid (for a review see \cite{BrownReview}); these actions usually involve several variables characterising various physical and thermodynamic properties of a fluid. In minisuperspace, however, the number of variables is greatly reduced: in cosmological models a perfect fluid is essentially only characterised by an energy density $\rho(t)$ and equation of state $p=p(\rho)$. If we again consider a closed FLRW universe, a perfect fluid action for this case can be defined by \cite{Brown1990}
\begin{equation}
S_{{\rm f}} := \int {\rm d}t\left(U\,\dot\varphi-2\pi^2 a^3\,N \rho\left(\frac{U}{2\pi^2 a^3}\right)\right)
\label{fluidlagrangian}
\end{equation}
which depends on the lapse $N$ and scale factor $a$ of the metric (\ref{flrwmetric}), as well as on the total particle number $U$ and its conjugate variable $\varphi$. The equation of state is implicit in the definition of the energy density $\rho$ as a function of the particle number density $n=U/(2\pi^2 a^3)$.

The variables $U$ and $\varphi$ can now be integrated out; since the only role of $\varphi$ is to enforce particle number conservation $\dot{U}=0$ one can remove $\varphi$ and treat $U$ as a fixed constant, analogous to a mass or energy. This is what we will do in the following. We will also specify to the case of a dust perfect fluid, where $\rho(n)\propto n$. The fluid action (\ref{fluidlagrangian}) then reduces to $-M\int {\rm d}t\, N$, where $M$ is the total mass of the dust particles in the system. The integral over $t$ then simply corresponds to the total proper time of the dust particles.\footnote{Notice that our previous $\Lambda$ model can also be obtained as a special case of the action (\ref{fluidlagrangian}) if we choose $\rho$ to be a constant, so that we obtain a term linear in the total four-volume rather than total proper time. Similarly, all other perfect fluids would correspond to a term linear in some other global geometric quantity.}

Using our discretisation with frusta we can model the dust by assuming that there is a dust particle with fixed mass $m$ moving along the central axis of each frustum. Each dust particle contributes to the action an amount $-m T$, where $T$ is the proper time measured by the particle. Since ${\rm d}T={\rm d}h_0=N\,{\rm d}t$, the Lagrangian for the shell discretisations is given by 
\ba
L_{\text{D-Shell}} \,\,:=\, \,L_{\text{Shell}} (\Lambda=0) - M\, N
\ea
where $M= n_\tau m$ is the total mass of the dust particles, and $L_{\text{Shell}}$ is defined in (\ref{LShell}). The dust Lagrangian is then analogous to the one in the continuum.

In the following we consider only the case $(a)$ and $(b)$ defined in Section \ref{Sec:Signatures}, that is the cases where the trapeziums are time-like.

In the same way as we discussed for the model with cosmological constant, we can truncate the Lagrangian $L_{\text{D-Shell}}$ to terms of up to second order in the extrinsic curvature, i.e., second order in $\dot l/N$. Replacing the length $l$ of the spatial edges with the scale factor in the same manner as discussed above (\ref{Lshelltr2}), we obtain
\ba\label{Lshelltr2b}
L^{\text{Tr}}_{\text{D-Shell}}(a,\dot a, N;M)&=&\frac{3\pi}{4G}\left(  \tilde c_1N a  - \tilde c_2  \frac{a \dot{a}^2}{N} \right)- M N \, ,
\ea
where $\tilde c_1,\tilde c_2$ are defined in Table \ref{Table3}.  This truncated Lagrangian agrees in its general form with the continuum Lagrangian, for which we have $\tilde c_1=1$ and $\tilde c_2=1$. To see in which regime the truncation is justified, we consider the Hamiltonian constraints, which can be obtained by varying  $L_{\text{D-Shell}}$, resp. $L^{\text{Tr}}_{\text{D-Shell}}$  with respect to the lapse parameter $N$. Without the truncation, we obtain the following constraint equation for $(\dot l/N)$:
\ba\label{dustdotl}
D((\dot l/N)^2):=\frac{1}{6\sqrt{2}} \left(8-(\dot l/N)^2 \right)^{-1/2} \left(  48 n_e \pi-       144 n_\tau \cos^{-1}\left(\frac{8+(\dot l/N)^2}{24-(\dot l/N)^2} \right) \right)  &=&  \frac{8\pi G M}{l} \, . \q\q
 \ea
The left-hand side of (\ref{dustdotl}) defines a function $D$ of $(\dot l/N)^2$ which, starting from $D(0)$, increases monotonically with growing $|\dot l/N|$ and diverges for $|\dot l /N|\rightarrow \infty$. But the right-hand side now scales with $1/l$ instead of $l^2$, as in the case of a cosmological constant (\ref{detdotl}).
\begin{figure}
\centering
\includegraphics[scale=0.96]{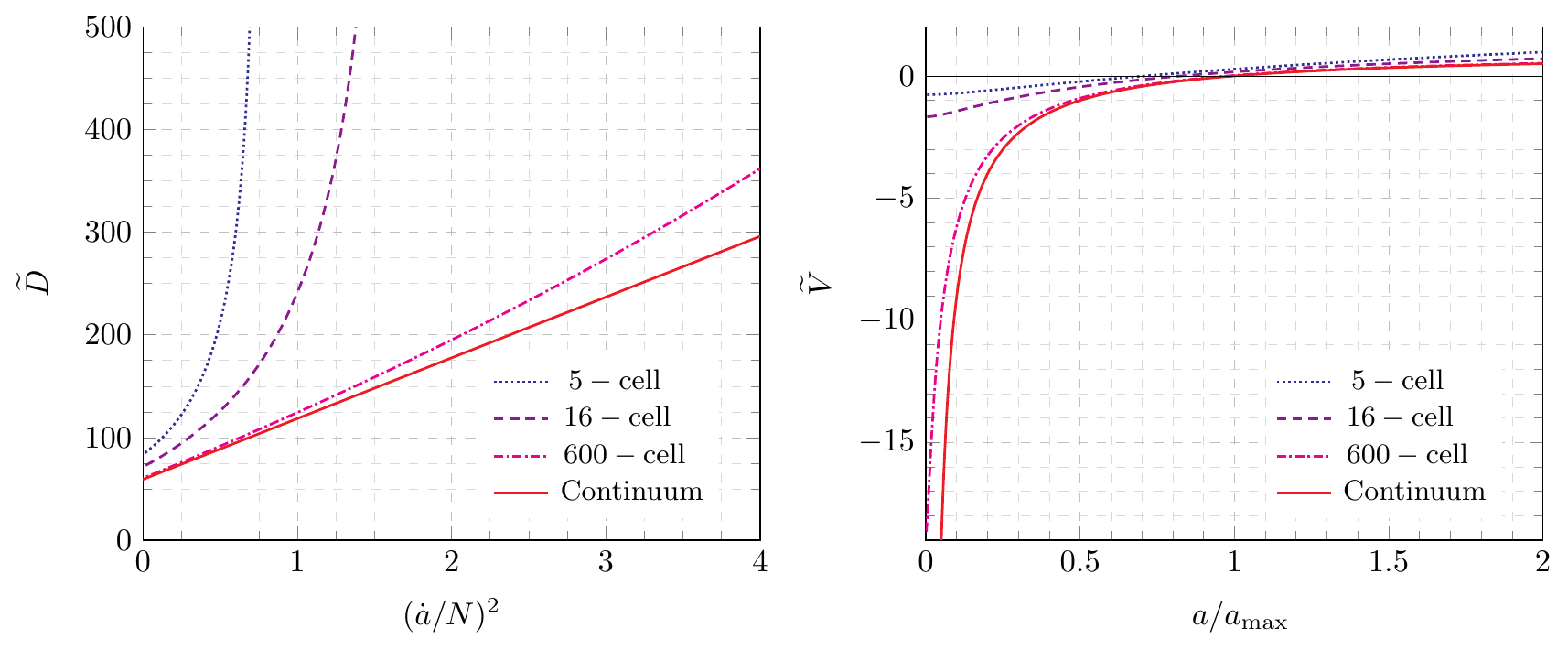}
\caption{Left panel: Rescaled versions $\widetilde{D} = \widetilde{D}(X)$ of $D$ as functions of $(\dot{a}/N)^2$ given by $\widetilde{D} \left((\dot{a}/N)^2 \right) := D \left((\nu(X) \dot{a}/N)^2\right)\cdot \nu(X)$. Recall that $X=5,16,600$ is the number of tetrahedra in the inner resp. outer $3$--spheres of the four--dimensional blocks, and $\nu(X)$ is the volume conversion factor for the $X$-cells given in Table \ref{Table2}. $a = l/\nu(X)$ is the triangulation-dependent scale factor. The continuous red line represents $\widetilde{D} ( (\dot{a}/N)^2)$ for the general relativistic continuum given by $\widetilde{D} = 6 \pi^2 (1+ (\dot{a}/N )^2 )$. Right panel: The potential $\widetilde{V}$ as a function of $a/a_{\text{max}}$ where $a_{\text{max}} := 8 \pi G M/(6 \pi^2)$ given by $\widetilde{V} (a/a_{\text{max}}) = - \widetilde{D}^{-1}( 6 \pi^2 a_{\text{max}}/a )$. The continuous red line is the potential for the general relativistic case given by $\widetilde{V} =1 - a_{\text{max}}/a$.} \label{fig:Figure-F-and-Potential-Dust-Full}
\end{figure}

As before, we can introduce the potential $V(l)=-D^{-1}(8\pi G M/l)$ such that the generalised velocity is determined by $(\dot l/N)^2=-V(l)$ (see Figure \ref{fig:Figure-F-and-Potential-Dust-Full}).  If we consider an evolution starting from a very small value for $l$ we obtain a corresponding value of $|\dot l/N|$ very near its maximally allowed value of $\sqrt{8}$. Assuming $\dot l/N>0$ and $N>0$ the expanding size of the universe leads to a decreasing value for $\dot l/N$ until we reach $\dot l/N=0$ and a maximal size $l_\text{max}=8\pi GM/D(0)$ for the universe. The universe then contracts, until it reaches $l=0$ again.  

\begin{figure}
\centering
\includegraphics[scale=0.96]{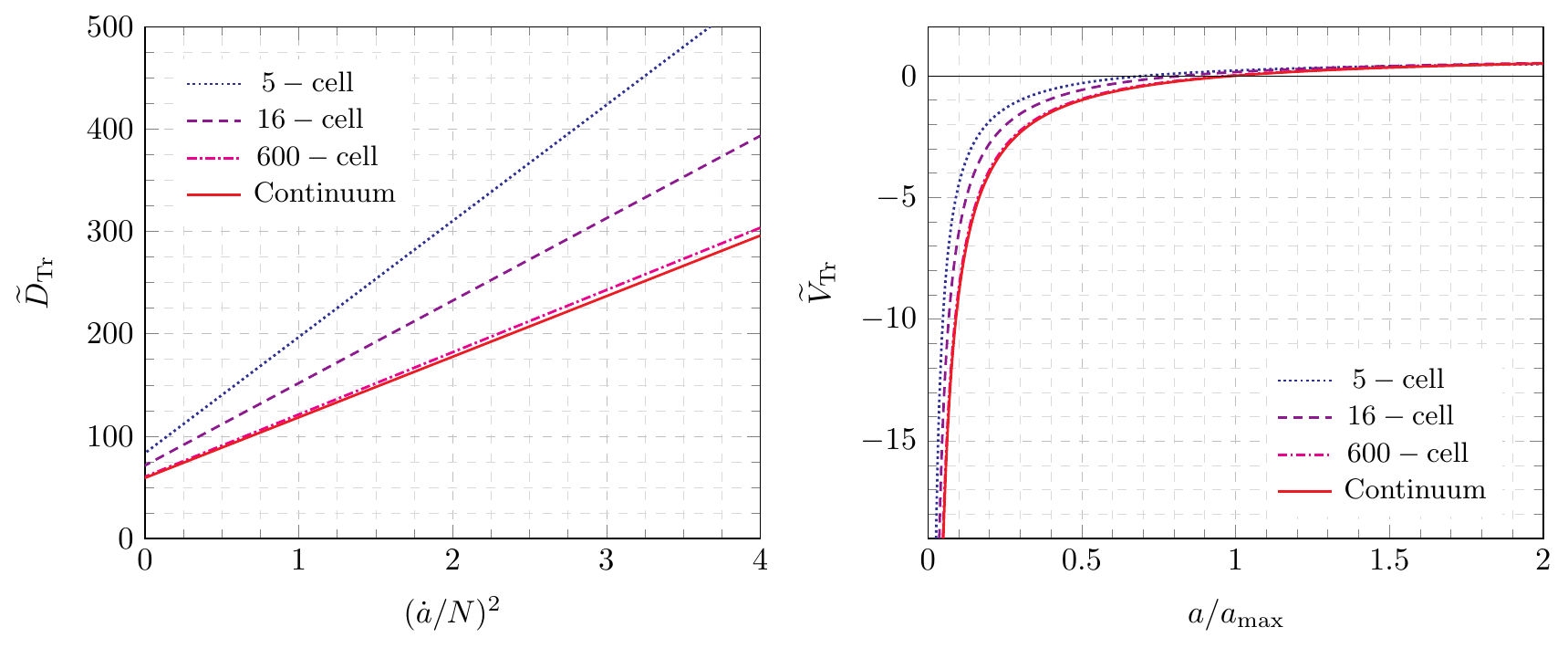}
\caption{Right and left panel: as in Fig.~\ref{fig:Figure-F-and-Potential-Dust-Full} but with $\tilde{D}$ replaced by $\widetilde{D}_{\text{Tr}}$ which is given by $\widetilde{D}$ truncated at second order in $(\dot{a}/N)$, and $\widetilde{V}$ replaced by $\widetilde{V}_{\text{Tr}}$ obtained in the same manner as in Fig.~\ref{fig:Figure-F-and-Potential-Dust-Full} but using the truncated function $\widetilde{D}_{\text{Tr}}$. In both panels, the continuous red lines represent the continuum case and are hence identical to the corresponding curves in Fig.~\ref{fig:Figure-F-and-Potential-Dust-Full}.} \label{fig:Figure-F-and-Potential-Dust-Truncated}
\end{figure}

We hence encounter small values for $|\dot l/N|$ around the maximal radius of the universe, whereas we obtain the maximal allowed values for $|\dot l/N|$ near $l=0$. We can thus expect that the time evolutions obtained respectively from the non-truncated and truncated Lagrangian differ in the regime of small scale factor. With the Lagrangian truncated to second order in $\dot l/N$, the Hamiltonian constraint is given by
\ba
8\pi G H_{\text{D-Shell}}^{\text{Tr}}:= -c_1 l - c_2 l (\dot l/N)^2 +8\pi G M \, ,
\ea
with $c_1$ and $c_2$ defined in (\ref{eq57}).
Thus we have $(\dot l/N)^2=  (c_2l)^{-1}8\pi G M- c_1/c_2$, and $(\dot l/N)$ diverges for $l\rightarrow 0$ (Figure \ref{fig:Figure-F-and-Potential-Dust-Truncated}).

\begin{figure}
\centering
\includegraphics[scale=1]{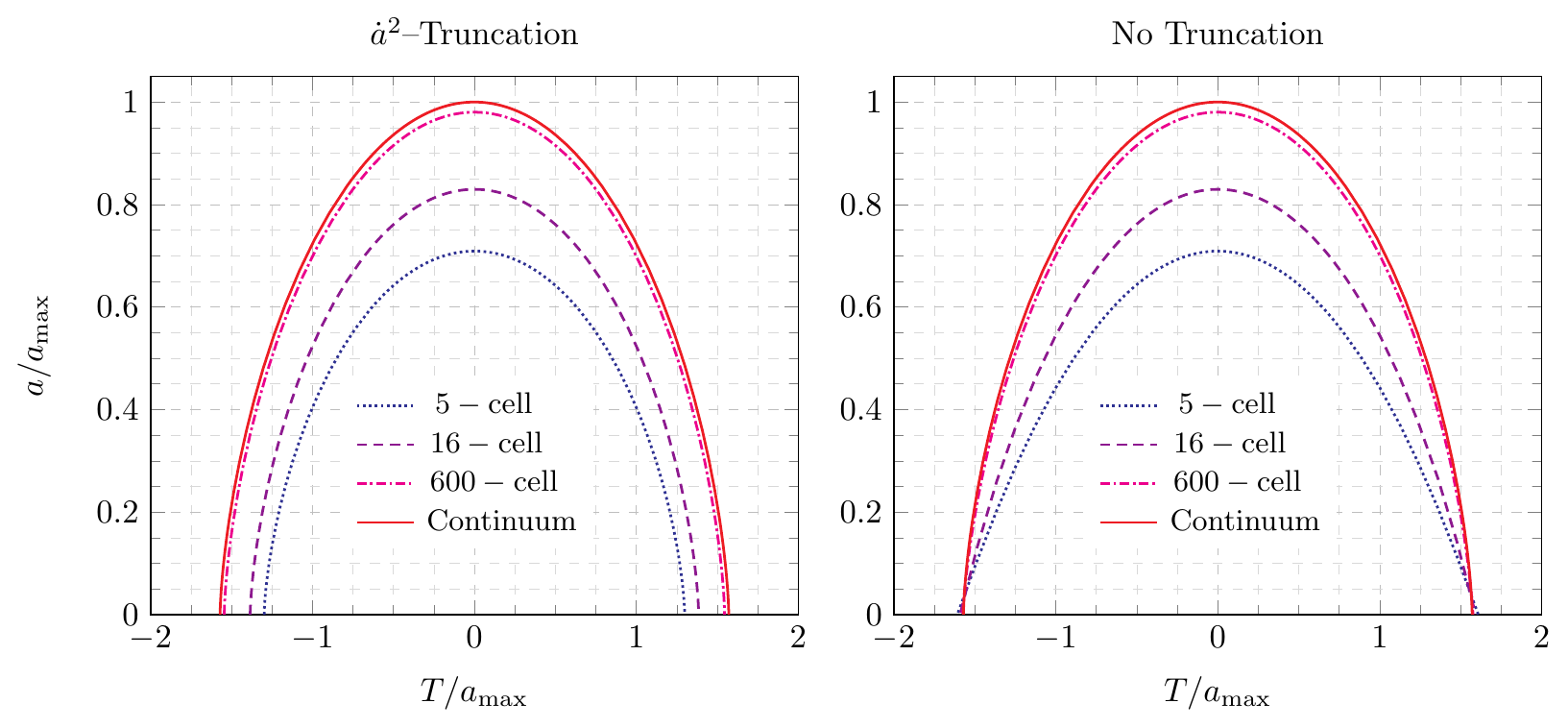}
\caption{Left panel: Analytical solutions $a(T)$ of the equations of motion obtained from the Regge Lagrangians $L^{\text{Tr}}_{\text{D-Shell}}$ for the $X = 5,16, 600$ boundary shells respectively, with dust matter of total mass $M$ and truncated at second order in $\dot{a}$. $T$ denotes proper time. To obtain dimensionless quantities we rescaled $a$ and $T$ by $a_\text{max}=4 G M/(3 \pi)$. Right panel: Numerical solutions $a(T)$ of the equations of motion obtained from the Regge Lagrangians $L_{\text{D-Shell}}$ with dust matter of mass $M$.
The continuous red lines in both panels represent the exact general relativistic solutions $a(T)$ given in parametric form by $a(\eta) = a_{\text{max}}(1-\cos(\eta - \pi))/2$ and $T(\eta) =a_{\text{max}} (\eta - \sin(\eta-\pi))/2$, where $\eta$ is conformal time.} \label{fig:Figure-Dust-Solutions}
\end{figure}

Figure \ref{fig:Figure-Dust-Solutions} compares the evolutions of the 5-, 16- and 600-cell universe with and without the truncation as well as with the continuum evolution.  We indeed see that the behaviour differs for small scale factors.

\subsection{Remarks on the path integral}

Let us also consider for the case with dust a path integral with boundary condition $a(t_0)=0$.  Different from the case with a positive cosmological constant, we do have a real solution starting from $a=0$, the Hamilton--Jacobi function is therefore real as long as we use boundary values $a_0, a_1\leq a_{\text{max}}$, where $a_{\text{max}}:=4GM/(3\pi)$ is the maximal value of the scale factor that can be reached by a classical Lorentzian solution.  But we encounter a singular behaviour for the continuum solution: as $(\dot a/N) \sim a^{-1/2}$ for $a\rightarrow 0$, the extrinsic curvature $K=3\dot{a}/(Na)$ blows up as $a^{-3/2}$. On the other hand, this singular behaviour is rather mild as the momentum $p=-(3\pi a\dot a)/(2G N)$ even goes to zero for $a\rightarrow 0$. 

The Hamilton--Jacobi function for the continuum is given by
\ba\label{HJDust}
\frac{S_{\rm HJ}( a_0,a_1)}{M a_{\text{max}}}
&=& \frac{\sigma_0}{2}\left(\sin^{-1}(\sqrt{\tilde a_0}) - (1-2\tilde a_0) \sqrt{ \tilde a_0-\tilde a_0^2}\right)\nn\\
&&
\,+\, \frac{\sigma_1}{2} \left(\sin^{-1}(\sqrt{\tilde a_1}) - (1-2\tilde a_1) \sqrt{ \tilde a_1-\tilde a_1^2}\right)
\ea
where $\tilde a_i =a_i/a_{\text{max}}$.  The sign ambiguities $\sigma_i=\pm$ arise for the same reason as for the case with positive cosmological constant: one source for the ambiguities is whether $a_0$ and $a_1$ both refer to moments where the universe expands or contracts, or not. The second source  results from the possibility to have either positive or negative lapse. 

When evaluating a path integral to determine a propagator or solution to the Wheeler--DeWitt equation, the four classical solutions (\ref{HJDust}) would be associated to four possible saddle points in an integral over the lapse, as we discussed for the model with cosmological constant in Section \ref{Sec:Cont}. Now all saddle points are on the real axis, and would naively all contribute to the path integral (if we integrate over positive and negative lapse; if we only allow for positive lapse we only have two saddle points). A more detailed analysis would however involve determining the steepest-descent contours and identifying the relevant saddle points. For a detailed discussion of these contours in a model which can have real or imaginary saddle point solutions see, e.g., \cite{Garayetal}. The issue of relevant saddle points was also discussed in \cite{GielenTurok} where the propagator for a model with a radiation perfect fluid was evaluated through a path integral. In that work, for the case of a closed universe the authors only used one of the saddle point solutions by demanding a continuous limit as the spatial curvature is taken to zero; in this limit some of the solutions, which correspond to a universe that first expands and then contracts, disappear.

The solutions to the discrete evolution equations are very similar to the continuum ones; we therefore have the same features  (and sign ambiguities) for the discrete Hamilton--Jacobi function as for the continuum one. In the discrete case,  in contrast to the continuum, we do not have a singular behaviour for $\dot l$ when $l\rightarrow 0$. This is due to the bound $(\dot l/N)^2 \leq 8$ on the generalised velocity for the length $l$. (On the other hand, the momentum $p$ conjugated to $l$ goes to a finite value for $l\rightarrow 0$ in both the continuum and discrete theories.) But we remind the reader that, if we translate this bound to the scale factor variable, it becomes explicitly triangulation dependent: $(\dot a/N)^2 \leq 8/\nu(X)^2$. In particular, since $\nu(X)$ is smallest for the 600-cell this bound is already weaker for the 600-cell compared to the other triangulations. To make definite conclusions about a singularity avoiding behaviour one should therefore consider an infinite refinement limit, which requires more general triangulations than the $X$-cells we have considered in this work. 

\begin{figure}[H]
\centering
\includegraphics[scale=1]{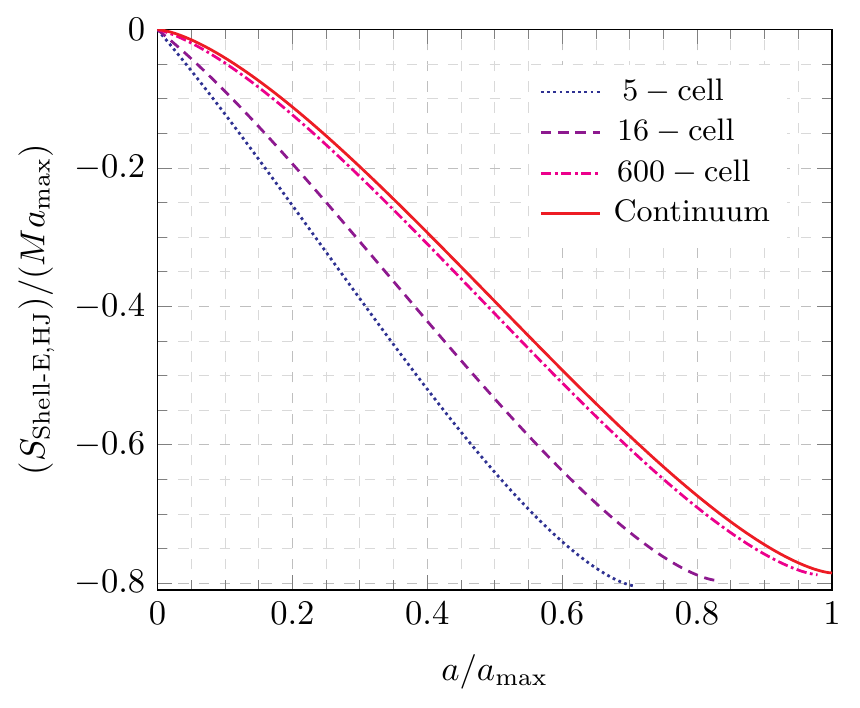}
\caption{Rescaled Hamilton--Jacobi function $S_{\text{Shell-E,HJ}}/(M a_\text{max})$ as a function of $a/a_\text{max}=l/(\nu(X) a_\text{max})$ for the three $X$-cell based shell triangulations compared with the continuum result.} \label{fig:DustHJ}
\end{figure}

The Hamilton--Jacobi function can be computed numerically, in the same manner as described in Section \ref{Sec:NBP}.  Figure \ref{fig:DustHJ} shows the Hamilton--Jacobi functions for different $X$-cell based triangulations as well as for the continuum for a solution that goes from $a_0=0$ to $a_1=a$ with $a\leq a_{\text{max}}$. 

We see that the discrete models approximate the continuum quite well. In particular, the Hamilton--Jacobi functions for the  600-cell based  model and the continuum match closely.

 \section{Discussion and outlook}\label{Sec:Discuss}
 
In this work we introduced  Lorentzian quantum cosmology models based on Regge discretisations.  Lorentzian models in Regge calculus have not been studied extensively so far. We have seen that already quite simple discretisations come with surprising features, which do not appear for the Euclidean theory. These simple models could therefore inform us about  important choices for the path integral, in particular which kind of configurations to allow and which to exclude. 

One class of such configurations are the ones with irregular light cone structure. These seem to appear generically and are likely to dominate the path integral, see also \cite{EffSF3} for an explicit numerical evaluation of the path integral with and without such configurations. One possibility to avoid such a dominance is to only allow configurations with regular light cone structure, as is also done in Causal Dynamical Triangulations \cite{CDTReview}.   

On the other hand we have seen that the path integral might have saddle points in the complex plane which amount to an effective Wick rotation to Euclidean configurations. This raises the question of whether to allow  in the Regge path integral also 4-simplices with Euclidean signature, but associated with an exponentially suppressed amplitude. The argument for including these configurations is also supported by the semiclassical analysis of spin foam amplitudes, in which such exponentially suppressed contributions appear \cite{BarrettFoxon,HanLiu}.

Another feature appearing in the Lorentzian shell models is the bound $(\dot l/N)^2 \leq 8$ for the generalised velocities associated to the length of the spatial edges. This bound is connected to the change of signature  from time-like to space-like for the bulk triangles (or trapeziums) in a given shell, which also leads to the appearance of configurations with irregular light cone structure. Interestingly, at least in the limit of infinitesimal small time steps, the regimes with regular and irregular light cone structure are dynamically disconnected. 

This bound $(\dot l/N)^2 \leq 8$ is the main cause for the deviations in the dynamics of the discrete shell models from the continuum dynamics (where the velocity is not bounded from above). It is likely to change if more general and refined triangulations are used; notice that already when translating from $l$ to the cosmological scale factor $a$, this bound becomes explicitly discretisation-dependent. But some kind of bound connected to the change of signature for the triangles (or more generally bones) in the discretisation might survive and have interesting physical consequences.

We discussed a range of models based on different discretisations, and analysed to which degree these models reproduce the continuum FLRW model.  The models are sufficiently simple to be accessible to a non-perturbative path integral approach. For example,  the recently introduced effective spin foam approach \cite{EffSF1,EffSF2,EffSF3}  comes with a much improved numerical efficiency which will allow treating the models presented here.  If one takes the limitations of these discrete models into account, a study of the non-perturbative path integrals promises important insights:
 \begin{itemize}
 \item For the model based on the subdivided 4-polytope discussed in Section \ref{Sec:Ball} we only have one integration variable, which represents the lapse. We have seen that this discretisation can only qualitatively reproduce the early Euclidean phase of the continuum. But studying the non-perturbative path integral in this regime already offers a deeper understanding of Lorentzian quantum cosmology: although there is only one integration variable, the integral is over an infinite range. The first question is therefore whether, with an appropriate choice of measure, this integral (which in the spin foam approach actually amounts to a summation) can be made well-defined. Assuming this is the case, a numerical evaluation could then reveal which of the saddle points are actually relevant, and thus whether the Euclidean phase does contribute with an exponentially suppressing or enhancing factor. 
 \item The path integral for one (finite) time step in the shell model, discussed in Section \ref{Sec:Shells}, also involves only one integration.  The boundary values for the scale factor can be chosen such that the solutions interpolating between them are either in the Euclidean phase, transitioning from the Euclidean to Lorentzian phase, or inside the Lorentzian phase. We can thus again gain insights about the role of the various, possibly complex, saddle points.  
 
 Another important feature of the shell model is the restoration of time reparametrisation invariance in the limit of vanishing time steps. It will be interesting to see whether such a  reparametrisation invariance can be also obtained in the quantum theory. Whether this happens might depend on the choice of measure, and can therefore validate a particular choice. Studying the restoration of reparametrisation invariance will be a crucial test for the spin foam approach: in spin foams geometric observables such as lengths, areas and volumes are discrete. One could therefore expect a minimal size for the time steps, which could potentially interfere with the restoration of time reparametrisation invariance.

 \item By coupling gravity to matter, in our case a simple dust model, we obtain a Lorentzian beginning of the universe. We have seen that the discrete evolution equations show singularity avoidance due to the bound $(\dot l/N)^2 \leq 8$ on the generalised velocity associated to the lengths of the spatial edges, but that this bound becomes discretisation-dependent if we translate it to the generalised velocity for the scale factor.  As spin foam quantisation implements discrete spectra for geometric variables such as areas and three-volumes, it can provide a different mechanism for singularity resolution. Here it would be again important to understand to which degree this mechanism depends on the choice of discretisation. 
 
 \end{itemize}

 Even the spin foam study of the subdivided 4-polytope would go beyond the calculations available so far for spin foam cosmology, where previous studies \cite{SFC1,SFC2} considered a so-called dipole discretisation. The dipole describes the dual graph of the boundary triangulation of a non-subdivided 4-polytope, given by two tetrahedra identified with each other. Such a 4-polytope can however be considered as degenerate, and its 4-volume vanishes.\footnote{The classical limit discussed in \cite{SFC1,SFC2} does not describe a closed, but a spatially flat FLRW universe. The conclusion in \cite{SFC2} seems to be that more interesting dynamics (including spatial curvature) would require more complicated boundary graphs. A cosmological constant introduced by modifying the spin foam amplitude \cite{SFC3} was based on the nonvanishing 3-volume, rather than the vanishing 4-volume. \cite{SFC2} also includes a discussion of the Hamiltonian dynamics (for dipole graphs) which can be compared with our discussion of the shells and the limit of taking infinitesimal time steps. For further discussion of the choices made in these previous models of spin foam cosmology see also \cite{SFC4}.} The set-up can be compared to our discussion in Section \ref{NSD} based on non-subdivided, but also non-degenerate polytopes.
 
  One main cause for the limitation of the previous results are the high computational demands for evaluating spin foam amplitudes. The newly introduced effective spin foam models \cite{EffSF1,EffSF2,EffSF3} very much reduce the required computational effort but as, e.g., the study of subdivided 4-polytopes and shells requires integrations or summations over an infinite range, this effort would still be considerable. This is why we discussed a range of options but also concentrated on the simplest choices, such as regular 4-polytopes.

 There are of course many ways in which the models we have studied can be generalised. This includes the coupling to different matter sources, e.g., more general perfect fluids \cite{BrownReview}. The models can be also easily generalised to include anisotropies and inhomogeneities \footnote{The simplest model allowing for an anisotropy is based on the 5-cell, where one allows the appearance of two different boundary lengths. This can be done so that each vertex neighbourhood has still the same geometry, and one can choose all bulk edge lengths to be equal.}. Their (non-perturbative) study would  require more numerical resources, but already quite simple models could reveal more about potential instabilities for the dynamics of inhomogeneities in Lorentzian quantum cosmology \cite{Debate}.  Another interesting extension is to modify the shell model and introduce a time-dependent discretisation for the shells \cite{DittrichHoehn,DittrichSteinhaus13}, allowing for the generation of new modes during time evolution.
 
 \begin{acknowledgments}
We are very thankful to Hal Haggard for collaboration at an early stage for this work.
Research at Perimeter Institute is supported in part by the Government of Canada through the Department of Innovation, Science and Economic Development Canada and by the Province of Ontario through the Ministry of Colleges and Universities. 
The work of SG was funded by the Royal Society through a University Research Fellowship (UF160622) and a Research Grant  (RGF\textbackslash R1\textbackslash 180030).
\end{acknowledgments}

 \appendix
 \section{Triangulation for the shells}\label{Tshells}
 
 Here we will discuss how to construct triangulations of  four-dimensional spherical shells, such that their inner and outer boundary is given by the boundary of the 5-cell, 16-cell or 600-cell. To this end we will first consider a triangulation of the four-dimensional auxiliary building blocks (frusta) of topology $\tau\times [0,1]$. We denote the vertices of the tetrahedron $\tau \times \{0\}$ by $\{1,2,3,4\}$ and the vertices of the tetrahedron $\tau \times \{1\}$ by $\{1',2',3',4'\}$. We then define the triangulation of ${\cal B}=\tau\times [0,1]$ to consist of the four-simplices
 \ba\label{cB}
 {\cal B}=\{  \{1,2,3,4,4'\},  \{1,2,3,3',4'\},  \{1,2,2',3',4'\}, \{1,1',2',3',4'\}\}  \, .
\ea
We define the lengths of edges in $\tau \times \{0\}$ to be $l_0$ and the lengths of edges in $\tau \times \{1\}$ to be $l_1$. Edges of the form $e=(v,v')$ (e.g., $e=(1,1')$) are called struts; their length squared is defined to be $m_0^2$. We furthermore have edges $e=(v,w')$ with $v<w$ (e.g., $e=(1,2')$). These are the diagonals in the trapeziums $(v,w,v',w')$ and are defined to have length squared $d_0^2$.
 
To construct the triangulation for the entire shell, we glue a four-dimensional building block ${\cal B}$ on top of every tetrahedron in the triangulation of the three-sphere. Consider two such tetrahedra $\tau_1,\tau_2$ sharing a triangle $t$. To be able to glue ${\cal B}_1=\tau_1 \times [0,1]$ to ${\cal B}_2=\tau_2 \times [0,2]$ we must require that all the diagonals in $t \times [0,1]$, coming respectively from the triangulation of ${\cal B}_1$  and ${\cal B}_2$, agree.  To ensure this we can proceed as follows: enumerate the vertices in the triangulation of the three-sphere. The triangulation of the four-dimensional shell has a vertex set given by the union of $\{1,2,\ldots, n_v\}$ and $\{1',2',\ldots, n'_v\}$, where $n_v$ is the number of vertices in the triangulated three-sphere. We order the vertices for each tetrahedron $\tau$ in the three-shell. For a given tetrahedron $\tau= \{v_1,v_2,v_3,v_4\}$ with $v_1<v_2<v_3<v_4$  we map the ordered set $\{v_1,v_2,v_3,v_4\}$ to the ordered set $\{1,2,3,4\}$ and define an associated  triangulated frustum ${\cal B}$ as in (\ref{cB}).

This ensures that for a given set of trapeziums $\{v,w,v',w'\}$, which are all identified to each other under gluing, the diagonals agree in orientation: the diagonal edge is given by $\{v,w'\}$ if $v<w$ and by $\{w,v'\}$ if $w<v$. 

As an example, we provide here the set of four-simplices in the triangulated shell based on the 5-cell partitioned into five frusta ${\cal B}_i, i=1,\ldots,5$:
\ba
 {\cal B}_1&=&\{  \{1,2,3,4,4'\},  \{1,2,3,3',4'\},  \{1,2,2',3',4'\}, \{1,1',2',3',4'\}\} \,,\nn\\
 {\cal B}_2&=&\{ \{1,2,3,5,5'\}, \{1,2,3,3',5'\},  \{1,2,2',3',5'\}, \{1,1',2',3',5'\}\}\,, \nn\\
 {\cal B}_3&=&\{ \{1,2,4,5,5'\}, \{1,2,4,4',5'\},  \{1,2,2',4',5'\}, \{1,1',2',4',5'\}\}\,, \nn\\
 {\cal B}_4&=&\{ \{1,3,4,5,5'\}, \{1,3,4,4',5'\},  \{1,3,3',4',5'\}, \{1,1',3',4',5'\}\} \,,\nn\\
{\cal B}_5&=&\{ \{2,3,4,5,5'\}, \{2,3,4,4',5'\},  \{2,3,3',4',5'\}, \{2,2',3',4',5'\}\} \,.
\ea

\providecommand{\href}[2]{#2}
\begingroup
\endgroup


\begin{thebibliography}{100}

\bibitem{EuclideanQG} S.~W.~Hawking, ``Euclidean Quantum Gravity,'' in L\'evy M., Deser S. (eds.), {\it Recent Developments in Gravitation: Carg\`ese 1978, NATO Advanced Study Institutes Series}, Series B: Physics, vol. 44 (Springer, 2011); G.~W.~Gibbons, S.~W.~Hawking (eds.), {\it Euclidean Quantum Gravity} (World Scientific, 1993).

\bibitem{unbounded}
G.~W.~Gibbons,
``The Einstein action of Riemannian metrics and its relation to quantum gravity and thermodynamics,''
Phys.\ Lett.\ A \textbf{61} (1977), 3-5.

\bibitem{HartleHawk} J.~B.~Hartle and S.~W.~Hawking,
``Wave function of the Universe,''
Phys.\ Rev.\ D\ \textbf{28} (1983), 2960-2975.

\bibitem{HalliwellLouko}
J.~J.~Halliwell and J.~Louko,
``Steepest-descent contours in the path-integral approach to quantum cosmology. I. The de Sitter minisuperspace model,''
Phys.\ Rev.\ D \textbf{39} (1989), 2206-2215.

\bibitem{Debate}
J.~Feldbrugge, J.~L.~Lehners and N.~Turok,
``No Smooth Beginning for Spacetime,''
Phys.\ Rev.\ Lett.\ \textbf{119} (2017), 171301
[arXiv:1705.00192 [hep-th]];
J.~Diaz Dorronsoro, J.~J.~Halliwell, J.~B.~Hartle, T.~Hertog and O.~Janssen,
``Real no-boundary wave function in Lorentzian quantum cosmology,''
Phys.\ Rev.\ D \textbf{96} (2017), 043505
[arXiv:1705.05340 [gr-qc]];
J.~Feldbrugge, J.~L.~Lehners and N.~Turok,
``No rescue for the no boundary proposal: Pointers to the future of quantum cosmology,''
Phys.\ Rev.\ D \textbf{97} (2018), 023509
[arXiv:1708.05104 [hep-th]].

\bibitem{BrownMartinez}
J.~D.~Brown and E.~A.~Martinez,
``Lorentzian path integral for minisuperspace cosmology,''
Phys.\ Rev.\ D \textbf{42} (1990), 1931-1943.

\bibitem{Feldbrugge}
J.~Feldbrugge, J.~L.~Lehners and N.~Turok,
``Lorentzian quantum cosmology,''
Phys.\ Rev.\ D \textbf{95} (2017), 103508,
[arXiv:1703.02076 [hep-th]].


\bibitem{Regge} T.~Regge,
  ``General Relativity without Coordinates,''
Nuovo Cim.  {\bf 19} (1961) 558--571.

\bibitem{Sorkin2019} R.~D.~Sorkin,
``Lorentzian angles and trigonometry including lightlike vectors,''
[arXiv:1908.10022 [gr-qc]].

\bibitem{Perez} C.~Rovelli,
\textit{Quantum Gravity}
(Cambridge University Press, Cambridge, 2004);
A.~Perez,
  ``The Spin-Foam Approach to Quantum Gravity,''
  Living Rev.\ Rel.  {\bf 16} (2013) 3.


\bibitem{Collins1973}
P.~A.~Collins and R.~M.~Williams, ``Dynamics of the Friedmann Universe Using Regge Calculus'', Phys.\ Rev.\ D {\bf 7} (1973), 965-971.


\bibitem{Hartle1985}
J.~B.~Hartle,
``Simplicial minisuperspace I. General discussion,''
J.\ Math.\ Phys. \textbf{26} (1985), 804-814;
J.~B.~Hartle,
``Simplicial minisuperspace. II. Some classical solutions on simple triangulations,''
J.\ Math.\ Phys. \textbf{27} (1986), 287-295;
J.~B.~Hartle,
``Simplicial minisuperspace. III. Integration contours in a five-simplex model,''
J.\ Math.\ Phys. \textbf{30} (1989), 452-460.

\bibitem{Brewin}
L.~Brewin,
``Friedmann cosmologies via the Regge calculus,"
Class.\ Quant.\ Grav. \textbf{4} (1987), 889-928.


\bibitem{Liu2015}
R.~G.~Liu and R.~M.~Williams, ``Regge calculus models of the closed vacuum $\Lambda$--FLRW universe,'' Phys.\ Rev.\ D \textbf{93} (2016), 024032

\bibitem{Tsuda}
R.~Tsuda and T.~Fujiwara,
``Oscillating 4-Polytopal Universe in Regge Calculus,''
Prog.\ Theor.\ Exp.\ Phys. \textbf{2021} (2021) 083E01
[arXiv:2011.04120 [gr-qc]].



\bibitem{EffSF1} S.~K.~Asante, B.~Dittrich and H.~M.~Haggard,
``Effective Spin Foam Models for Four-Dimensional Quantum Gravity,''
Phys.\ Rev.\ Lett. \textbf{125} (2020), 231301
[arXiv:2004.07013 [gr-qc]].

\bibitem{EffSF2} S.~K.~Asante, B.~Dittrich and H.~M.~Haggard,
``Discrete gravity dynamics from effective spin foams,''
Class.\ Quant.\ Grav. \textbf{38} (2021), 145023
[arXiv:2011.14468 [gr-qc]].

\bibitem{EffSF3}
S.~K.~Asante, B.~Dittrich and J.~Padua-Arg\"uelles,
``Effective spin foam models for Lorentzian quantum gravity,''
Class.\ Quant.\ Grav. in press (2021),
[arXiv:2104.00485 [gr-qc]].


\bibitem{DiTucci}
A.~Di Tucci, J.~L.~Lehners and L.~Sberna,
``No-boundary prescriptions in Lorentzian quantum cosmology,''
Phys.\ Rev.\ D \textbf{100} (2019), 123543
[arXiv:1911.06701 [hep-th]].

\bibitem{GHY}
J.~W.~York, Jr.,
``Role of Conformal Three-Geometry in the Dynamics of Gravitation,''
Phys.\ Rev.\ Lett. \textbf{28} (1972), 1082-1085;
G.~W.~Gibbons and S.~W.~Hawking,
``Action integrals and partition functions in quantum gravity,''
Phys.\ Rev.\ D \textbf{15} (1977), 2752-2756.

\bibitem{Halliwell}
J.~J.~Halliwell,
``Derivation of the Wheeler--De Witt Equation from a path integral for minisuperspace models,''
Phys.\ Rev.\ D \textbf{38} (1988), 2468-2481.

\bibitem{BFV}
E.~S.~Fradkin and G.~A.~Vilkovisky,
``Quantization of relativistic systems with constraints,''
Phys.\ Lett.\ B \textbf{55} (1975), 224-226;
I.~A.~Batalin and G.~A.~Vilkovisky,
``Relativistic S-matrix of dynamical systems with boson and fermion constraints,''
Phys.\ Lett.\ B \textbf{69} (1977), 309-312.


\bibitem{NewRegge}
B.~Bahr and B.~Dittrich,
``Regge calculus from a new angle,''
New\ J.\ Phys. \textbf{12} (2010), 033010
[arXiv:0907.4325 [gr-qc]].

\bibitem{Improved}
B.~Bahr and B.~Dittrich,
``Improved and perfect actions in discrete gravity,''
Phys.\ Rev.\ D \textbf{80} (2009), 124030
[arXiv:0907.4323 [gr-qc]].

\bibitem{AreaAngle}
B.~Dittrich and S.~Speziale,
``Area-angle variables for general relativity,''
New\ J.\ Phys. \textbf{10} (2008), 083006
[arXiv:0802.0864 [gr-qc]].

\bibitem{Sorkin74} R.~Sorkin,
``Time-evolution problem in regge calculus,''
Phys.\ Rev.\ D \textbf{12} (1975), 385-396
[Erratum: Phys.\ Rev.\ D \textbf{23} (1981), 565-565]

\bibitem{Cheeger1984}
J.~Cheeger, W.~M\"uller and R.~Schrader, ``On the curvature of piecewise flat spaces,'' Commun.\ Math.\ Phys. \textbf{92} (1984), 405-454.

\bibitem{WilliamsReview}
R.~Loll,
``Discrete Approaches to Quantum Gravity in Four Dimensions,''
Living\ Rev.\ Rel. \textbf{1} (1998), 13
[arXiv:gr-qc/9805049 [gr-qc]];
T.~Regge and R.~M.~Williams,
``Discrete structures in gravity,''
J.\ Math.\ Phys. \textbf{41} (2000), 3964-3984
[arXiv:gr-qc/0012035 [gr-qc]];
H.~W.~Hamber,
{\it Quantum Gravitation: The Feynman Path Integral Approach}
(Springer, 2009).

\bibitem{Menotti}
P.~Menotti and P.~P.~Peirano,
``Diffeomorphism invariant measure for finite-dimensional geometries,''
Nucl.\ Phys.\ B \textbf{488} (1997), 719-734
[arXiv:hep-th/9607071 [hep-th]].

\bibitem{DittrichSteinhausM}
B.~Dittrich and S.~Steinhaus,
  ``Path integral measure and triangulation independence in discrete gravity,''
  Phys.\ Rev.\ D {\bf 85} (2012) 044032
  [arXiv:1110.6866 [gr-qc]];
   B.~Dittrich, W.~Kaminski and S.~Steinhaus,
  ``Discretization independence implies non-locality in 4D discrete quantum gravity,''
  Class.\ Quant.\ Grav.\  {\bf 31} (2014),  245009
  [arXiv:1404.5288 [gr-qc]];
     B.~Bahr and S.~Steinhaus,
  ``Numerical evidence for a phase transition in 4d spin foam quantum gravity,''
  Phys.\ Rev.\ Lett.\  {\bf 117} (2016),  141302
  [arXiv:1605.07649 [gr-qc]].


\bibitem{CDTReview} J.~Ambj{\o}rn, J.~Jurkiewicz and R.~Loll,
``Nonperturbative Lorentzian Path Integral for Gravity,''
Phys.\ Rev.\ Lett. \textbf{85} (2000), 924-927
[arXiv:hep-th/0002050 [hep-th]].

\bibitem{LollJordan} S.~Jordan and R.~Loll,
``Causal Dynamical Triangulations without preferred foliation,''
Phys.\ Lett.\ B \textbf{724} (2013), 155-159
[arXiv:1305.4582 [hep-th]].

\bibitem{OritiFeynman}
D.~Oriti,
``Feynman propagator for Spin Foam Quantum Gravity,''
Phys.\ Rev.\ Lett. \textbf{94} (2005), 111301
[arXiv:gr-qc/0410134 [gr-qc]].

\bibitem{satisfyconstraints}
J.~J.~Halliwell and J.~B.~Hartle,
``Wave functions constructed from an invariant sum over histories satisfy constraints,''
Phys.\ Rev.\ D \textbf{43} (1991), 1170-1194.

\bibitem{Coxeter} 
H.~S.~M.~Coxeter, {\it Regular Polytopes} (Dover Publications, 3rd revised edition, 1973).

\bibitem{600cell}
Wikipedia, ``600-cell, Visualization'', Wikimedia Foundation, 06 December 2021, https://en.wikipedia.org/wiki/600-cell\#Visualization.

\bibitem{120cell}
Wikipedia, ``120-cell, Projections'', Wikimedia Foundation, 06 December 2021, https://en.wikipedia.org/wiki/120-cell\#Projections.

\bibitem{DittrichSteinhaus13}
B.~Dittrich and S.~Steinhaus,
``Time evolution as refining, coarse graining and entangling,''
New\ J.\ Phys. \textbf{16} (2014), 123041
[arXiv:1311.7565 [gr-qc]].

\bibitem{DittrichHoehn}
B.~Dittrich and P.~A.~H\"ohn,
``Canonical simplicial gravity,''
Class.\ Quant.\ Grav. \textbf{29} (2012), 115009
[arXiv:1108.1974 [gr-qc]];
B.~Dittrich and P.~A.~H\"ohn,
``Constraint analysis for variational discrete systems,''
J.\ Math.\ Phys. \textbf{54} (2013), 093505
[arXiv:1303.4294 [math-ph]].

\bibitem{GozVid}
F.~Gozzini and F.~Vidotto,
``Primordial fluctuations from quantum gravity,''
Front.\ Astron.\ Space\ Sci. \textbf{7} (2021), 629466
[arXiv:1906.02211 [gr-qc]].

\bibitem{BarrettFoxon} J.~W.~Barrett and T.~J.~Foxon,
``Semiclassical limits of simplicial quantum gravity,''
Class.\ Quant.\ Grav. \textbf{11} (1994), 543-556
[arXiv:gr-qc/9310016 [gr-qc]].

\bibitem{HanLiu}
 M.~Han and H.~Liu, ``Analytic continuation of spinfoam models,''
Phys.\ Rev.\ D \textbf{105} (2022), 024012
[arXiv:2104.06902 [gr-qc]].

\bibitem{Engle} J.~Engle,
  ``Proposed proper Engle--Pereira--Rovelli--Livine vertex amplitude,''
  Phys.\ Rev.\ D {\bf 87} (2013),  084048
  [arXiv:1111.2865 [gr-qc]];
  J.~Engle,
  ``A spin-foam vertex amplitude with the correct semiclassical limit,''
  Phys.\ Lett.\ B {\bf 724} (2013) 333-337
  [arXiv:1201.2187 [gr-qc]].


\bibitem{DittrichDiff} 
 B.~Dittrich,
``How to construct diffeomorphism symmetry on the lattice,''
PoS \textbf{QGQGS2011} (2011), 012
[arXiv:1201.3840 [gr-qc]];
B.~Dittrich,
  ``The Continuum Limit of Loop Quantum Gravity: A Framework for Solving the Theory,''
  in Ashtekar A., Pullin J. (eds.), {\it Loop Quantum Gravity: The First 30 Years}  (World Scientific, 2017),
  [arXiv:1409.1450 [gr-qc]].

\bibitem{DittrichLoll2}
B.~Dittrich and R.~Loll,
``Counting a black hole in Lorentzian product triangulations,''
Class.\ Quant.\ Grav. \textbf{23} (2006), 3849-3878
[arXiv:gr-qc/0506035 [gr-qc]].

\bibitem{Mikov}
A.~Mikovic,
``Piecewise Flat Metrics and Quantum Gravity,''
[arXiv:2001.11439 [gr-qc]].

\bibitem{BahrDittrich09a} B.~Bahr and B.~Dittrich,
``(Broken) Gauge symmetries and constraints in Regge calculus,''
Class.\ Quant.\ Grav. \textbf{26} (2009), 225011
[arXiv:0905.1670 [gr-qc]];
B.~Bahr and B.~Dittrich,
``Breaking and Restoring of Diffeomorphism Symmetry in Discrete Gravity,''
AIP\ Conf.\ Proc. \textbf{1196} (2009), 10-17
[arXiv:0909.5688 [gr-qc]].

\bibitem{PerfectPI} B.~Bahr, B.~Dittrich and S.~Steinhaus,
``Perfect discretization of reparametrization invariant path integrals,''
Phys.\ Rev.\ D \textbf{83} (2011), 105026
[arXiv:1101.4775 [gr-qc]].

\bibitem{BrownReview}
J.~D.~Brown,
``Action functionals for relativistic perfect fluids,''
Class.\ Quant.\ Grav. \textbf{10} (1993), 1579-1606
[arXiv:gr-qc/9304026 [gr-qc]].

\bibitem{Brown1990}
J.~D.~Brown,
``Tunneling in perfect-fluid (minisuperspace) quantum cosmology,''
Phys.\ Rev.\ D \textbf{41} (1990), 1125-1141.

\bibitem{Garayetal}
L.~J.~Garay, J.~J.~Halliwell and G.~A.~Mena Marug\'an,
``Path-integral quantum cosmology: A class of exactly soluble scalar field minisuperspace models with exponential potentials,''
Phys. Rev. D \textbf{43} (1991), 2572-2589.

\bibitem{GielenTurok}
S.~Gielen and N.~Turok,
``Quantum propagation across cosmological singularities,''
Phys.\ Rev.\ D \textbf{95} (2017), 103510
[arXiv:1612.02792 [gr-qc]].

\bibitem{SFC1}
E.~Bianchi, C.~Rovelli and F.~Vidotto,
``Towards spinfoam cosmology,''
Phys.\ Rev.\ D \textbf{82} (2010), 084035
[arXiv:1003.3483 [gr-qc]].

\bibitem{SFC2}
E.~R.~Livine and M.~Mart{\'i}n-Benito,
``Classical setting and effective dynamics for spinfoam cosmology,''
Class.\ Quant.\ Grav. \textbf{30} (2013), 035006
[arXiv:1111.2867 [gr-qc]].

\bibitem{SFC3}
E.~Bianchi, T.~Krajewski, C.~Rovelli and F.~Vidotto,
``Cosmological constant in spinfoam cosmology,''
Phys.\ Rev.\ D \textbf{83} (2011), 104015
[arXiv:1101.4049 [gr-qc]].

\bibitem{SFC4}
F.~Hellmann,
``Expansions in spin foam cosmology,''
Phys.\ Rev.\ D \textbf{84} (2011), 103516
[arXiv:1105.1334 [gr-qc]].


\end{thebibliography}
\end{document}